\documentclass[useAMs,usenatbib,a4paper]{mn2e}
\usepackage{savesym}
\usepackage{graphicx}
\expandafter\let\csname equation*\endcsname\relax
  \expandafter\let\csname endequation*\endcsname\relax 
\usepackage{subfig}
\usepackage{amsmath}
\usepackage{amssymb}
\usepackage{verbatim}
\usepackage{array}
\usepackage{times}
\usepackage[total={17.8cm,24.0cm},centering]{geometry} 
\usepackage{rotating}

\newcommand{\be}{\begin{equation}}
\newcommand{\ee}{\end{equation}}
\newcommand{\eea}{\end{eqnarray}}
\newcommand{\bea}{\begin{eqnarray}}
\newcommand{\pd}{\partial}

\newcommand{\m}{\mathrm}

\title[Radiative losses at the base of black hole jets]{Using radiative energy losses to constrain the magnetisation and magnetic reconnection rate at the base of black hole jets}
\author[William J. Potter]{William J. Potter\thanks{E-mail:
will.potter@astro.ox.ac.uk (WJP)}
\\
Oxford Astrophysics. University of Oxford, Denys Wilkinson Building, Keble Road, Oxford, OX1 3RH, United Kingdom}
\begin{document}

\date{}

\pagerange{\pageref{firstpage}--\pageref{lastpage}} \pubyear{2015}

\maketitle

\label{firstpage}

\begin{abstract}
We calculate the severe radiative energy losses which occur at the base of black hole jets using a relativistic fluid jet model, including in-situ acceleration of non-thermal leptons by magnetic reconnection. Our results demonstrate that including a self-consistent treatment of radiative energy losses is necessary to perform accurate MHD simulations of powerful jets and that jet spectra calculated via post-processing are liable to vastly overestimate the amount of non-thermal emission. If no more than $95\%$ of the initial total jet power is radiated away by the plasma travels as it travels along the length of the jet, we can place a lower bound on the magnetisation of the jet plasma at the base of the jet. For typical powerful jets, we find that the plasma at the jet base is required to be highly magnetised, with at least 10,000 times more energy contained in magnetic fields than in non-thermal leptons. Using a simple power-law model of magnetic reconnection, motivated by simulations of collisionless reconnection, we determine the allowed range of the large-scale average reconnection rate along the jet, by restricting the total radiative energy losses incurred and the distance at which the jet first comes into equipartition. We calculate analytic expressions for the cumulative radiative energy losses due to synchrotron and inverse-Compton emission along jets, and derive analytic formulae for the constraint on the initial magnetisation.    
\end{abstract}


\section{Introduction}

Accreting black holes are observed to produce powerful extended relativistic plasma jets. It is thought that the energy supplied to these jets originates from the rotational energy of the black hole \citep{1977MNRAS.179..433B}. A spinning black hole induces the rotation of space-time in its vicinity which twists the nearby magnetic field that has been dragged in by the accreting plasma disc. The black hole's rotational energy is thereby converted into magnetic energy by stretching and twisting the magnetic field, increasing its strength. The resulting enhanced magnetic pressure gradient away from the hole is sufficient to accelerate the jet plasma to relativistic velocities as demonstrated by numerical simulations (see for example \citealt{2004ApJ...611..977M}, \citealt{2006ApJ...641..103H}, \citealt{2006MNRAS.368.1561M}, \citealt{2007MNRAS.380...51K} and \citealt{2010ApJ...711...50T}). Polarised synchrotron emission is observed along the jet and this requires a high energy non-thermal electron population to be present (\citealt{1992ApJ...398..454W}, \citealt{2000ApJ...541...66L} and \citealt{2010ApJ...710L.126M} etc.). 

In relativistic magnetohydrodynamic (MHD) simulations of jets the effect of radiative energy losses are neglected and the magnetisation or equipartition fraction at the base of the jet (the ratio of energy contained in the magnetic field to the energy contained in non-thermal particles) is set artificially, by either numerically imposing the initial magnetisation within the jet, or by setting a lower limit, or floor, for the mass density of the jet plasma. In this paper we demonstrate the importance of the severe radiative energy losses of non-thermal electrons close to the jet base in constraining the jet magnetisation. This initial magnetisation is very important because it determines: the terminal bulk Lorentz factor of the jet (e.g. \citealt{2009MNRAS.394.1182K}); the distance over which the jet accelerates before it reaches equipartition, stops accelerating and becomes conical (e.g. \citealt{2006MNRAS.368.1561M} and \citealt{2007MNRAS.380...51K}); and the susceptibility of the jet to the kink instability (which occurs if the jet has a large, dominant toroidal field, e.g. \citealt{2010MNRAS.402....7M}, \citealt{2015MNRAS.452.1089P} and \citealt{2016MNRAS.tmpL..44T}). Previous research into modelling the synchrotron and inverse-Compton emission from the non-thermal particles in the jet have generally focused on static, physically disconnected spherical emitting regions at large distances along the jet where the majority of the high energy emission is thought to originate (\citealt{1993ApJ...416..458D}, \citealt{1998ApJ...509..608T}, \citealt{2007A&A...476.1151T} etc.). In this paper we focus on the radiative energy losses which occur close to the base of the jet. By calculating the radiative energy losses using a relativistic fluid jet model which conserves energy-momentum, we show that these losses are, in fact, so severe that they can be used to place tight constraints on the initial magnetisation of the jet plasma. 

The paper starts with an introduction to our fluid jet model in section \ref{section2}. We set out our assumptions and derive the equations of energy conservation and particle number conservation allowing for a jet shape, bulk Lorentz factor and in-situ non-thermal particle acceleration which depend on distance. In section \ref{section3} we embark on preliminary calculations in order to demonstrate the severity of the radiative energy losses and justify our assumption that the jet plasma is composed of a magnetised non-thermal electron-positron pair-plasma (hereafter we use electron as short for electron-positron). In sections \ref{section4}-\ref{section6} we derive analytic expressions which determine the cumulative radiative energy losses of the jet plasma by synchrotron, synchrotron self-Compton (SSC) or inverse-Compton scattering of external radiation fields (referred to as external-Compton, or EC). Analytic and numerical solutions of the radiative energy loss equations are then used to place important constraints on the magnetisation of the jet base in both supermassive and stellar mass black hole systems, in section \ref{section7}, by requiring that a substantial fraction of the initial jet energy is retained to large distances and not radiated away. In section \ref{section8}, we then consider the effect of in-situ magnetic reconnection occurring along the jet, dissipating magnetic energy and replenishing the non-thermal electron population. We impose constraints on the total radiative energy losses sustainable and the distance at which the jet plasma reaches equipartition to determine a set of favourable parameters describing the large-scale average reconnection rate. The purpose of this work is to illustrate the physical importance of radiative energy losses on the dynamics of jets and show that a self-consistent implementation of radiative energy losses is a necessary component for realistic numerical simulations of jets.

\section{Fluid Jet Model}  \label{section2}

In this section we introduce the 1D relativistic fluid jet model which we will use to calculate the radiative energy losses of the jet plasma as it travels along the base of the jet (this model is based on the inhomogeneous jet emission model presented in \citealt{2012MNRAS.423..756P}, \citealt{2013MNRAS.429.1189P}, \citealt{2013MNRAS.431.1840P}, \citealt{2013MNRAS.436..304P} and \citealt{2015MNRAS.453.4070P}). The jet is allowed to have a variable shape, bulk Lorentz factor and equipartition fraction. Relativistic energy-momentum is conserved along the jet allowing for a detailed, self-consistent treatment of the effect of radiative energy losses, in-situ particle acceleration to the non-thermal electron population and the conversion of magnetic energy to bulk kinetic energy as the jet accelerates. We have chosen to use a 1D fluid model due to the computational expense of calculating the radiative energy losses to the electron population (this is the primary reason that a self-consistent calculation of radiative energy losses is neglected in 2D and 3D relativistic MHD simulations of jets).

As a first step we need to calculate the total energy density at the jet base, for a jet with initial total power $P_{\m{j}}$. Throughout the paper we use standard cylindrical coordinates where, $z$, is the length along the jet axis, $R$, is the cylindrical radius and $\phi$, is the azimuthal angle. We use primed coordinates for quantities measured in the plasma rest frame and unprimed for the lab frame (in which the black hole is at rest). To calculate the initial energy density of the jet plasma, consider the energy flux and volume flux of plasma passing through a surface whose normal is parallel to the jet axis and is located at a fixed spatial position in the lab frame at the base of the jet. The volume flux passing through this surface will be given by the cross-sectional area, $\pi R^{2}$, of the jet multiplied by the lab frame bulk velocity of the plasma, $v_{\m{bulk}}(z)=\beta_{\m{bulk}}(z)c$. The energy flux passing through this surface is simply the total jet power, $P_{\m{j}}$, so the energy density of the jet plasma measured in the lab frame is ratio of these two fluxes
\be 
U_{\m{tot}\,0}(z)=\frac{P_{\m{j}}}{\pi R^{2}(z) \beta_{\m{bulk}}(z)c}.  \label{Utot0}
\ee
It is usually assumed that the base of the jet is Poynting flux or electromagnetically dominated, i.e. the energy in magnetic field is much larger than in non-thermal particles. In this work we consider a leptonic jet model in which the jet plasma is composed of a magnetised non-thermal electron-positron pair plasma. We outline the arguments justifying this assumption in section \ref{section3.4}. We wish to calculate the evolution of the magnetisation or equipartition fraction of the plasma due to radiative losses and so we introduce the parameter $f_{\m{B}}$, which is the fraction of the total energy of the plasma contained in magnetic energy, $f_{\m{B}}(z)=U_{\m{B}}/U_{\m{tot}}$, where, $U_{\m{B}}$, is the magnetic energy density and $U_{\m{tot}}=U_{\m{B}}+U_{\m{e}\pm}$, the total energy density, with, $U_{\m{e}\pm}$, the non-thermal electron-positron energy density. The fractional magnetic energy is related to the equipartition fraction, or magnetisation, $\sigma$, which we define as, $\sigma=U_{\m{B}}/U_{\m{e}\pm}$, with, $f_{\m{B}}=\sigma/(1+\sigma)$. The initial magnetisation of the jet plasma is not known and is effectively an inputted parameter in most relativistic MHD simulations, however, the magnetisation can have a profound effect on the results of these simulations. In the simple case where no radiative losses are taken into account and the jet is not disrupted prematurely by instabilities, theoretical expectations and the results of MHD simulations are that the jet will accelerate until its magnetisation approaches unity and the majority of magnetic energy is converted into bulk kinetic energy (\citealt{2003ApJ...596.1080V}, \citealt{2006MNRAS.368.1561M} and \citealt{2007MNRAS.380...51K}). We can use this simple assumption to estimate the maximum bulk Lorentz factor of the jet plasma by a simple energy argument, provided that our jet plasma contains only cold, non-radiating particles and magnetic fields at the base. Consider a single blob of jet plasma with a lab frame energy, $E_{\m{tot}}$, initially composed of magnetic, $E_{\m{B}\,0}$, and cold particle components, $E_{\m{p}\,0}$
\be
E_{\m{tot}}=E_{\m{B}\,0}+E_{\m{p}\,0}=E_{\m{p}\,0}(\sigma_{0}+1), \label{est_gam}
\ee 
where in the last equality we have substituted the initial magnetisation, $\sigma_{0}$, at the base of the jet, $z=z_{0}$. Under the assumption of no energy losses, the total energy of the blob will be conserved and the maximum bulk Lorentz factor, $\gamma_{\m{max}}$ will occur when all of the initial magnetic energy has been converted into accelerating the cold particles to a high bulk Lorentz factor, until the jet reaches equipartition i.e. $\sigma(z)\approx 1$
\be
E_{\m{tot}}=2\gamma_{\m{max}}E_{\m{p}\,0}, \qquad \gamma_{\m{max}}=\frac{\sigma_{0}+1}{2},
\ee
where we have related the final energy of the particles moving with a bulk Lorentz factor, $\gamma_{\m{max}}$, to the initial total energy. This demonstrates how the initial equipartition fraction or magnetisation, $\sigma_{0}$, plays an important role in determining the terminal bulk Lorentz factor of the jet. In reality, it is artificial to assume that all of the magnetic energy is converted into bulk kinetic energy (in simulations magnetic acceleration ceases being efficient at roughly equipartition $\sigma=1$, \citealt{2006MNRAS.368.1561M} and \citealt{2009MNRAS.394.1182K}) and this neglects energy losses from radiation, magnetic reconnection and work done on the environment. It does, however, provide an estimate of the upper bound on the maximum terminal Lorentz factor and illustrates the importance of understanding the initial magnetisation. The simple estimate above illustrates why magnetisations of $10-20$ are typically assumed at the jet base for AGN jets \citep{2007MNRAS.380...51K}, since using these initial magnetisations simulated jets possess terminal bulk Lorentz factors typical of the observed superluminal speeds $1\gtrsim\gamma_{bulk}\gtrsim30$ (e.g. \citealt{2009A&A...494..527H}).

\subsection{The stress-energy tensor of the jet plasma}
At large distances where the jet plasma becomes relativistic we need to calculate the rest frame magnetic and particle energy densities. This has been calculated in \citealt{2013MNRAS.429.1189P} for a small scale homogeneous isotropic tangled magnetic field and isotropic particle velocity distribution in the rest frame, which can be described by a relativistic perfect fluid. The assumption of a homogeneous plasma is commonplace in the literature and is reasonable since we do not expect a large transverse pressure gradient, or equivalently energy density gradient. This is because these inhomogeneities would tend to be smoothed out by the pressure gradient after a few sound crossing timescales (since the jet remains in causal contact with itself throughout the accelerating parabolic base, \citealt{2008ApJ...679..990Z}). Under these assumptions the rest frame energy densities can be written as a relativistic perfect fluid
\be
T'^{\mu \nu}=\begin{pmatrix} \rho' & 0 &0 &0 \\ 0 & \frac{\rho'}{3} &0 &0\\ 0 &0 &\frac{\rho'}{3} &0\\ 0 &0 &0 &\frac{\rho'}{3}\end{pmatrix}, \label{perfectfluid}
\ee 
where, $\rho'$, is the total rest frame energy density and we have assumed a relativistic equation of state, $p'=\rho'/3$, appropriate for a high energy non-thermal electron population. To find the stress-energy tensor in the lab frame we Lorentz transform the rest frame tensor
\bea
&&T^{\mu \nu}(z)=\Lambda^{\mu}_{\,\, \sigma} T'^{\sigma \rho } \Lambda _{\,\,\rho}^{ \nu}=... \nonumber \\  &&\hspace{-0.5cm} \begin{pmatrix} \gamma_{\m{bulk}}^{2}(1+\beta^{2}_{\m{bulk}}/3) \rho' & (4/3)\gamma_{\m{bulk}}^{2}\beta_{\m{bulk}} \rho' &0 &0\\ (4/3)\gamma_{\m{bulk}}^{2}\beta_{\m{bulk}} \rho' &\gamma_{\m{bulk}}^{2}(1/3+\beta^{2}_{\m{bulk}}) \rho' &0 &0\\0&0&\frac{\rho'}{3}&0\\0&0&0&\frac{\rho'}{3}\end{pmatrix}, \nonumber \\ \label{LTEM}
\eea
\bea
\Lambda^{\mu}_{\,\,\nu} =&& \begin{pmatrix} \gamma_{\m{bulk}} & \gamma_{\m{bulk}}\beta_{\m{bulk}} &0 &0\\ \gamma_{\m{bulk}}\beta_{\m{bulk}} &\gamma_{\m{bulk}} &0 &0\\0&0&1&0\\0&0&0&1\end{pmatrix} ,
\eea
where the bulk Lorentz factor, $\gamma_{\m{bulk}}(z)$, and bulk velocity are functions of, $z$, the lab frame distance along the jet axis. We expect almost the entirety of the jet base to be moving with at least a mildly relativistic bulk velocity in which case we can make the approximation, $\beta_{\m{bulk}}\approx 1$ (we shall assume $\beta_{\m{bulk}}\sim 1$ throughout the rest of the paper, unless otherwise stated). The lab frame energy density is given by 
\be
\qquad T^{00}=U_{\m{tot}}\approx \frac{4}{3}\gamma_{\m{bulk}}^{2}U'_{\m{tot}}, \qquad T^{00}=T^{01}=T^{10}=T^{11},   \label{U'}
\ee

\subsection{Conservation of Energy-Momentum}
In this paper we wish to understand and calculate the effect of radiative energy losses and in-situ magnetic dissipation/reconnection (the conversion of magnetic energy into non-thermal particle energy) on the jet plasma. It is known that the radiative lifetimes of the emitting electrons are short compared to the time taken to travel along the jet and so in order for the jet to remain bright along its entire length some form of in-situ particle acceleration must occur (\citealt{2001A&A...373..447J} and \citealt{2005A&A...431..477J}). \citealt{2001A&A...373..447J} found no indications of the compact, bright structures at optical wavelengths which would usually be associated with strong shocks but instead observed approximately continuous emission along the jet. This is evidence for a continuous in-situ acceleration process in operation (\citealt{2001A&A...373..447J} and \citealt{2005A&A...431..477J}). Since the jet is believed to be launched via electromagnetic forces and its initial energy predominantly magnetic, we consider the possibility of conversion of this magnetic energy into accelerating non-thermal particles. Resistive dissipation of magnetic fields (magnetic reconnection) has been shown to result in the acceleration of non-thermal particles (see for example \citealt{2001ApJ...562L..63Z}, \citealt{2012ApJ...754L..33C} and \citealt{2015MNRAS.450..183S}) and so it seems a likely mechanism for the continuous in-situ acceleration process operating in jets. Whilst detailed small-scale simulations of reconnection have been performed which demonstrate non-thermal particle acceleration, we wish to consider the viability of such a process in a large-scale jet model which includes radiative energy losses. Our hope is to understand and constrain the average rate of reconnection occurring along the jet by considering the radiative energy losses to the jet plasma and the evolution of the magnetisation or equipartition fraction along the jet. These constraints can then be used to inform general relativistic magnetohydrodynamic (GRMHD) jet simulations and discriminate between different models of reconnection. 

We start with the equations of conservation of energy momentum and conservation of particle number flux
\be
\nabla_{\mu}T^{\mu \nu}=0, \qquad \nabla_{\mu}J^{\mu}_{\m{p}}=0 \label{cons_eq}
\ee 
where $T^{\mu\nu}$ is the stress-energy tensor, $J^{\mu}_{\m{p}}=n'_{\m{e}}U^{\mu}_{\m{bulk}}$ is the total particle number flux (electron and positron), $n'_{\m{e}}$ is the rest frame non-thermal particle number density and $U^{\mu}_{\m{bulk}}$ is the fluid 4-velocity. We assume that the distribution of particle and magnetic fields are homogeneous and isotropic on small-scales and so the stress-energy tensor will again take the form of a perfect fluid (\ref{perfectfluid}). We choose to decompose the stress-energy tensor into an electron-positron component, a magnetic component, a term representing the radiative energy losses and a term representing the in-situ particle acceleration (we shall focus on the possibility of in-situ particle acceleration by magnetic reconnection in this paper). Following \citealt{2015MNRAS.453.4070P} we can integrate these conservation equations over the invariant 4-volume of the jet and use the divergence theorem in 4-dimensions to convert this to a series of integrals over 3-dimensional hypersurfaces. Integrating the equation for energy-momentum we find
\bea
&&\hspace{-1.0cm}\int \nabla_{\mu}T^{\mu \nu} d^{4}V=\int T^{\mu \nu} d^{3}S^{\mu}= \int_{z}^{z+dz}\int_{0}^{2\pi}\int_{0}^{R} T^{0\nu} RdRd\phi dz \nonumber\\
&&\hspace{-1.0cm}+\int_{t}^{t+dt}\int_{0}^{2\pi}\int_{0}^{R} T^{1\nu} RdRd\phi dt +\int_{t}^{t+dt}\int_{z}^{z+dz}\int_{0}^{2\pi} T^{2\nu} Rd\phi dzdt+ \nonumber\\ &&\hspace{-1.0cm}+\int_{t}^{t+dt}\int_{z}^{z+dz}\int_{0}^{R} T^{3\nu} dRdzdt=0.\nonumber\\ \label{divE}
\eea
We assume that the jet is in a steady-state (i.e. time-independent) and so all terms containing an integral over time will be equal to zero. This gives us our equation for conservation of energy-momentum in the relativistic limit $\beta_{\m{bulk}}\approx1$
\be
\frac{\partial}{\partial z}\left(\frac{4}{3}\gamma_{\m{bulk}}(z)^{2}\pi R^{2}(z)\rho'(z)\right)=0.
\label{ce}
\ee
In the relativistic limit, which we assume in this work, the equations for energy and the $z$-component of momentum are both given by equation \ref{ce} above and so both can be simultaneously satisfied. Clearly, since we are primarily interested in ensuring conservation of energy along the jet and we are using a 1-dimensional model in order to calculate radiative energy losses in detail, we do not solve the full relativistic 3D MHD equations for the jet and, in general, a full 3D solution of the relativistic MHD equations is required if we wish to satisfy the energy and momentum conservation equations simultaneously. Decomposing the total energy density in (\ref{ce}) we find
\be
\frac{\partial}{\partial z}\left(\frac{4}{3}\gamma_{\m{bulk}}(z)^{2}\pi R^{2}(z)[U'_{\m{B}}+U'_{\m{e}\pm}+U'_{\m{rad}}]\right)=0.  \label{Urad}
\ee
where we have also included a term representing the effective cumulative radiative energy losses $U'_{\m{rad}}$. This term is artificial in the sense that real radiative energy losses will be emitted and eventually escape the jet, so will not remain as an energy density within the jet volume. It is the total radiated energy, $\pi R^{2}U_{\m{rad}}dz$, which has been emitted by a propagating cylindrical slab which is the true physically meaningful quantity we are interested in and we have simply chosen this to be represented artificially in the form of an energy density, in (\ref{Urad}), for mathematical convenience. We define the total remaining jet energy density as $U'_{\m{tot}}(z)$, and the fraction of the total initial plasma energy which remains in the jet and has not been radiated as $f_{\m{loss}}(z)$ (the fraction of total energy which has been radiated is $1-f_{\m{loss}}$). 
\be
U'_{\m{tot}}=U'_{\m{e}\pm}+U'_{\m{B}}, \qquad U'_{\m{tot}}=f_{\m{loss}}U'_{\m{tot}\,0},
\ee
 where $U_{\m{tot}\, 0}(z)$ is defined in (\ref{Utot0}) and the rest frame energy density is related to the lab frame energy density via (\ref{U'})
\be
U'_{\m{tot}}=\frac{3f_{\m{loss}}U_{\m{tot}\,0}}{4\gamma_{\m{bulk}}^{2}}=\frac{3f_{\m{loss}}P_{\m{j}}}{4\pi cR^{2}\gamma_{\m{bulk}}^{2}}.\label{U'tot}
\ee
Differentiating $U'_{\m{tot}}$ w.r.t. $z$ we find
\be
\frac{\partial U'_{\m{tot}}}{\partial z}=U'_{\m{tot}\,0}\frac{\partial f_{\m{loss}}}{\partial z}+f_{\m{loss}}\frac{\partial U'_{\m{tot}\,0}}{\partial z}.
\ee
Using the same method as in (\ref{divE}) we integrate the equation for conservation of particle number flux in a time-independent jet (\ref{cons_eq}) to find
\bea
\int \nabla_{\mu}J^{\mu}_{\m{p}} d^{4}V=\int J^{\mu}_{\m{p}} d^{3}S^{\mu}=\frac{\partial}{\partial z}[\pi R^{2}(z)n'_{\m{e}}(z)U^{0}_{\m{bulk}}(z)]=0,\nonumber \\
\frac{\partial}{\partial z}[\pi R^{2}(z)\gamma_{\m{bulk}}(z)cn'_{\m{e}}(z)]=0,
\label{cc}
\eea
where the zero component of the bulk 4-velocity is $U^{0}_{\m{bulk}}\simeq \gamma_{\m{bulk}} c$ and the result above has been calculated by writing out the four 3-dimensional hypersurface integrals as in equation \ref{divE} and setting the three time-dependent integrals to zero. 

Let us now consider the effect of particle number conservation on the evolution of the particle energy density in the case of a varying bulk Lorentz factor, but in the absence of any explicit reacceleration or radiative energy losses. Taking the equation for particle number conservation (\ref{cc}), multiplying by the bulk Lorentz factor, rearranging using Leibniz's rule and using $n'_{\m{e}}\propto U'_{\m{e}\pm}$ we find
\be 
2\gamma_{\m{bulk}}\frac{\partial(R^{2}\gamma_{\m{bulk}}U'_{\m{e}\pm})}{\partial z}=0,
\ee
\be
\frac{\partial (R^{2}\gamma_{\m{bulk}}^{2}U'_{\m{e}\pm})}{\partial z}=2R^{2}\gamma_{\m{bulk}}U'_{\m{e}\pm}\frac{\partial \gamma_{\m{bulk}}}{\partial z}.
\ee
This is the required equation for the evolution of the particle energy density in which the particle number flux remains constant, and in which an electron population remains unaltered by radiative losses or reacceleration as it travels along an accelerating jet. Comparison to equation \ref{Urad}, having set the radiative loss term to zero, shows that the second term must be equal to the rate of change of the magnetic energy density. This can be understood simply as the amount of magnetic energy expended in order to accelerate the bulk velocity of the particles in the jet (in the case of deceleration it can either be interpreted as the conversion of bulk particle energy into magnetic energy, heating or the acceleration of non-thermal electrons). This is consistent with the result in (\ref{est_gam}), that a jet can only accelerate whilst there is the magnetic energy available to do so. We also wish to consider in-situ particle acceleration by allowing the conversion of magnetic energy into accelerating additional non-thermal electrons via magnetic reconnection. We choose to include this in-situ particle acceleration via a term, $U_{\m{rec}}$, representing the energy density transferred from magnetic fields to non-thermal particles. Similarly to the radiative loss energy density term $U_{\m{rad}}$, the reconnection energy density is a mathematical convenience used to represent the transfer of energy from the magnetic to particle energy density and should not be interpreted as a real energy density which is present in the jet. The equation for the evolution of the particle energy density then becomes
\be
\frac{\partial (R^{2}\gamma_{\m{bulk}}^{2}U'_{\m{e}\pm})}{\partial z}=2\gamma_{\m{bulk}} R^{2}U'_{\m{e}\pm}\frac{\partial \gamma_{\m{bulk}}}{\partial z}+\frac{\partial (\gamma^{2}_{\m{bulk}}R^{2}[U'_{\m{rec}}-U'_{\m{rad}}])}{\partial z},
\label{particleeq}
\ee
where $U'_{\m{rad}}$ and $U'_{\m{rec}}$ are the terms corresponding to the radiation energy losses experienced by the particles and the particle energy injected by magnetic reconnection. Comparison with the equation \ref{Urad} allows us to write down the evolution equation for the magnetic energy density.
\be
\frac{\partial (R^{2}\gamma_{\m{bulk}}^{2}U'_{\m{B}})}{\partial z}=-2\gamma_{\m{bulk}}R^{2}U'_{\m{e}\pm}\frac{\partial \gamma_{\m{bulk}}}{\partial z}-\frac{\partial(R^{2} \gamma^{2}_{\m{bulk}}U'_{\m{rec}})}{\partial z}. \label{UB_evol}
\ee 
The first term on the right of the equation is the energy lost by the magnetic field by doing work accelerating the bulk velocity of the plasma, the second term represents the magnetic energy dissipated by magnetic reconnection. This equation can be simplified by introducing a new variable, $f_{\m{B}}(z)$, which we define as the fraction of total energy contained in magnetic fields
\be
U'_{\m{B}}=f_{\m{B}} U'_{\m{tot}}, \qquad f_{\m{B}}=\frac{U'_{\m{B}}}{U'_{\m{B}}+U'_{\m{e}\pm}}. \label{UBprimed}
\ee 
Substituting into (\ref{UB_evol}), using (\ref{U'tot}) and cancelling the constant terms we find
\be
\frac{\partial (f_{\m{B}} f_{\m{loss}})}{\partial z}=-2(1-f_{\m{B}})f_{\m{loss}} \frac{\partial\ln \gamma_{\m{bulk}}}{\partial z}-\frac{4\pi c}{3P_{\m{j}}}\frac{\partial(R^{2}\gamma_{\m{bulk}}^{2} U'_{\m{rec}})}{\partial z}.
\ee
 Finally, we wish to find an equation for the evolution of the fractional magnetic energy as a function of distance along the jet. Taking the equation above, expanding the derivatives with the chain rule and rearranging we find
\be
\frac{\partial f_{\m{B}}}{\partial z}=-f_{\m{B}}\frac{\partial \ln f_{\m{loss}}}{\partial z} -2(1-f_{\m{B}})\frac{\partial  \ln \gamma}{\partial z} -\frac{4\pi c}{3P_{\m{j}}f_{\m{loss}}}\frac{\partial(R^{2}\gamma_{\m{bulk}}^{2} U'_{\m{rec}})}{\partial z}, \label{magevol}
\ee 
where the terms on the RHS from left to right represent: the increase in the fractional magnetic energy due to radiative energy losses acting on the particle energy density, the change in fractional magnetic energy due to the acceleration of the bulk velocity of the jet by magnetic forces, and the decrease in fractional magnetic energy due to resistive dissipation of the magnetic fields. This equation for the evolution of the fractional magnetic energy can then be solved simultaneously with the equation for the fractional radiative energy losses determining, $f_{\m{loss}}$, which will be the subject of sections \ref{section4}-\ref{section6} and an equation determining the rate of magnetic reconnection along the jet, which will be the subject of section \ref{section8}. 

In order to calculate the power radiated by the plasma we will need to assume an electron energy distribution. We assume a simple power-law form for the distribution since this is generically produced in shock acceleration and magnetic reconnection (\citealt{1978MNRAS.182..147B}, \citealt{2011MNRAS.tmp.1506B} and \citealt{2001ApJ...562L..63Z}) and importantly will also allow us to calculate analytic expressions for the energy losses as a function of the jet parameters in section \ref{section6}. The number of electrons per unit volume, per unit energy measured in the plasma rest frame is 
\bea
n_{\m{e}}(E_{\m{e}})&=&AE_{\m{e}}^{-\alpha} \qquad \m{for}\,\,\,E_{\m{min}}>E_{\m{e}}>E_{\m{max}}, \nonumber \\
n_{\m{e}}(E_{\m{e}})&=&0 \qquad \qquad \,\, \m{else}.  \label{ne}
\eea 
It will be useful in later calculations to define the following moments of the electron energy distribution 
\be
\langle E_{\m{e}}^{p} \rangle=\int_{E_{\m{min}}}^{E_{\m{max}}}E_{\m{e}}^{p-\alpha}dE_{\m{e}},
\ee
\bea
\hspace{1cm}&\langle E_{\m{e}}^{p} \rangle=\dfrac{(E_{\m{max}}^{1+p-\alpha}-E_{\m{min}}^{1+p-\alpha})}{1+p-\alpha}& \m{for}\,\,\alpha\ne p,\nonumber \\
&\langle E_{\m{e}}^{p} \rangle=\ln \left(\dfrac{E_{\m{max}}}{E_{\m{min}}}\right)& \m{for}\,\,\alpha=1+p, \label{momentEe}
\eea
where $p$ is a constant. The results of this paper can be easily extended to a series of broken power laws if a more complicated electron distribution function is required. 

\section{Preliminary Calculations}\label{section3}
\subsection{Synchrotron and Inverse-Compton Emission}
We do not know the structure of the magnetic field in jets on small scales and so throughout this paper we shall make the standard assumption that in the rest frame of the plasma both the electron velocity distribution and the small-scale magnetic field are isotropic and homogeneous. The average synchrotron power emitted by an electron or positron of energy $E_{\m{e}}=\gamma m_{\m{e}}c^{2}$ moving in a tangled homogeneous isotropic magnetic field of strength $B$ is \citep{2011hea..book.....L}
\be
p_{\m{synch}}=\frac{4}{3}\sigma_{\m{T}}c\gamma^{2}\beta^{2}U_{\m{B}}, \qquad U_{\m{B}}=\frac{B^{2}}{2\mu_{0}}. \label{UB}
\ee 
where $\sigma_{\m{T}}$ is the Thomson cross section, the electron velocity is $v=\beta c$ and $B$ and $U_{\m{B}}$ are the magnetic field strength and magnetic energy density respectively. The synchrotron emission is sharply peaked around a critical frequency $\nu_{c}$ given by \citep{2011hea..book.....L}
\be
\nu_{c}=\frac{3\gamma^{2}eB}{4\pi m_{\m{e}}}, \qquad \epsilon\nu_{c}=E_{\m{e}}^{2}, \qquad \epsilon=\frac{4\pi m_{\m{e}}^{3}c^{4}}{3eB}. \label{nuc}
\ee 
The inverse-Compton emitted power of an electron in the case of an isotropic electron and photon velocity distribution is given by \citep{2011hea..book.....L}
\be
p_{IC}=\frac{4}{3}\sigma_{\m{T}}c\gamma^{2}\beta^{2}U_{\gamma}, \label{pIC}
\ee
Where $U_\gamma$ is the local photon energy density. In the case where the photon energy becomes large compared to the electron rest mass energy in the rest frame of the electron, $\gamma^{2}E_{\gamma}\gtrsim E_{\m{e}}$, the electron recoil becomes significant and it is no longer appropriate to use the Thomson cross-section and the full Klein-Nishina cross-section should be used here instead, see \cite{1970RvMP...42..237B}. 

\subsection{Estimating the Synchrotron Radiative Lifetime at the Base of the Jet} \label{section3.3}

Let us now briefly demonstrate the importance of synchrotron radiative losses to the non-thermal electron-positron plasma at the base of a jet, in order to motivate the more detailed calculations in later sections. For an electron with Lorentz factor $\gamma$, we wish to calculate the energy lost via synchrotron emission as a function of the distance travelled. To estimate if these losses are significant, we calculate the characteristic radiative lifetime of the electrons:
\be
t_{\m{synch}}=\frac{E_{\m{e}}}{p_{\m{synch}}}=\frac{3m_{\m{e}}c}{4\sigma_{\m{T}}c\beta^{2}\gamma U_{\m{B}}}.
\ee
Calculating $U_{\m{B}}$ using equation \ref{Utot0} and substituting we find
\be
t_{\m{synch}}=\frac{3\pi m_{\m{e}}c^{2}R^{2}}{4\sigma_{\m{T}}\beta\gamma^{2}f_{\m{B}}f_{\m{loss}} P_{\m{j}}}.
\ee
It is more useful to calculate the characteristic energy loss distance, $d_{\m{synch}}\approx ct_{\m{synch}}$, along the jet over which the electrons can propagate before losing a significant fraction of their initial energy. 
\be
\frac{d_{\m{synch}}}{R}=\frac{3\pi m_{\m{e}}c^{3}R}{4\sigma_{\m{T}}\beta\gamma^{2}f_{\m{B}}f_{\m{loss}} P_{\m{j}}}.
\ee
Writing this equation in terms of dimensionless quantities: the black hole Mass in units of solar mass $M^{*}=M/M_{\odot}$,  the jet power in units of Eddington luminosity $f_{\m{Edd}}=P_{\m{j}}/L_{\m{Edd}}$ and the cylindrical radius in units of Schwarzschild radii $R^{*}=R/r_{\m{s}}$
\be
\frac{d_{\m{synch}}}{R}=2.0\times10^{-4} \frac{R^{*}}{\beta \gamma^{2} f_{\m{B}} f_{\m{loss}}f_{\m{Edd}}} \label{dsynch}
\ee 
Since we expect $\beta_{\m{bulk}}\sim \beta \sim 1$ and $f_{\m{B}}\approx f_{\m{loss}} \approx 1$ (the jet is magnetically dominated at the base and has not yet suffered severe radiative energy losses), it is clear that even mildly relativistic electrons will cool very rapidly by emitting radiation at the base of the jet. This means that in the absence of any electron reacceleration the energy density contained in non-thermal particles at the base of the jet will rapidly decrease due to radiation losses and the jet plasma will become almost entirely magnetised. This means that simulations in which non-thermal emission from electrons are calculated via post-processing (added after the completion of the simulation) could easily overestimate the power of high energy synchrotron emission by large factors of tens to thousands. This is significant because it demonstrates that a continuous reacceleration process must be occurring throughout the base of the jet in order to produce the continuous radio emission observed (e.g. \citealt{2012ApJ...745L..28A} and \citealt{2013ApJ...775...70H}). We will include the effects of reacceleration due to magnetic reconnection in section \ref{section8}. 

\subsection{Estimating a lower bound on the maximum electron energy from radio observations}
\begin{table*}
\centering
\begin{tabular}{| c | c | c |}
\hline
Frequency (Ghz) & Radius ($r_{\m{s}}$) & $E_e/m_{\m{e}}c^{2}$  \\ \hline 
230 & 2.3 & 45 \\ \hline
86.4 & 10 & 21 \\ \hline
43.2& 15 & 18 \\ \hline
23.8 & 30 & 19 \\ \hline
15.2 & 40 & 17 \\ \hline
8.4 & 60 & 16 \\ \hline
5.0 & 80 & 14 \\ \hline
2.3 & 100 & 11 \\ \hline
\end{tabular}
\caption{Estimates of the electron energies required to produce the observed synchrotron emission at different distances along the jet in M87 as measured by \citealt{2013ApJ...775...70H}. These estimates form a lower bound on the maximum electron energy and show that relativistic non-thermal electrons are present throughout the base of the jet (see section \ref{section3.3} for further discussion).  }
\label{Table1}
\end{table*}

So far we have not made any assumptions about the parameters describing our power-law electron energy spectrum. One of the most crucial parameters in determining the radiative losses to the population will be the maximum electron energy, $E_{\m{max}}$, because the power radiated by an individual electron is proportional to at least the square of the electron energy in synchrotron or EC (or roughly the third power for SSC). We shall now estimate a lower bound on the maximum electron energy present at the base of the jet by calculating the electron energy required to produce the observed radio emission from M87 at a given distance and frequency. The magnetic field strength is given by (\ref{UBprimed})
\be
B'\approx \left(\frac{3\mu_{0}f_{\m{B}}f_{\m{loss}} P_{\m{j}}}{2\pi R^{2}c\gamma^{2}_{\m{bulk}}}\right)^{1/2}. \label{est_B}
\ee 
Radio VLBI observations tracking the motion of bright components of the plasma suggest that the velocities in most of the base region are only mildly relativistic or sub-relativistic \citep{2014ApJ...781L...2A} and so we shall take the Lorentz transformation factor (\ref{LTEM}) to be $(1+\beta^{2}_{\m{bulk}}/3)\gamma^{2}\approx 1$. It is worth pointing out that since larger bulk Lorentz factors will result in a smaller rest frame magnetic field strength and because the emitted synchrotron frequency $\nu'\propto E_{\m{e}}^{2}B'$, our assumption of a mildly relativistic bulk Lorentz factor will tend to underestimate the energy of electrons emitting a given observed frequency. Using equation \ref{nuc} we estimate the electron energy responsible for the observed emission at a frequency $\nu$. We assume that the jet is highly magnetised at the base $f_{\m{B}} \approx 1$, this is required in order for the jet material to be accelerated to relativistic terminal bulk Lorentz factors (see for example \citealt{2009MNRAS.394.1182K}) and we assume that the base is not sufficiently relativistic for the Doppler shift to significantly change the observed frequency of the photons compared to the rest frame emitted frequency. 
\be
E_{\m{e}}\approx\left(\frac{4\pi m_{\m{e}}^{3}c^{4} \nu_{\m{obs}}}{3eB'}\right)^{1/2}, \label{M87Ee}
\ee 
In M87 synchrotron emission has been observed at the base of the jet at frequencies of 230Ghz ($\lambda\approx$ 1.3mm) and originating from a region of size $2R=5.5r_{\m{s}}$, assuming $M\approx 6.2\times10^{9}M_{\odot}$ and an estimated jet power $3\times 10^{37}$W \citep{2012Sci...338..355D}. If we make the assumption of a mildly relativistic, highly magnetised jet plasma at the base and use equation \ref{est_B}, we estimate $B=0.0027$T, this is comparable to the estimate of $0.0050-0.0124$T by \cite{2015ApJ...803...30K}, who used a similar method to estimate the magnetic field strength but assumed a bulk velocity factor of $\beta_{\m{bulk}}=1/3$. Using equation \ref{M87Ee} we estimate the Lorentz factor of the electrons responsible for the observed radio emission at 230Ghz originating from 2R=$5.5r_{\m{s}}$, in M87, is $\gamma\approx 45$. This provides an estimate of the lower bound of maximum electron energies present at the base of the jet. However, it is important to emphasise that this does not mean that higher energy electrons emitting at higher frequencies are not also present at the base. This is because higher frequency emission, if present, is not yet spatially resolvable on these length scales, with current instruments. 

The jet in M87 has also been observed at a range of lower radio frequencies with estimates of the radius of the emitting region (\citealt{2013ApJ...775...70H}, note that in this work the black hole mass of M87 was taken to be $M_{BH}=6\times 10^{9}M_{\odot}$).  Using (\ref{est_B}) to calculate the magnetic field strength we can estimate the electron energy corresponding to these observations. In Table \ref{Table1} we show our estimates of the lower bound to the maximum electron energies present at different distances along the jet. These effectively form estimates of the lower bound of the maximum electron energy out to a distance of $z\approx 400r_{\m{s}}$. The important point to note is that a substantial number of relativistic electrons are observed throughout the base. We would suggest that since the presence of any non-thermal electrons implies that a non-thermal acceleration process is active in these regions (due to the short cooling lengths, see equation \ref{dsynch}), it is likely that higher energy electrons with energies comparable to those seen in blazar jets (GeV$>E_{\m{max}}>$TeV) are also likely to be present alongside the lower energy radio emitting electrons.  

\subsection{Initial electron distribution from spark gap} \label{section3.4}

In the previous subsection we demonstrated that relativistic electrons are required to exist throughout the base of jets in order to be compatible with high-resolution radio observations. In this subsection we shall outline arguments justifying the assumption that the base of the jet plasma should be composed of a dominant magnetic field and a relativistic electron-positron plasma. These arguments were originally presented by \citealt{1969ApJ...157..869G} and \citealt{1977MNRAS.179..433B} to justify the assumption of force-free magnetospheres surrounding pulsars and black holes respectively. In force-free dynamics (where the electromagnetic forces dominate and overwhelm the inertia of the fluid), which is thought to be applicable to black hole magnetospheres, the force-free equation usually adopted is 
\be
F^{\mu \nu}J_{\mu}=0,
\ee
where $F^{\mu \nu}$ is the electromagnetic field tensor and $J_{\nu}$ the 4-current \citep{1977MNRAS.179..433B}. The generalised Ohm's law in the case of a perfect conductor is
\be
{\bf E}+{\bf v}\times {\bf B}=\eta{\bf J}\approx 0, \qquad  {\bf E}=-{\bf v}\times {\bf B}. \label{Ohm}
\ee 
where $\eta$ is the usual electrical resistivity. In the force-free magnetosphere it is usually assumed that the plasma has a large conductivity, $\eta\sim0$, such that the electric field measured in the rest frame of the fluid is negligible compared to the magnetic field. In order for this assumption to be satisfied there is a minimum charge density required to screen/short-circuit the electric field. Using Maxwell's equations this is approximately given by the Goldreich-Julian charge density $\rho_{\m{GJ}}$ \citep{1969ApJ...157..869G}
\be
\nabla.\bf{E}=\frac{ \rho_{\m{GJ}}}{\epsilon_{0}},
\ee
\be
\rho_{\m{GJ}}\approx-\frac{\epsilon_{0}}{\delta}{\bf v}\times {\bf B},
\ee
where we have used (\ref{Ohm}) to substitute the magnetic field, ${\bf B}$, and fluid velocity, ${\bf v}$, for the electric field $E$ and we have assumed that the magnetic field changes significantly over a length-scale, $\delta$. Clearly, due to the immense gravitational field close to the black hole, it is likely to be difficult to provide a source of charged particles to the base of the jets through direct supply from the accretion disc alone, and so it is envisaged that, much like in a pulsar magnetosphere, a spark gap will form and alleviate this charge shortage. In the absence of sufficient charged particles a strong induced electric field will be present of order $\sim {\bf v}\times {\bf B}$. In \citealt{1977MNRAS.179..433B} and more recent work (\citealt{1998ApJ...497..563H} and \citealt{2015ApJ...809...97B}) the authors have estimated that such large electric fields can be created close to the black hole that they will lead to the strong acceleration of any electrons present to such high energies that they to emit a sufficient number of high energy photons to produce a pair cascade of electrons and positrons. This mechanism will continue to increase the charge density until there are sufficient charge carriers to screen the strong electric fields, this is referred to as a {\lq}spark gap{\rq}. This mechanism then justifies the treatment of the highly magnetised region using force-free electrodynamics and suggests the jet will, initially, be primarily composed of a non-thermal electron-positron plasma.

The Lorentz factor to which these electron-positron pairs are accelerated can be estimated by balancing the accelerating electric force against the radiative drag from emitting synchrotron, inverse-Compton and curvature radiation. These Lorentz factors have been calculated by \citealt{2015ApJ...809...97B}, for example, to be highly relativistic with maximum electron Lorentz factors $\gamma\sim 10^{9}$ estimated for M87. Whilst this estimate of the electron Lorentz factors may possess large uncertainties due to difficulties in precisely determining the energy density of the radiation field and structure of the magnetic field close to the black hole, it is clear that the electron-positron pairs produced will be relativistic and non-thermal (these relativistic energies will also help to prevent rapid re-annihilation of the pairs and emission of 511keV photons). Whether or not a proton or hadronic component exists alongside the non-thermal pair plasma is as yet unclear, although if such a hadronic component exists it seems to carry a smaller fraction of the total energy than the electron-positron component (see for example \citealt{1980Natur.288..149K}, \citealt{1996MNRAS.283..873R}, \citealt{1998Natur.395..457W}, \citealt{1998MNRAS.293..288C}, \citealt{2000ApJ...545..100H} and \citealt{2016MNRAS.457.1124K}). 

A small component of energy contained in cold protons would not significantly alter the results of this paper because we have defined the jet magnetisation as the ratio of magnetic to non-thermal lepton energies $U_{B}/U_{e\,\pm}$, however, clearly the inclusion of an additional non-radiating particle component would decrease the total jet magnetisation defined in terms of the ratio of magnetic to total particle energies. Addition of a small cold particle component would also slightly increase the amount of magnetic energy required in accelerating the jet to relativistic bulk velocities. GRMHD simulations of the jet base show the region to be essentially free of matter due, in part, to the effective centrifugal barrier as you approach the spin axis of the black hole (e.g. \cite{2006ApJ...641..103H} and \citealt{2006MNRAS.368.1561M}). This supports the idea of a spark gap or other method of photon-photon pair production as the means of providing charged particles at the jet base and suggests that introduction of protons to the magnetised jet occurs via gradual entrainment at larger distances from the black hole. These arguments justify our assumption of a magnetically dominated non-thermal electron positron pair plasma at the base of the jet. 

\section{Radiative Energy Losses} \label{section4}

We shall now turn our attention to calculating the radiative energy losses due to optically thick and thin synchrotron emission, synchrotron self-Compton emission and inverse-Compton scattering of external photon fields in our 1D fluid jet. 
\subsection{Synchrotron emission}

The synchrotron emissivity of a homogeneous isotropic relativistic electron plasma ($\beta\approx 1$), measured in the rest frame is
\be
j_{\m{E}}(E_{\m{e}})=n_{\m{e}}(E_{\m{e}})p_{\m{synch}}(E_{\m{e}})=\frac{4\sigma_{\m{T}}AE_{\m{e}}^{2-\alpha}U'_{\m{B}}}{3m_{\m{e}}^{2}c^{3}},\label{jE}
\ee
where we have used equations \ref{ne} and \ref{UB}. We wish to express the emissivity in terms of the input parameters for the jet using the following equations
\be
A=\frac{U'_{\m{e}\pm}}{\langle E_{\m{e}}\rangle}, \qquad U_{\m{tot}}=4\gamma_{\m{bulk}}^{2}U_{\m{tot}}'/3,
\ee
\be
U_{\m{tot}}=U_{\m{B}}+U_{\m{e}\pm}, \qquad U_{\m{B}}=f_{\m{B}} U_{\m{tot}}, \qquad U_{\m{e}\pm}=U_{\m{tot}}(1-f_{\m{B}}),  \label{UB2}
\ee 
\be
U'_{\m{B}}U'_{\m{e}\pm}=9\frac{U^{2}_{\m{tot}}f_{\m{B}}(1-f_{\m{B}})}{16\gamma_{\m{bulk}}^{4}}, \qquad U_{\m{tot}}=\frac{P_{\m{j}}f_{\m{loss}}}{\pi R^{2}c} \label{UB3}
\ee 
where $\langle E_{\m{e}}\rangle$ is defined in (\ref{momentEe}), we have assumed $\beta_{\m{bulk}}\approx 1$ and we use the fractional energy loss function, $f_{\m{loss}}$, which is the fraction of remaining energy in the jet plasma compared to its initial energy, as a function of distance along the jet i.e. $f_{\m{loss}}(z)=U_{\m{tot}}(z)/U_{\m{tot}\,0}(z)$. Substituting these expressions into equation \ref{jE}
\be
j_{\m{E}}=\frac{\sigma_{\m{T}}f_{\m{B}}(1-f_{\m{B}})(f_{\m{loss}}P_{\m{j}})^{2}E_{\m{e}}^{2-\alpha}}{m_{\m{e}}^{2}c^{5}\langle E_{\m{e}}\rangle\pi^{2} R^{4}\gamma_{\m{bulk}}^{4}}.
\ee
When calculating the synchrotron self-absorption it can be more useful to write the emissivity in terms of emitted synchrotron frequency, $\nu$, in which case we can convert from electron energy, $E_{\m{e}}$, to $\nu$ and $\epsilon$, using (\ref{nuc})
\be
j_{\nu}=j_{\m{E}}\frac{dE_{\m{e}}}{d\nu}=\frac{1}{2}\epsilon^{1/2}\nu^{-1/2}j_{\m{E}},
\ee
\be
j_{\nu}=\frac{\sigma_{\m{T}}f_{\m{B}}(1-f_{\m{B}})(f_{\m{loss}}P_{\m{j}})^{2}\epsilon^{(3-\alpha)/2}\nu^{(1-\alpha)/2}}{2m_{\m{e}}^{2}c^{5}\langle E_{\m{e}}\rangle \pi^{2} R^{4}\gamma_{\m{bulk}}^{4}}.
\ee
Our approximate criteria for the plasma becoming optically thick is determined by the condition that the brightness temperature of the synchrotron emission cannot exceed that of a blackbody with the emitting electron temperature $(\gamma-1)m_{\m{e}}c^{2}\approx E_{\m{e}}=k_{b}T$ \citep{2011hea..book.....L}
\be
j_{\nu}R\gtrsim \frac{4\pi\epsilon^{1/2}\nu^{5/2}}{c^{2}}, \label{SSAcriterion}
\ee 
in this self-absorbed regime the emissivity is then limited to the equivalent black-body emission in the Rayleigh-Jeans limit
\be
j^{\m{SSA}}_{\nu}\approx\frac{4\pi \epsilon^{1/2}\nu^{5/2}}{Rc^{2}}.
\ee
The synchrotron emission will be self-absorbed for frequencies $\nu\lesssim\nu_{\m{SSA}}$ and become optically thin for frequencies $\nu\gtrsim\nu_{\m{SSA}}$. We can calculate the maximum self-absorbed frequency $\nu_{\m{SSA}}$ by setting the two sides of equation \ref{SSAcriterion} to be equal
\bea
\hspace{0.5cm}\nu_{\m{SSA}}&\approx&\left[\frac{3\sigma_{\m{T}}f_{\m{B}}(1-f_{\m{B}})f_{\m{loss}}^{2}P_{\m{j}}^{2}\epsilon^{(2-\alpha)/2}}{32\pi^{3}m_{\m{e}}^{2}c^{3}\langle E_{\m{e}}\rangle R^{3}\gamma_{\m{bulk}}^{4}}\right]^{\dfrac{2}{4-\alpha}},\nonumber \\ \nonumber\\
\epsilon&=&\left[\frac{4\pi m_{\m{e}}^{3}c^{4}}{3e}\left(\frac{2\pi R^{2}\gamma_{\m{bulk}}^{2}c}{3\mu_{0}f_{\m{B}}f_{\m{loss}}P_{\m{j}}}\right)^{1/2}\right]^{\dfrac{2-\alpha}{2}}. \nonumber \\
\eea
We shall outline in section \ref{section6.2} why SSA is unlikely to be an important contribution to the radiative energy losses and so it does not matter that here we use an analytic approximation to the full synchrotron opacity (\ref{SSAcriterion}). Let us assume that the form of the particle distribution remains constant along the jet and allow for a decrease in the total energy density of particles by changing the particle number normalisation as energy losses occur, $A\propto U'_{\m{e}\pm}\propto (1-f_{\m{B}})P_{\m{j}}f_{\m{loss}}$. Whilst using a fixed form for the electron energy spectrum is slightly contrived, the physics governing the acceleration of particles in jets is not yet known in any detail. The evolution of the electron energy distribution at the base of the jet is also poorly constrained by observations and so the simple power-law form (\ref{ne}) is a reasonable choice for the purpose of our calculations. Also, because the radiated power is usually dominated by the highest energy electrons present, our results will be largely insensitive to the other precise details of the electron distribution. 

The power emitted by synchrotron radiation from a cylindrical slab of unit width in the $z$ direction, measured in the rest frame is then given by 
\be
P'_{\m{synch}}(z)=\int_{\nu_{\m{min}}}^{\nu_{\m{SSA}}}  \pi R^{2}j_{\nu}^{\m{SSA}}d\nu+\int_{\nu_{\m{SSA}}}^{\nu_{\m{max}}} \pi R^{2}j_{\nu}d\nu.
\ee
If the synchrotron losses are dominated by the optically thin contribution this becomes
\be
P'_{\m{synch}}(z)=\frac{3\sigma_{\m{T}}\langle E_{\m{e}}^{2}\rangle f_{\m{B}}(1-f_{\m{B}})(f_{\m{loss}}P_{\m{j}})^{2}}{4\pi m_{\m{e}}^{2}c^{5}\langle E_{\m{e}} \rangle R^{2}\gamma_{\m{bulk}}^{4}}. \label{P'synch}
\ee
where $\langle E_{\m{e}}^{2}\rangle$ is defined in equation \ref{momentEe}.
\subsection{Synchrotron Self-Compton}

We shall now calculate the synchrotron self-Compton (SSC) emission for the power-law electron distribution given by (\ref{ne}). In order to obtain analytic expressions we shall assume that the scattering is in the Thomson regime ($\gamma^{2}E_{\gamma}<<E_{\m{e}}$) and the distribution of electron and photon velocities are isotropic in the rest frame of the plasma. To calculate the emitted power we need the photon energy density contained in synchrotron photons measured the rest frame, as a function of distance along the jet. We calculate the synchrotron photon energy density following \citealt{2012MNRAS.423..756P}. The optically thick synchrotron emission is assumed to have a photon energy distribution, which, at a given frequency is equal to the blackbody distribution in the Rayleigh-Jeans limit at the electron temperature $U_{\nu}=4\pi B_{\nu}/c$.
\be
U'^{\m{SSC}}_{\gamma}=\int_{\nu_{\m{min}}}^{\nu_{\m{SSA}}}U_{\nu}d\nu=\frac{16\pi\epsilon^{1/2}}{7c^{3}}(\nu_{\m{SSA}}^{7/2}-\nu^{7/2}).
\ee
In the optically thin regime the emitted photon energy density in a thin annulus surrounding the jet is a good approximation to the average density i.e.
\be
U'^{\m{SSC}}_{\gamma}=\frac{P'_{\m{synch}} dz'}{2\pi R dz' c}.  \label{USSC}
\ee 
The power emitted in the rest frame of the plasma by a cylindrical slab of unit width in the $z$-direction is then given by equations \ref{ne} and \ref{pIC}
\bea
&&\hspace{-0.5cm}P'_{\m{SSC}}=\int_{E_{\m{min}}}^{E_{\m{max}}}\pi R^{2}n_{\m{e}}p_{IC}dE_{\m{e}}=...\nonumber\\
&&\hspace{2.4cm}...\frac{\sigma_{\m{T}}\langle E_{\m{e}}^{2}\rangle(1-f_{\m{B}})f_{\m{loss}}P_{\m{j}}}{m_{\m{e}}^{2}c^{4}\langle E_{\m{e}}\rangle\gamma_{\m{bulk}}^{2}}U'^{\m{SSC}}_{\gamma}\label{P'SSC}
\eea

\subsection{External Compton}

Finally we consider the power radiated by inverse-Compton scattering external photons (external-Compton or EC). In this case we shall assume that the external photon field is isotropically distributed in the lab frame and can be described by a power law form (for a detailed calculation of the external photon fields in AGN relevant for jet emission see for example \citealt{2013MNRAS.429.1189P})
\be
U_{\gamma}^{\m{EC}}=U_{\gamma0}^{\m{EC}} \left(\frac{z}{r_{\m{s}}}\right)^{a}.
\ee
If the form is instead well-described by a series of different power laws, the power radiated can be calculated by stitching together the analytic solutions in section \ref{section6.4}. We wish to calculate the photon energy density in the rest frame of the jet, which requires Doppler-boosting the photon distribution. This corresponds to Lorentz transforming the energy density using the method in (\ref{LTEM}), which results in an average energy increase in the photon energy density of $\sim 4\gamma_{\m{bulk}}^{2}/3$. 
\be
U'^{\m{EC}}_{\gamma}=\frac{4}{3}\gamma_{\m{bulk}}^{2}U^{\m{EC}}_{\gamma}.
\ee
The photon distribution will no longer be isotropic in the plasma rest frame but will be preferentially beamed along the jet axis in the direction of the black hole with a characteristic opening angle $\theta\approx 1/\gamma_{\m{bulk}}$. Since the electron velocity distribution is assumed to be isotropic, this beaming does not change the total power emitted from the isotropic photon case, however, it does mean that the inverse-Compton emission is anisotropically emitted in the rest frame and preferentially beamed along the jet axis (but away from the black hole since head-on electron photon collisions transfer the maximum energy from the electron to the photon).

With these considerations it is easy to calculate the power emitted by EC in the plasma rest frame by a cylindrical slab of unit width using equation \ref{pIC} (or simply by substituting $U'^{\m{EC}}_{\gamma}$ for $U'^{\m{SSC}}_{\gamma}$ in equation \ref{P'SSC})
\be
P'_{\m{EC}}=\frac{\sigma_{\m{T}}\langle E_{\m{e}}^{2}\rangle(1-f_{\m{B}})f_{\m{loss}}P_{\m{j}}}{m_{\m{e}}^{2}c^{4}\langle E_{\m{e}}\rangle\gamma_{\m{bulk}}^{2}}U'^{\m{EC}}_{\gamma}.\label{P'EC}
\ee  
\section{Radiative Energy Loss Equation}\label{section5}
Now that we have calculated the power radiated by the dominant non-thermal processes we can calculate the effect of these energy losses on the total energy of the plasma as it travels along the jet. The energy emitted by a cylindrical slab of infinitesimal width $dw$ (measured parallel to the jet axis $z$) travelling a distance $dz$ in time $dt$ along the jet in the lab frame will correspond to a Lorentz transformed slab of width $dw'$, which is stationary and emits for a time $dt'$ in the rest frame with
\be
dw'=\gamma_{\m{bulk}}dw, \qquad dt=\gamma_{\m{bulk}}dt', \qquad dz=\beta_{\m{bulk}}cdt, \label{LT2}
\ee
where the Lorentz contraction of the jet length and slab width occur in the opposite sense because in the lab frame the jet structure is stationary whilst the plasma itself is moving. In the lab frame the energy contained in a cylindrical slab of width $dw$ is $E_{\m{j}}$ and in the rest frame we need to take into account the Lorentz contraction of the slab width to calculate the corresponding rest frame energy $E'_{\m{j}}$. 
\be
E_{\m{j}}=U_{\m{tot}}\pi R^{2}dw, \qquad E'_{\m{j}}=U'_{\m{tot}}\pi R^{2}dw', \nonumber
\ee
\be
E'_{\m{j}}=\frac{3E_{\m{j}}}{4\gamma_{\m{bulk}}},\qquad E'_{\m{j}}=\frac{3f_{\m{loss}}P_{\m{j}}dw}{4c\gamma_{\m{bulk}}} \label{E'j}
\ee
where we have used equations \ref{U'tot} and \ref{LT2}, and $\beta_{\m{bulk}}\approx1$. In the absence of any energy losses and choosing a lab frame width $dw$ such that $E_{\m{j}}$ is constant (i.e. $dw\propto \beta_{\m{bulk}}$, or in the relativistic case that we consider, $dw$ is constant), the change in the rest frame energy density is given by differentiating the equation for $E'_{\m{j}}$ in terms of $E_{\m{j}}$ given above.
\be
dE'_{\m{j}}=-E'_{\m{j}} d\ln\gamma_{\m{bulk}}(z)
\ee
We calculate the cumulative energy emitted by the slab of rest frame width $dw'$, $dE_{\m{rad}}$, by integrating the radiated power, $P'_{\m{rad}}$, over the time taken to travel along the jet in the rest frame and using the equation above
\be
dE'_{\m{rad}}=-E'_{\m{rad}} d\ln\gamma_{\m{bulk}}(z)+P'_{\m{rad}}(z)dw'dt'
\ee
\be
\frac{\partial E'_{\m{rad}}}{\pd z}=-E'_{\m{rad}} \frac{\pd\ln\gamma_{\m{bulk}}(z)}{\pd z}+\frac{P'_{\m{rad}}(z)dw}{c}
\ee
where we have used (\ref{LT2}). We wish to calculate the value of $f_{\m{loss}}=U_{\m{tot}}/U_{\m{tot}\,0}$, the ratio of the current energy in the plasma, to the energy in the plasma if no energy losses had occurred. This can also be expressed in terms of the radiation energy density as, $U_{\m{rad}}=(1-f_{\m{loss}})U_{\m{tot}\,0}$, e.g. $f_{\m{loss}}=1$ corresponds to no energy losses and $f_{\m{loss}}=0.2$ corresponds to 80$\%$ of the initial energy having been radiated away. The radiated energy in the rest frame $E'_{\m{rad}}$ is related to the radiated energy density $U'_{\m{rad}}$ by
\be
E'_{\m{rad}}=\pi R^{2}U'_{\m{rad}}dw', \qquad E_{\m{rad}}=\frac{4}{3}\gamma_{\m{bulk}}E'_{\m{rad}} \label{Erad}
\ee
where we have used the formula in (\ref{E'j}) for the Lorentz transformation of the energy in a slab. Using the above equations we can convert the equation for the evolution of the radiated energy from an equation in $E'_{\m{rad}}$ to an equation in $E_{\m{rad}}$
\bea
&&\hspace{-0.5cm}\frac{3}{4\gamma_{\m{bulk}}}\frac{\partial E_{\m{rad}}}{\pd z}=\frac{1}{\gamma_{\m{bulk}}}\frac{(\gamma_{\m{bulk}}\partial E'_{\m{rad}})}{\pd z}= \nonumber \\
&&\hspace{1cm}...\frac{\partial E'_{\m{rad}}}{\pd z}+E'_{\m{rad}} \frac{\pd\ln\gamma_{\m{bulk}}(z)}{\pd z}=\frac{P'_{\m{rad}}(z)dw}{c} \label{dErad}
\eea
Expressing $E_{\m{rad}}$ in terms of $U'_{\m{rad}}$ using equation \ref{Erad} and cancelling the common factor of the constant lab frame width $dw$ on both sides of the equation we find.
\be
\frac{\pd (\pi R^{2}\gamma^{2}_{\m{bulk}}U'_{\m{rad}})}{\pd z}=\frac{\gamma_{\m{bulk}}P'_{\m{rad}}(z)}{c} \label{U'rad}
\ee
To calculate the evolution of the fractional energy losses due to radiation we restate $f_{\m{loss}}$ in terms of $E_{\m{rad}}$
\be
E_{\m{rad}}=(1-f_{\m{loss}})E_{\m{j}\,0}, \qquad f_{\m{loss}}=1-\frac{E_{\m{rad}}}{E_{\m{j}\,0}}, \nonumber
\ee
\be
E_{\m{j}\,0}=\pi R^{2} U_{\m{tot}\,0}dw, \qquad E_{\m{j}\,0}=\frac{P_{\m{j}}dw}{c},
\ee
where $E_{\m{j}\,0}$ is a constant along the jet. Differentiating the equation for $f_{\m{loss}}$ with respect to $z$ we find
\be
\frac{\pd f_{\m{loss}}}{\pd z}=-\frac{1}{E_{\m{j}\,0}}\frac{\pd E_{\m{rad}}}{\pd z}
\ee
\be
\frac{\pd f_{\m{loss}}}{\pd z}=-\frac{\gamma_{\m{bulk}}P'_{\m{rad}}}{P_{\m{j}}}\nonumber
\ee
\be 
 P'_{\m{rad}}=P'_{\m{synch}}+P'_{\m{SSC}}+P'_{\m{EC}}.   \label{floss}
\ee
where the total radiative energy losses are the sum of the synchrotron, SSC and EC losses, and we have used equation \ref{Utot0} for the total initial jet power $P_{\m{j}}$. The advantage of the approximations we have made is that this equation has analytic solutions in the four regimes where one of optically thick synchrotron, optically thin synchrotron, SSC or EC dominates over the other radiative losses.

\section{Analytic solutions for radiative energy losses}\label{section6}

In order to make progress we first need to parameterise the bulk properties of our jet fluid, which we allow to have a variable shape and bulk Lorentz factor given by
\be
R(z)=R_{0}r_{\m{s}}\left(\frac{z}{r_{\m{s}}}\right)^{b}, \qquad \gamma_{\m{bulk}}=\gamma_{0}\left(\frac{z}{r_{\m{s}}}\right)^{c}.
\ee
The jet structure at the base of M87 has been observed by radio VLBI measurements to be parabolic with a form, $b\approx0.58$ \citep{2012ApJ...745L..28A}. In the ultra-relativistic regime the bulk Lorentz factor is expected to scale with $c\sim1-b$ using axisymmetric analytic solutions to the special relativistic MHD equations (e.g. \citealt{2003ApJ...596.1080V} and \citealt{2009MNRAS.394.1182K}). However, the empirical results of \citealt{2015MNRAS.453.4070P} found a dependence of the bulk Lorentz factor of approximately $\gamma_{0}\approx0.8$ and $c\approx 0.25$ by fitting a fluid jet emission model to the spectra of a sample of 42 blazar jets and so we shall use these values in our numerical calculations, when required. In order to calculate analytic expressions we shall assume that both the electron and the jet bulk velocities are, at least, mildly relativistic i.e. $\beta=\beta_{\m{bulk}}\approx 1$. Let us now consider the four different regimes. 
\subsection{Optically thin synchrotron regime}
We expect optically thin synchrotron emission to dominate in the base region when $U'_{\m{B}}>>U'_{\gamma}$ for frequencies greater than the synchrotron self-absorption frequency (most of the radiated synchrotron power is emitted at high frequencies for electron distributions with $n_{\m{e}}(E_{\m{e}})\propto E_{\m{e}}^{-\alpha}$, $\alpha <3$). Substituting (\ref{P'synch}) into the equation for radiative energy losses (\ref{floss}), we find 
\be
\frac{df_{\m{loss}}}{dz}=-A_{\m{synch}}f_{\m{loss}}^{2}\frac{f_{\m{B}}(1-f_{\m{B}})P_{\m{j}}}{R^{2}\gamma_{\m{bulk}}^{3}}, \label{synch_loss}
\ee
\be
A_{\m{synch}}=\frac{3\sigma_{\m{T}}\langle E_{\m{e}}^{2}\rangle}{4\pi m_{\m{e}}^{2}c^{5}\langle E_{\m{e}}\rangle}.
\ee
Integrating the equation by separating variables and using the condition $f_{\m{loss}}(z_{0})=1$ we find
\bea
&&\hspace{-0.5cm}f_{\m{loss}}=\nonumber\\ &&\hspace{-0.5cm}\left[1+\frac{C_{\m{synch}}f_{\m{B}}(1-f_{\m{B}})f_{\m{Edd}}(X^{1-2b-3c}-X_{0}^{1-2b-3c})}{(1-2b-3c)R_{0}^{2}\gamma_{0}^{3}}\right]^{-1}, \nonumber
\eea
\be
C_{\m{synch}}=A_{\m{synch}}\frac{L_{\m{Edd}}}{r_{\m{s}}}, \qquad X=\frac{z}{r_{\m{s}}}, \qquad f_{\m{Edd}}=\frac{P_{\m{j}}}{L_{\m{Edd}}}.\label{synchloss}
\ee
where $L_{\m{Edd}}$ is the Eddington luminosity and $f_{\m{Edd}}$ the fractional Eddington luminosity of the jet. In the final expression we have converted all quantities into dimensionless units to highlight the independence of the result on black hole mass. 
\subsection{Synchrotron self-absorption regime} \label{section6.2}
Synchrotron self-absorption will be the dominant regime if the maximum electron energy is very low ($\nu_{\m{max}}<\nu_{\m{SSA}}$) so that essentially all synchrotron emission is self-absorbed. Since the maximum electron energy must be small, this also indicates that the radiative losses in this regime are unlikely to be significant since the emitted synchrotron power is proportional to the square of the electron Lorentz factor. This regime could be valid at the very base of the jet if the electron energies are small and the bulk motion is slow such that the magnetic energy density is much larger than that of the Doppler-boosted external photon field. Since $\nu_{\m{SSA}}$ decreases rapidly along the jet as the radius expands it is not a good approximation to assume SSA dominates the radiative losses throughout the base region for a fixed form of the electron population, as in equation \ref{ne}, since this assumption will lead to unphysical results (i.e. the more energy is lost from the plasma, the more powerful the SSA emission becomes, since the rest frame B-field has decreased, the effective temperature of the electrons emitting at a given frequency increases). For these reasons the analytic expression is not useful and so we do not include it here. 

\subsection{Synchrotron self-Compton regime}

We expect the SSC emission to dominate at the base of the jet if high energy electrons are present due to the high density of synchrotron seed photons i.e. provided $U'^{\m{SSC}}_{\gamma}>>U'_{\m{B}}$ and $U'^{\m{SSC}}_{\gamma}>>U'^{\m{EC}}_{\gamma}$. However, at larger distances along the jet the SSC emission becomes subdominant as the cross sectional area of the jet increases and the magnetic field decreases (see equations \ref{UB2} and \ref{UB3}). Since the majority of synchrotron power is emitted at the highest frequencies (for $\alpha<3$) where we expect the synchrotron emission to be optically thin, we take the synchrotron photon energy density to be given by equation \ref{USSC}. In the case where SSC dominates, using (\ref{P'SSC}), equation \ref{floss} becomes
\be
\frac{\partial f_{\m{loss}}}{\partial z}=-A_{\m{SSC}}f_{\m{loss}}^{3}\frac{f_{\m{B}}(1-f_{\m{B}})^{2}P^{2}_{\m{j}}}{R^{3}\gamma_{\m{bulk}}^{5}},
\ee
\be
A_{\m{SSC}}=\frac{3}{8}\left[\frac{\sigma_{\m{T}}\langle E_{\m{e}}^{2}\rangle}{\pi m_{\m{e}}^{2}c^{5}\langle E_{\m{e}}\rangle}\right]^{2}.
\ee
Integrating by separating variables, using the condition $f_{\m{loss}}(z_{0})=1$ and converting to dimensionless units we calculate
\bea
&&\hspace{-0.5cm}f_{\m{loss}}=\nonumber \\&&\hspace{-0.5cm} \left[1+C_{\m{SSC}}\frac{f_{\m{B}}(1-f_{\m{B}})^{2}f_{\m{Edd}}^{2}(X^{1-3b-5c}-X_{0}^{1-3b-5c})}{R_{0}^{3}\gamma_{0}^{5}(1-3b-5c)}\right]^{-1/2}, \nonumber
\eea
\be
C_{\m{SSC}}=2A_{\m{SSC}}\left(\frac{L_{\m{Edd}}}{r_{\m{s}}}\right)^{2},\qquad X=\frac{z}{r_{\m{s}}}, \qquad f_{\m{Edd}}=\frac{P_{\m{j}}}{L_{\m{Edd}}}.\label{SSCloss}
\ee
where again it is worth highlighting the independence of the result on black hole mass.
\subsection{External Compton regime} \label{section6.4}
The precise external photon field is model dependent and will depend on whether one considers stellar mass or supermassive black holes (for an example of a AGN external photon field see Figure 2 of \citealt{2015MNRAS.453.4070P}). In the regime where $U'^{\m{EC}}_{\gamma}$ is much larger than $U'_{\m{B}}$ and $U'^{\m{SSC}}_{\gamma}$, we calculate the fractional energy losses using equations \ref{P'EC} and \ref{floss}
\be
\frac{\partial f_{\m{loss}}}{\partial z}=-A_{\m{EC}}(1-f_{\m{B}})U_{\gamma 0}^{\m{EC}}f_{\m{loss}}\gamma_{0}X^{a+c},
\ee
\be
A_{\m{EC}}=\frac{4\sigma_{\m{T}}\langle E_{\m{e}}^{2}\rangle}{3m_{\m{e}}^{2}c^{4}\langle E_{\m{e}}\rangle}.
\ee
Again we integrate the equation by separating variables and using the condition $f_{\m{loss}}(z_{0})=1$
\be
f_{\m{loss}}=\exp\left[-\frac{A_{\m{EC}}(1-f_{\m{B}})U_{\gamma 0}^{\m{EC}}\gamma_{0}r_{\m{s}}(X^{1+a+c}-X_{0}^{1+a+c})}{1+a+c}\right].
\ee
Since the external photon energy density can be defined independently of the jet parameters, the energy losses are proportional to the particle energy density, and the expression takes the form of a simple exponential decline. We have chosen to retain the explicit factor of $r_{\m{s}}$ in the numerator as a reminder to the reader that the external photon field will have an indirect dependence on the black hole mass due to the different environments of stellar and supermassive black holes.
\subsection{SSC and EC domination}
In the case where the majority of radiative energy losses occur from inverse-Compton emission (i.e. SSC and EC losses exceed synchrotron losses) we are also able to solve (\ref{floss}) analytically using equations \ref{P'SSC} and \ref{P'EC}. 
\be
\frac{\partial f_{\m{loss}}}{\partial X}=-\frac{D_{\m{EC}}}{2}X^{a}f_{\m{loss}}-\frac{D_{\m{SSC}}}{2}X^{-3b-5c} f_{\m{loss}}^{3},
\ee
\be
D_{\m{EC}}=2A_{\m{EC}}r_{\m{s}}\gamma_{0}U_{\gamma \,0}(1-f_{\m{B}}), \nonumber 
\ee
\be
D_{\m{SSC}}=\frac{2A_{\m{SSC}}f_{\m{B}}(1-f_{\m{B}})^{2}(f_{\m{Edd}}L_{\m{Edd}})^{2}}{r_{\m{s}}^{2}R_{0}^{3}\gamma_{0}^{5}}.
\ee
Substituting $y=f_{\m{loss}}^{-2}$ this reduces to the first order differential equation
\be
\frac{\partial y}{\partial X}=D_{\m{EC}}X^{a}y+D_{\m{SSC}}X^{-3b-5c}.
\ee
Which has the solution
\bea
&&\hspace{-1.0cm}f_{\m{loss}}=\left[ C\exp\left(\frac{D_{\m{EC}}X^{a'}}{a'}\right)-\frac{D_{\m{SSC}}}{a'}X^{b'}\exp\left(\frac{D_{\m{EC}}X^{a'}}{a'}\right) ...\right. \nonumber \\&&\left. \times \left(\frac{D_{\m{EC}}X^{a'}}{a'}\right)^{-b'/a'}\Gamma\left(\frac{b'}{a'},\frac{D_{\m{EC}}X^{a'}}{a'}\right) \right]^{-1/2}, \label{SSC+ECloss}
\eea 
\be
a'=a+1, \qquad b'=1-3b-5c,
\ee
\bea
&&\hspace{-1.0cm}C=\exp\left(-\frac{D_{\m{EC}}X_{0}^{a'}}{a'}\right)\left[1+\frac{D_{\m{SSC}}}{a'}X_{0}^{b'}\exp\left(\frac{D_{\m{EC}}X_{0}^{a'}}{a'}\right)\right. .... \nonumber \\&& \left.\times \left(\frac{D_{\m{EC}}X_{0}^{a'}}{a'}\right)^{-b'/a'}\Gamma\left(\frac{b'}{a'},\frac{D_{\m{EC}}X_{0}^{a'}}{a'}\right) \right],
\eea
where $\Gamma(a,x)$ is the upper incomplete gamma function defined by
\be
\Gamma(a,x)=\int_{x}^{\infty}x'^{a-1}e^{-x'}dx'.
\ee
and the constant $C$ has been determined by the boundary condition $f_{\m{loss}}(z=z_{0})=1$. The expression above, although slightly more cumbersome than those for a single dominant term, is likely to be a good approximation to the radiative losses experienced in high power blazar jets where SSC is dominant at the base and EC is dominant at large distances.
\begin{figure}
	\centering
            \includegraphics[height=8cm, angle=270,clip=true, trim=0cm 0cm 0cm 0cm]{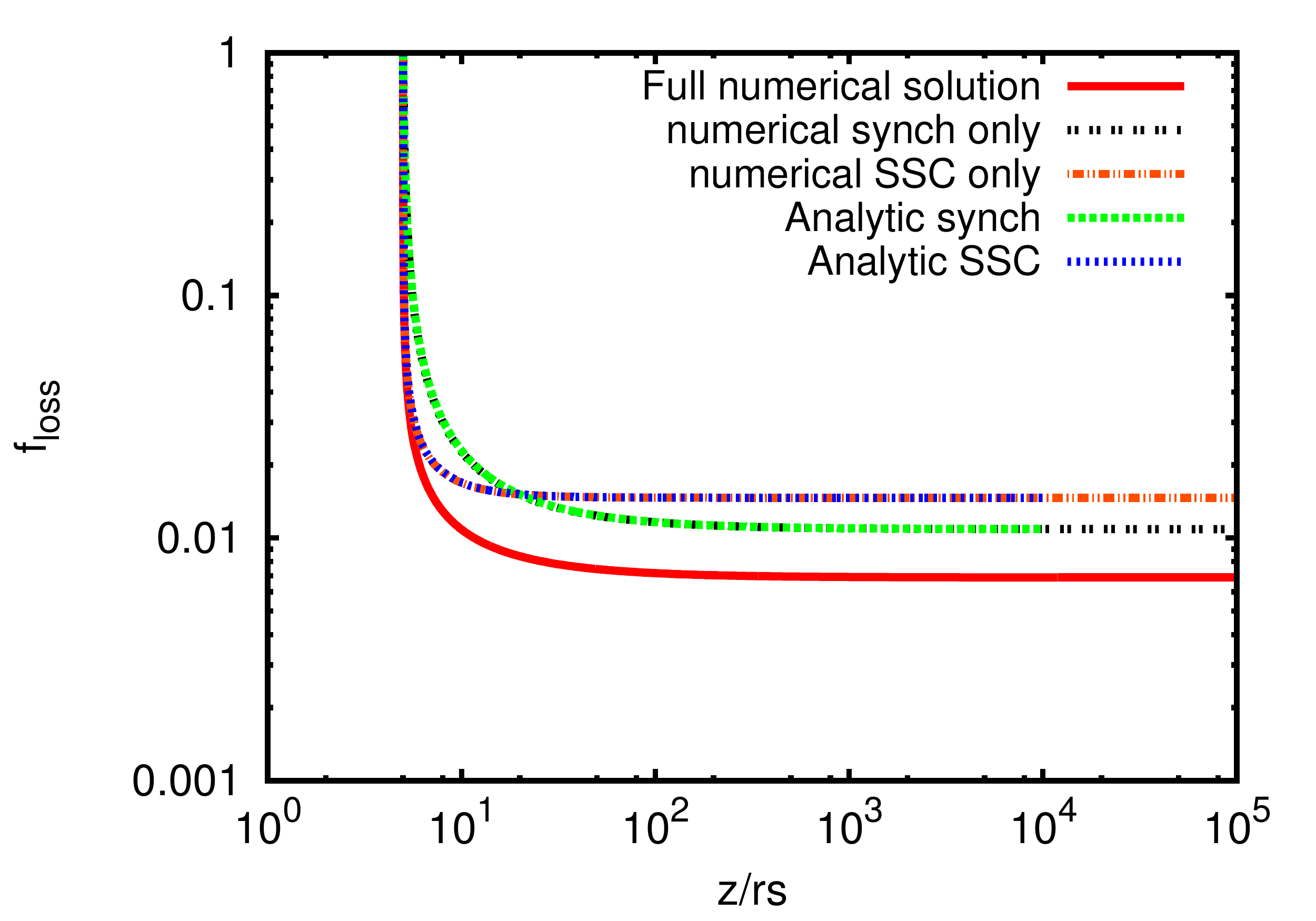} 				
	\caption{Comparison of the analytic formulae for synchrotron and SSC energy losses to a numerical solution of the full loss equation \ref{floss}, for Model A with parameters given in Table \ref{Table2}. The initial energy losses are dominated by SSC, with synchrotron losses dominating at larger distances. We show the full numerical solution and both numerical and analytic solutions including only synchrotron or SSC losses. We only plot the analytic solutions out to $z=10^{4}r_{\m{s}}$, in order to clearly show the consistency between the numerical and analytic solutions. }
\label{fig1}
\end{figure}

\begin{table*}
\centering
\begin{tabular}{| c | c | c | c | c | c | c | c | c | c | c | c | c | c | c |}
\hline\vspace{-0.25cm}
& & & & & & & & & & & & & & \\ 
Model & $E_{\m{min}}/m_{\m{e}}c^{2}$ & $E_{\m{max}}/m_{\m{e}}c^{2}$ & $\alpha$ & $f_{\m{Edd}}$ & $f_{\m{B}}$ & $R_{0}$ & $\gamma_{0}$ & $z_{0}$ & $U_{\gamma}'^{\m{EC}}$ &$a$ & $b$ & $c$ & $A_{\m{rec}}$ & $d$ \\ \hline 
A & 10 & $10^{4}$ & 1.9 & 0.1 & 0.99 & 2.2 & 0.8 & $5r_{\m{s}}$ & 0 & n.a. & 0.58 & 0.25 & n.a. & n.a.\\ \hline
B & 10 & variable & 1.9 & variable & variable & 2.2 & 0.8 & $5r_{\m{s}}$ & 0 & n.a. & 0.58 & 0.25 & n.a. & n.a. \\ \hline
C & 10 & $10^{4}$ & 1.9 & variable & variable & 2.2 & 0.8 &  $5r_{\m{s}}$ &  0 & n.a. & 0.58 & 0.25 & variable & variable\\ \hline
D & 10 & $10^{4}$ & 1.9 & 0.01 & variable & 2.2 & 0.8 &  $5r_{\m{s}}$ & 0 & n.a. & 0.58 & 0.25 & variable & variable\\ \hline
\end{tabular}
\caption{The values of the jet model parameters used in our numerical solutions. We have chosen typical parameters found by modelling the emission of a large sample of blazars \citep{2015MNRAS.453.4070P}. We have used {\lq}n.a.{\rq} to indicate a parameter which is not applicable to the given model (in all cases we have deliberately chosen not to specify a source of external photons since this would make our results less widely applicable).   }
\label{Table2}
\end{table*}

\subsection{Accuracy of analytic formulae}

In Figure \ref{fig1}, a comparison of the analytic formulae is shown relative to the numerical integral of equation \ref{floss}. The analytic and numerical solutions agree as we expect. We see that the dominant radiative energy losses come from SSC emission at small distances along the jet with synchrotron losses becoming dominant at larger distances. 

\section{Constraining the magnetisation}\label{section7}

Let us use the results of the calculations in the previous section to calculate the radiative energy losses associated with maintaining a constant magnetisation along a typical black hole jet. Taking the jet to have a parabolic base as in M87 with $R(z)\approx2.2r_{\m{s}} X^{0.58}$ and using the dependence of the bulk Lorentz factor $\gamma_{\m{bulk}} \sim 0.8X^{0.25}$, from \citealt{2015MNRAS.453.4070P}. We calculate the radiative energy losses of the electron population assuming that the energy in the non-thermal electron population is replenished by the magnetic field via resistive dissipation, in order to maintain the initial magnetisation (we shall deal explicitly with in-situ magnetic reconnection in section \ref{section8}). Since the energy is constantly being lost via radiation this allows an important constraint to be placed on the minimum magnetisation which can be maintained along the base of the jet for a given electron energy distribution. This is because we require that a substantial fraction of the initial jet power should be retained in the plasma until large distances ($\sim 85-97\%$, corresponding to $f_{\m{loss}}=0.85-0.97$, see \citealt{2012Sci...338.1445N} and references therein) where it can be observed heating radio lobes in FRII AGN jets, for example.

\subsection{Analytic constraints}\label{section7.1}

The analytic formulae for the fractional radiative energy losses due to synchrotron and inverse-Compton emission derived in the previous section can be used to place constraints on the minimum allowable magnetisation of the jet. This constraint is a minimum because smaller magnetisations would be closer to equipartition and would therefore radiate more efficiently, incurring heavier radiative energy losses  (\ref{synch_loss}). To do this we rearrange equation \ref{synchloss} to find an equation for the fractional magnetic energy density. Starting with the equation for synchrotron energy losses we find the quadratic equation
\be
f_{B}^{2}-f_{B}+\frac{(1-2b-3c)R_{0}^{2}\gamma_{0}^{3}}{C_{\m{synch}}f_{\m{Edd}}(X^{1-2b-3c}-X_{0}^{1-2b-3c})}.\left[\frac{1}{f_{\m{loss}}}-1\right]=0. \label{quadratic1}
\ee
The solutions to this quadratic equation are given by
\be
f_{B}=\frac{1}{2}\left(1\pm\sqrt{1-\frac{Y}{f_{\m{Edd}}}.\left[\frac{1}{f_{\m{loss}}}-1\right]}\,\,\right), 
\ee
\be
Y=\frac{4(1-2b-3c)R_{0}^{2}\gamma_{0}^{3}}{C_{\m{synch}}(X^{1-2b-3c}-X_{0}^{1-2b-3c})}
\ee
where the two solutions reflect the symmetry of the synchrotron losses to changes in the equipartition fraction i.e. synchrotron emission is maximised when a plasma is at equipartition, $f_{B}=1/2$ and drops off symmetrically for more magnetic or particle dominated plasmas due to the $f_{B}(1-f_{B})$ factor in equation \ref{synch_loss}. In this work we assume a magnetic launching mechanism for the relativistic jet and so we are only interested in the \lq{}$+$\rq{} solutions representing magnetically dominated jets. Substituting the fractional magnetic energy $f_{B}$ for the magnetisation, $\sigma=U_{B}/U_{e\pm}$, we find
\be
\sigma=\dfrac{1+\sqrt{1-\dfrac{Y}{f_{\m{Edd}}}.\left[\dfrac{1}{f_{\m{loss}}}-1\right]}}{1-\sqrt{1-\dfrac{Y}{f_{\m{Edd}}}.\left[\dfrac{1}{f_{\m{loss}}}-1\right]}}\,, \qquad f_{B}=\frac{\sigma}{1+\sigma}.
\ee
In general this equation, and the resulting constraints, will depend on the variables: $\alpha$, $E_{\m{max}}$, $E_{\m{min}}$, $b$, $c$, $f_{\m{Edd}}$, $X$, $X_{0}$, $R_{0}$, $\gamma_{0}$ and $f_{\m{loss}}$, so some simplification is clearly required to obtain a useful result. We are primarily interested in the dependence on the fractional Eddington jet power, magnetisation and fractional energy remaining in the jet. We therefore choose representative values for the other parameters shown under Model B in Table \ref{Table2}. The values of $E_{\m{min}}$, $\alpha$, $\gamma_{0}$ and $c$ are taken as typical model fits to blazar spectra \citep{2015MNRAS.453.4070P}, whilst $R_{0}$, $z_{0}$ and $b$ are typical parameters found from radio VLBI observations of the inner structure of the jet in M87 \citep{2012ApJ...745L..28A}. With these values, $(1-2b-3c)=-0.91$, so assuming that the jet base extends to a distance of at least $\gtrsim 50r_{s}$, the distance dependent term in the denominator will be approximately independent of the total jet length i.e. $(X^{1-2b-3c}-X_{0}^{1-2b-3c})\approx X_{0}^{1-2b-3c}$. This is because the synchrotron energy losses occur predominantly at the jet base where the highest magnetic field strengths exist. Using these approximations the $Y$ parameter in equation \ref{quadratic1} simplifies to
\be
Y=1.2\times10^{-15}\frac{(3-\alpha)(E_{\m{max}}^{2-\alpha}-E_{\m{min}}^{2-\alpha})}{(2-\alpha)(E_{\m{max}}^{3-\alpha}-E_{\m{min}}^{3-\alpha})}.
\ee
%\be
%f_{B}^{2}-f_{B}+\frac{2.2\times10^{-16}}{f_{\m{Edd}}}\frac{\langle E_{e}\rangle}{\langle E_{e}^{2} \rangle}\left(\frac{1}{f_{\m{loss}}}-1\right)=0.
%\ee
%The solutions to this quadratic equation are
%\be
%f_{B}=\frac{1}{2}\left(1\pm\sqrt{1-\frac{8.7\times10^{-16} \langle{E_{e}}\rangle}{f_{\m{Edd}}\langle E_{e}^{2} \rangle}.\left[\frac{1}{f_{\m{loss}}}-1\right]}\,\,\right), \label{quadratic2}
%\ee
 %Unfortunately for the case $\alpha\approx2$ the electron moment $\langle E_{e}\rangle$ does not easily simplify and is sensitive to the values of $E_{\m{max}}$ and $E_{\m{min}}$. In the limit $f_{B}\approx1$ (which we later show to be appropriate for regimes of interest) the expression simplifies by taking the binomial expansion of the square root in equation \ref{quadratic2}
%\be
%f_{B}\approx1-\frac{4.4\times10^{-16}}{f_{\m{Edd}}}\frac{\langle E_{e}\rangle}{\langle E_{e}^{2} \rangle}\left(\frac{1}{f_{\m{loss}}}-1\right).
%\ee
%Converting from fractional magnetic energy to magnetisation $\sigma=f_{B}/(1-f_{B})$
%\be
%\sigma\approx2.3\times10^{15} f_{\m{Edd}}\frac{(3-\alpha)(E_{\m{max}}^{2-\alpha}-E_{\m{min}}^{2-\alpha})}{(2-\alpha)(E_{\m{max}}^{3-\alpha}-E_{\m{min}}^{3-\alpha})}\left(\frac{1}{f_{\m{loss}}}-1\right)-1,
%\ee
%Inserting our typical values of $E_{\m{max}}$, $E_{\m{min}}$ and $\alpha$ from table ?? we find
%\be
%\sigma\approx2.8\times10^{5} f_{\m{Edd}}\left(\frac{1}{f_{\m{loss}}}-1\right)-1.
%\ee
\begin{figure*}
	\centering

		\subfloat[$E_{\m{max}}/m_{\m{e}}c^{2}=100$]{ \includegraphics[width=9cm, clip=true, trim=1cm 0cm 0cm 1cm]{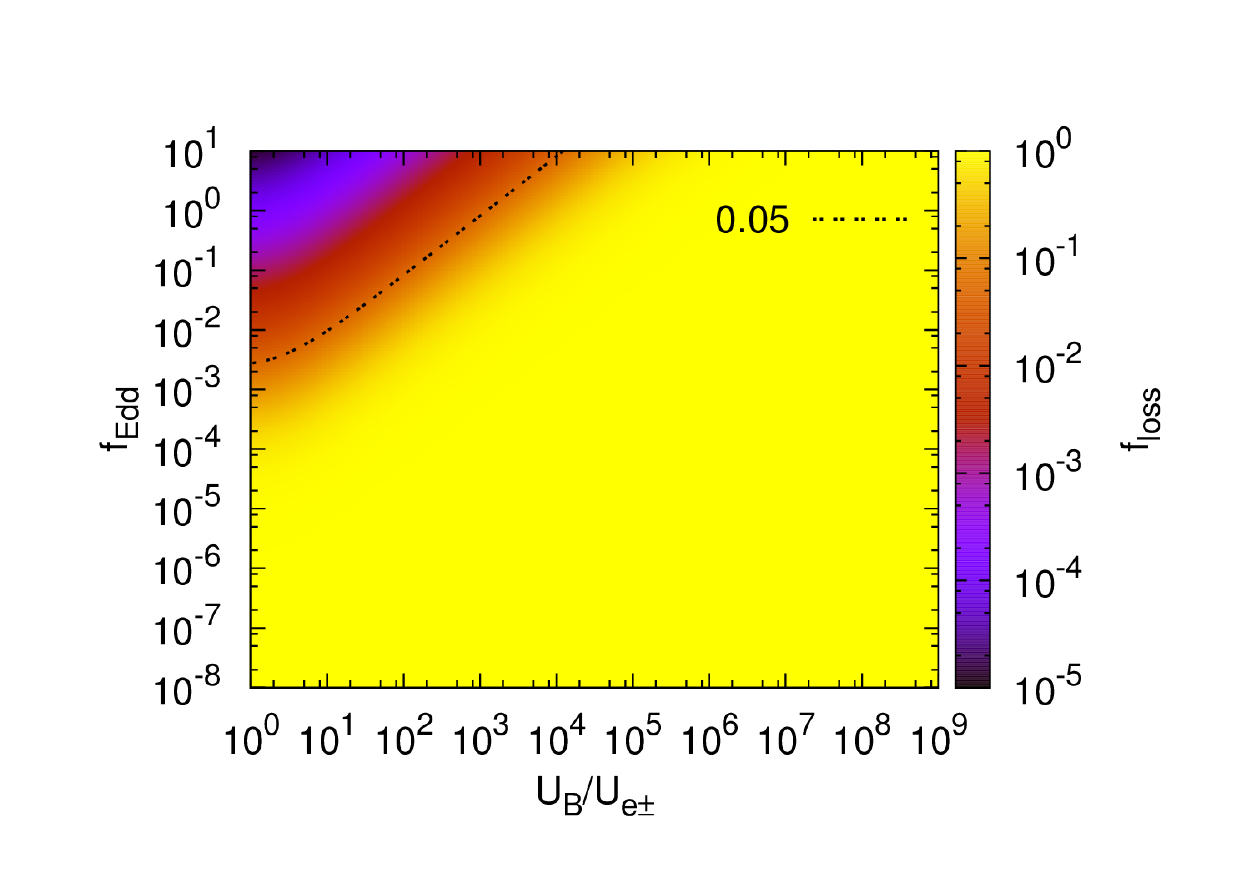} }
		\subfloat[$E_{\m{max}}/m_{\m{e}}c^{2}=1000$]{ \includegraphics[width=9cm, clip=true, trim=1cm 0cm 0cm 1cm]{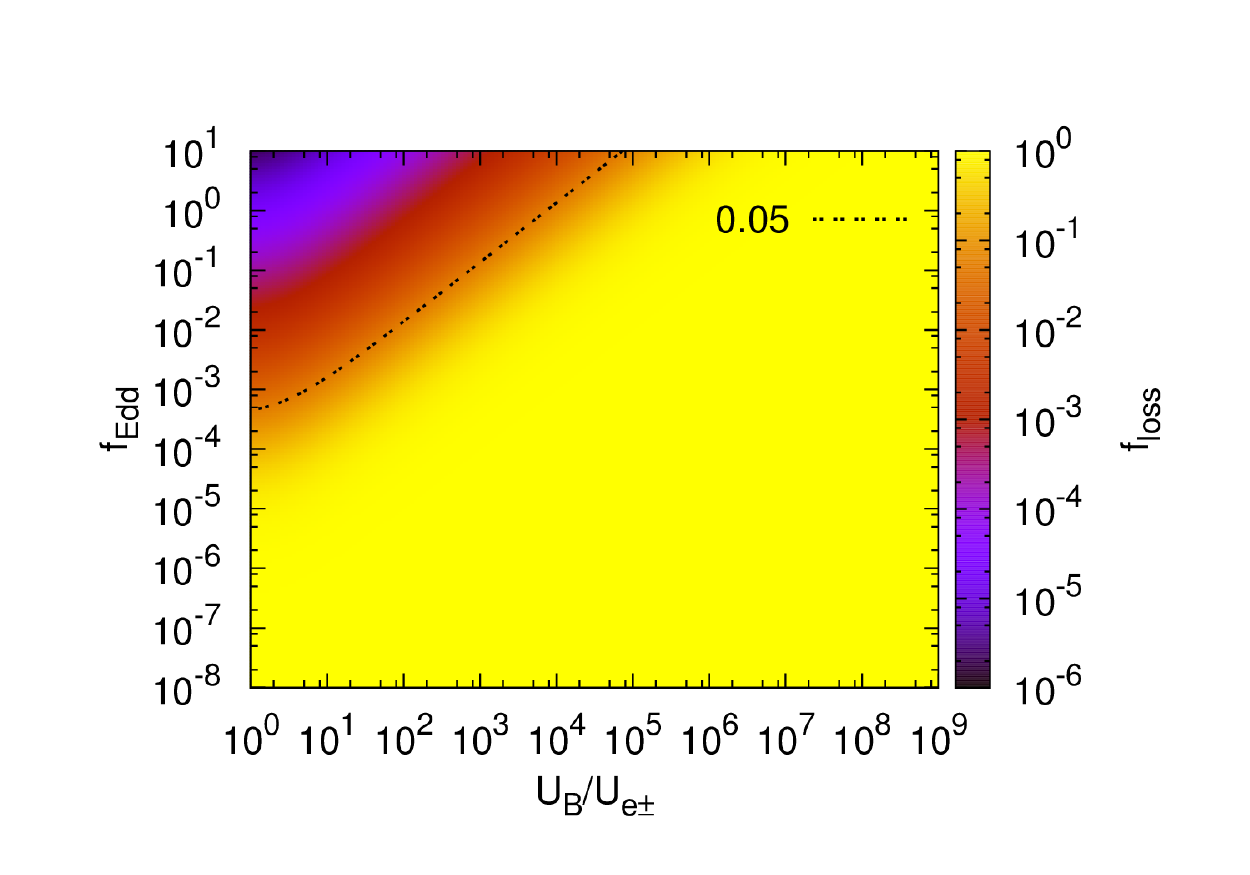} }	
\\
		\subfloat[$E_{\m{max}}/m_{\m{e}}c^{2}=10^{4}$]{ \includegraphics[width=9cm, clip=true, trim=1cm 0cm 0cm 1cm]{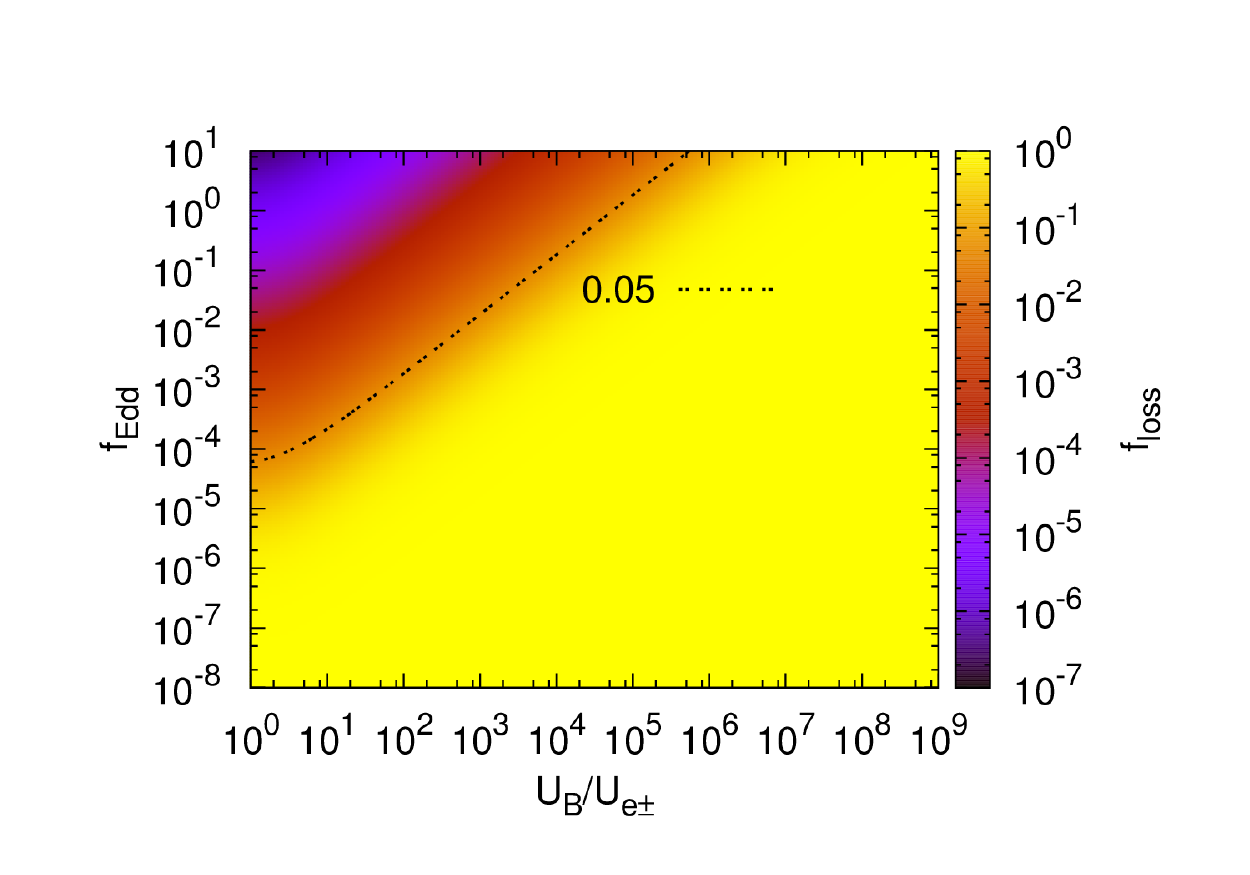} }
		\subfloat[$E_{\m{max}}/m_{\m{e}}c^{2}=10^{5}$]{ \includegraphics[width=9cm, clip=true, trim=1cm 0cm 0cm 1cm]{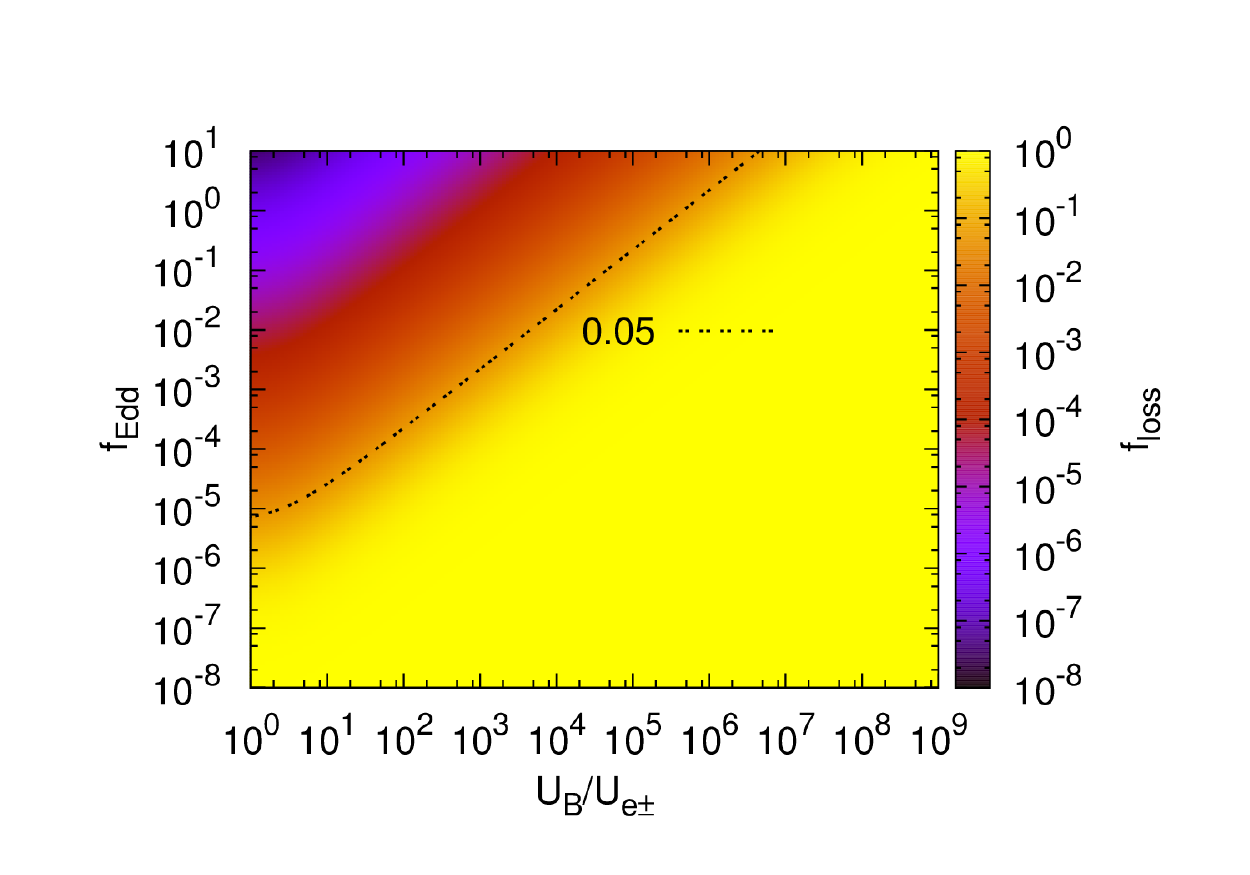} }	
\\
		\subfloat[$E_{\m{max}}/m_{\m{e}}c^{2}=10^{6}$]{ \includegraphics[width=9cm, clip=true, trim=1cm 0cm 0cm 1cm]{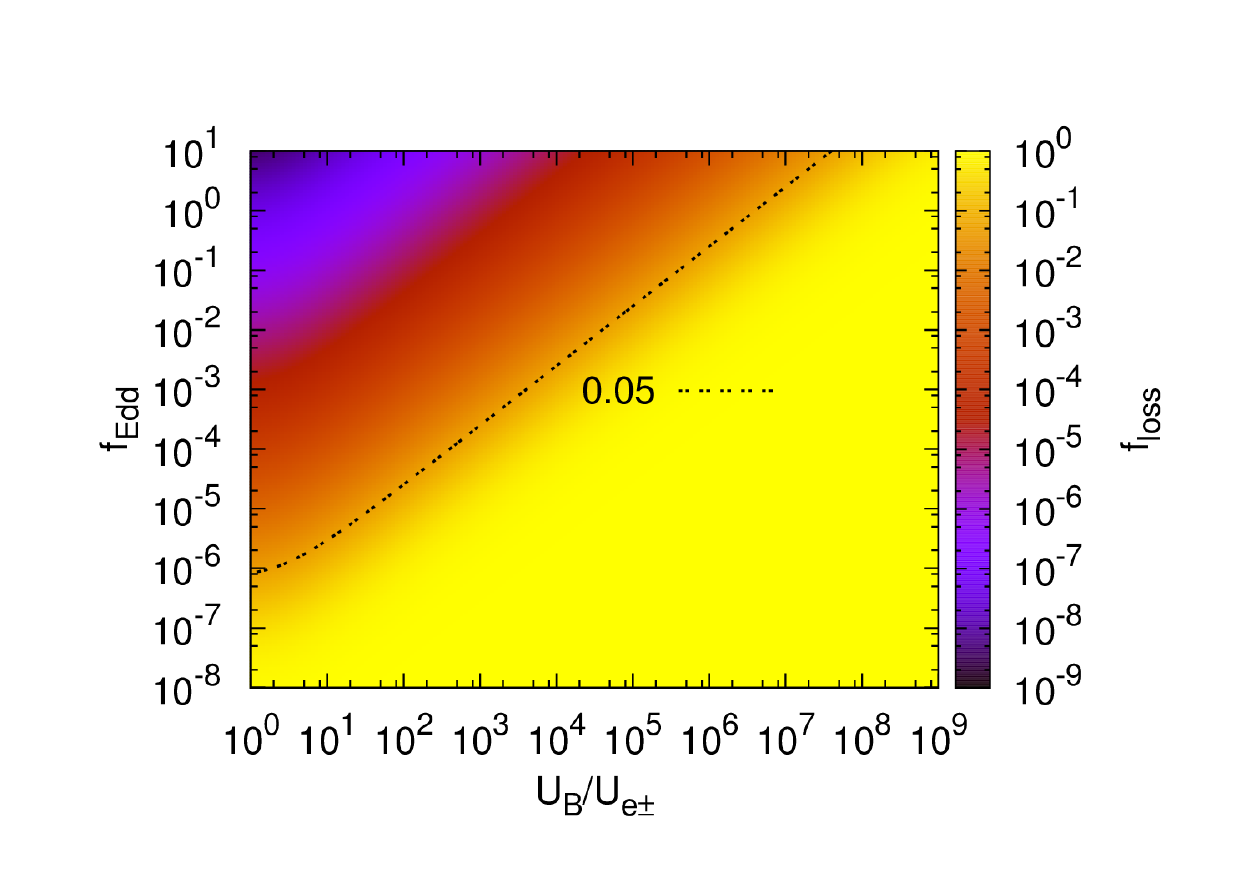} }
		\subfloat[$E_{\m{max}}/m_{\m{e}}c^{2}=10^{7}$]{ \includegraphics[width=9cm, clip=true, trim=1cm 0cm 0cm 1cm]{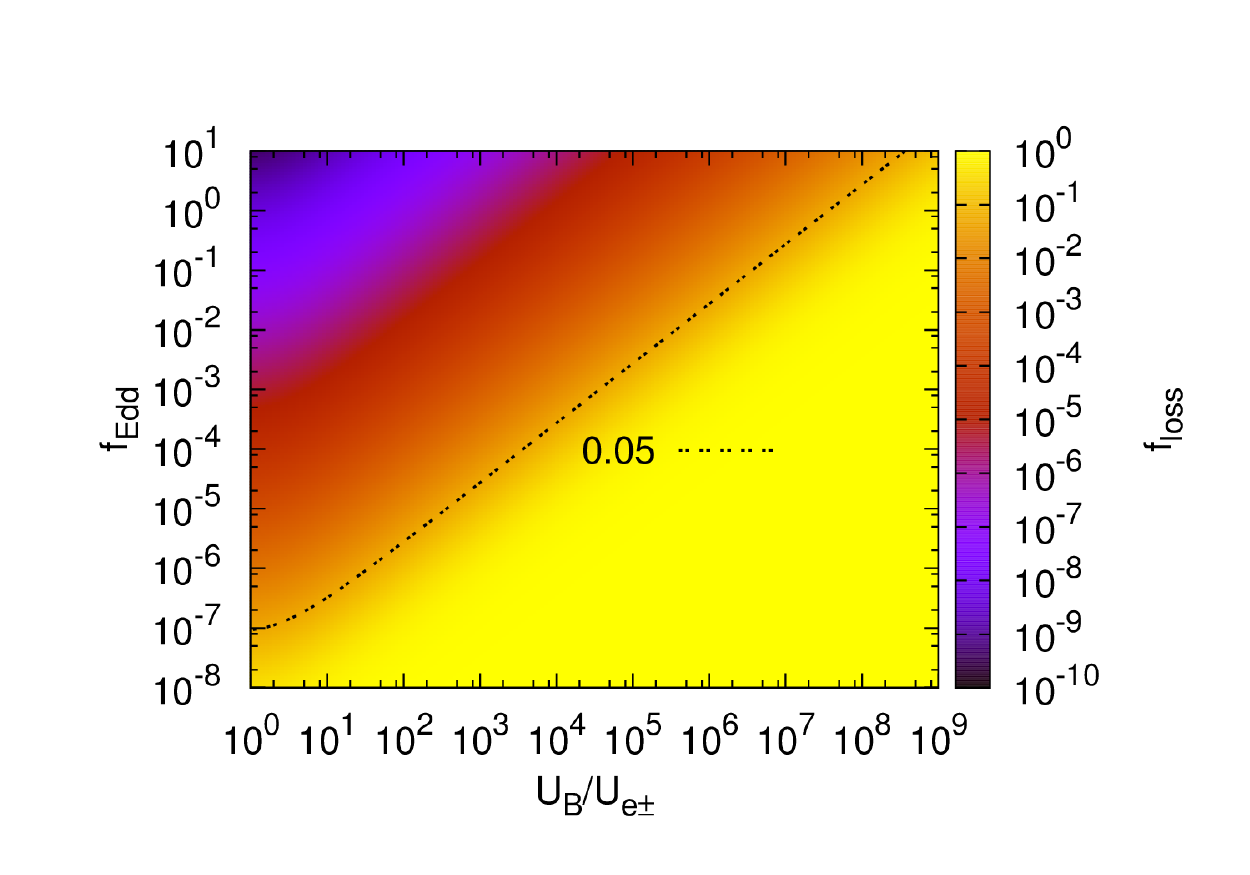} }

	\caption{The results of numerically integrating the radiative energy loss equation \ref{floss}, including synchrotron and SSC radiative losses for a variety of values of constant jet magnetisation, fractional Eddington power and maximum electron energy. The jet parameters used to obtain these results are shown as Model B in Table \ref{Table2}. We show that the base region of jets must be highly magnetised in order to avoid severe radiative energy losses which would otherwise radiate a large fraction of the initial power if the jet plasma were closer to equipartition. Larger maximum electron energies and higher fractional Eddington powers both increase the radiative losses. The black contour at $f_{\m{loss}}=0.05$ divides the region where $95\%$ of the jet power has been radiated away (only $5\%$ of the initial jet power remains in the plasma). Since a substantial jet power is observed to be retained by the jet to large distances ($\sim 85-97\%$ \citealt{2012Sci...338.1445N}), these calculations can be used to constrain the minimum magnetisation of the jet base. It is worth noting the symmetry of the graphs under a transformation $U_{\m{B}}/U_{\m{e}\pm}\rightarrow U_{\m{e}\pm}/U_{\m{B}}$. Since jets are currently believed to be launched electromagnetically we do not consider the case of a jet base which is particle dominated (a particle dominated jet base would both lack an obvious energy reservoir to accelerate the plasma to relativistic speeds and suffer from severe deceleration via Compton-drag).  }
\label{spectra}
\end{figure*}
\begin{figure}
	\centering
            \includegraphics[height=6.0cm, clip=true, trim=0cm 0cm 0cm 0cm]{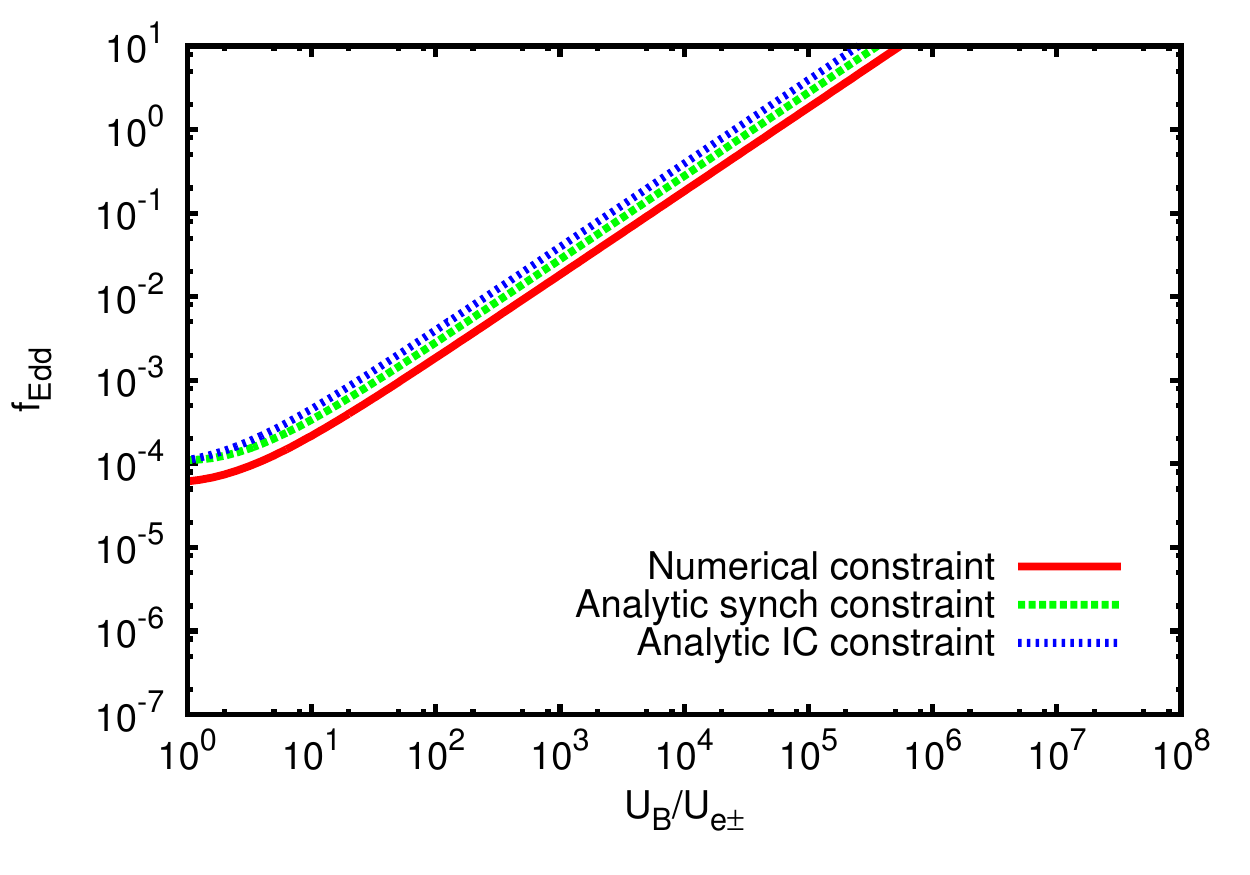}			
	\caption{A comparison between the analytic constraint on the maximum magnetisation derived in equation \ref{analyticconstraint} for synchrotron losses only and the full numerical constraint for $f_{\m{loss}}=0.05$, $E_{max}=10^{4}m_{e}c^{2}$ and other jet parameters given by Model B in Table \ref{Table2}. The analytic constraint from synchrotron losses is a good approximation and only differs by $\sim35\%$ from the full numerical results shown in Figure \ref{spectra} (this accuracy is independent of $E_{\m{max}}$). }
\label{analyticconstraint2}
\end{figure}

The derived constraint on the allowed range of magnetisations which permit a given final fractional energy loss $f_{\m{loss}}$ is then simply
\be
\sigma\geq\dfrac{1+\sqrt{1-\dfrac{Y}{f_{\m{Edd}}}.\left[\dfrac{1}{f_{\m{loss}}}-1\right]}}{1-\sqrt{1-\dfrac{Y}{f_{\m{Edd}}}.\left[\dfrac{1}{f_{\m{loss}}}-1\right]}}\,.\label{analyticconstraint}
\ee
It is also possible to use the equation for SSC energy losses (\ref{SSCloss}) to derive a constraint on the magnetisation. However, we find that the two results do not differ significantly (see Figure \ref{analyticconstraint2}) and since the constraint equation using SSC is cubic, we have chosen to use the simpler synchrotron constraint. We shall reserve commenting on the implications of these constraints until we have tested its accuracy (we already know it to be an underestimate since it neglects SSC losses) by comparing to a full numerical solution in the next section. 
    
\subsection{Numerical constraints}

In Figure \ref{spectra} we show the results of numerically integrating equation \ref{floss} using an adaptive Bulirsch-Stoer algorithm, for different values of the maximum electron energy, $E_{\m{max}}$, and fractional magnetic energy, $f_{\m{B}}$, with no external photon field. We choose to integrate these losses up to a distance of $10^{5}r_{\m{s}}$ after which we find that the radiative energy losses are no longer significant (this is the case for a constant magnetisation jet but will not remain true when we consider steady in-situ acceleration and a variable magnetisation in the next section). An important point to note is that these results are independent of the black hole mass and so hold for both X-ray binary and AGN systems. This symmetry would be broken by the addition of external photon fields which are different for stellar and supermassive black holes, and this is the reason we have chosen not to include an external-Compton component in these constraints. 

The results are surprisingly constraining and show that high Eddington power jets must be highly magnetised in order to avoid severe radiative energy losses which would otherwise drain the vast majority of the jet power. Since we observe jet powers which are comparable to the accretion power \citep{2009MNRAS.399.2041G} and the majority of this energy remains in the jet to large distances ($\sim 85\%-97\%$, \citealt{2012Sci...338.1445N}), we can make a conservative constraint on the minimum value of the magnetisation by assuming that at least $5\%$ of the total jet power is not radiated away in the jet base and is retained in the jet to large distances. In Figure \ref{spectra} we show this constraint as the dark black contour marking the miminum magnetisation (maximum non-thermal particle content) as a function of the fractional Eddington luminosity and maximum electron energy. These results are important because most MHD simulations the initial magnetisation of the jet is $\sigma \sim 10-100$ (\citealt{2006ApJ...641..103H}, \citealt{2007MNRAS.380...51K} and \citealt{2009MNRAS.394.1182K}). We have shown that this is likely to be unrealistically low, i.e. real jets have a much larger fraction of total energy contained in the magnetic field compared to non-thermal particles than this. 

In Figure \ref{analyticconstraint2} we compare the analytic constraints (\ref{analyticconstraint}), based only upon synchrotron energy losses, to the numerical constraint which includes both synchrotron and SSC energy losses. We find that the analytic constraint is a good approximation to the full result, typically only differing by $\sim35\%$. We also show the constraint on the minimum magnetisation derived from the analytic equation for SSC energy losses (\ref{SSCloss}). At the base of the jet, where the plasma is most dense, the inverse-Compton losses are comparable to those from synchrotron, however, at larger distances synchrotron losses dominate. These ratios of component energy losses are, in fact, independent of the choice of $E_{\m{max}}$. This justifies our choice to neglect the analytic constraint based on inverse-Compton losses in favour of that from synchrotron losses in section \ref{section7.1}, since we find the latter to be more accurate. 

If we were to include additional effects such as adiabatic energy losses, external-Compton emission and a more realistic limit on the maximum fraction of energy allowed to be radiated, these constraints would become even more severe. Adiabatic energy losses could be included via the addition of the following term to the r.h.s. of equation \ref{particleeq}
\be
-\frac{2b\gamma_{\m{bulk}}^{2}R^{2}U'_{e\pm}}{3z}
\ee
which originates from the usual adiabatic expansion loss formula for a fluid with a relativistic equation of state
\be
\frac{\partial \ln E_{e\pm}}{\partial \ln z}=-\frac{2b}{3}
\ee
Some caution is required, however, when using this standard formula for expansion losses since it is not clear that it is strictly applicable to a collisionless, non-thermal jet plasma expanding within a static confining funnel, formed from the ambient medium. The additional effects of adiabatic and external Compton losses are more sensitively dependent on our particular model parameters than the synchrotron and SSC emission, which necessarily occur if non-thermal electrons are present, and so we have chosen not to include them when calculating the constraints, in order to keep the result as trustworthy and widely applicable as possible. 

This is the main conclusion of this paper; due to severe radiative energy losses, the jet plasma is constrained to be highly magnetised at the base of the jet. We shall show in the next section that the effect of the severe radiative losses is effectively to remove the memory of the initial magnetisation over a radiative cooling lengthscale and the local magnetisation is then determined by the balance of in-situ reacceleration and radiative losses. These results allow us to place strong constraints on the particle acceleration processes happening at small distances along the jet and demonstrate the importance and necessity of including a detailed self-consistent treatment of radiative energy losses in realistic MHD simulations of jets.

\section{Modelling the power injected by magnetic reconnection} \label{section8}

Having started with a simple scenario, a jet with a constant magnetisation, let us now consider something more realistic: a jet in which the amount of particle acceleration is determined by the rate of magnetic reconnection. Since the jet base has such an abundance of magnetic energy, magnetic reconnection (the resistive dissipation of currents) is one of the most likely mechanisms which can convert this magnetic energy into accelerating non-thermal electrons. It is natural to expect a power-law electron energy distribution from reconnection since the resistive decay of a magnetic field induces a temporary electric field ($d{\bf B}/dt=-\nabla\times {\bf E}$) and this will preferentially accelerate charged particles with higher initial velocities to higher energies (since the power gained by an  accelerating electron will be proportional to its velocity), giving relatively steep electron spectral indices, $\alpha<2$, in numerical simulations (see for example \citealt{2001ApJ...562L..63Z}, \citealt{2007ApJ...670..702Z} and \citealt{2014ApJ...783L..21S}). In some simulations, in addition to acceleration by the induced electric field, a Fermi acceleration process also operates by particles scattering back and forth between reconnecting plasmoids which form in the reconnecting sheet and this can also efficiently accelerate non-thermal electrons \citep{2014ApJ...783L..21S}.

Magnetic reconnection is a complex non-linear process and remains the subject of intense research (\citealt{RevModPhys.82.603} and \citealt{Treumann2015}), so we do not pretend to capture the microscopic details of this process in our 1D fluid model. We can, however, put useful constraints on the power dissipated by magnetic reconnection which then goes into accelerating non-thermal electrons, by calculating their emission as they travel along the jet. This power is inevitably linked to the large-scale average reconnection timescale along the jet. Our approach is intended to be complementary to the detailed numerical simulations which focus on the small-scale reconnection physics but tend to neglect the large scale fluid flow and energy balance along the jet; in this work we instead focus on the average large-scale fluid flow and energy balance, at the expense of a detailed treatment of the microscopic reconnection physics. Our results will help to constrain and inform dedicated numerical simulations of the macroscopic reconnection rates required by considering radiative losses and energy constraints along the jet at large scales. 

In most astrophysical plasmas the amount of resistive dissipation of magnetic fields is negligible and occurs on a resistive timescale
\be
t_{\m{res}}=\frac{\mu_{0}\delta^{2}}{\eta},
\ee
where $\eta$ is the electrical resistivity (\ref{Ohm}) and $\delta$ is the characteristic lengthscale of the fluctuation in the magnetic field. This timescale is usually far longer than the typical observed timescale associated with the release of energy via magnetic reconnection in solar flares for example \citep{JGRA:JGRA7974}. It is found in MHD simulations and particle-in-cell (PIC) simulations of reconnection that once the process of resistive dissipation or magnetic reconnection begins to occur in a region with a strong gradient in the magnetic field, the magnetic energy is released in the form of heating and bulk acceleration of an outflow of plasma from the region. This creates a low density, low pressure reconnection region which then draws in additional plasma from both sides of the reconnecting current sheet (the magnetic field changes rapidly across the reconnecting region and this necessitates a concentrated thin current sheet since $\nabla \times {\bf B}\approx\mu_{0}{\bf J}$). The velocities of the outflow and inflow are, typically, close to the Alfv\`{e}n speed, $v_{\m{A}}$, with the precise relationship depending on the geometry of the reconnecting region and the proportion of magnetic energy which can be dissipated by the process. In collisionless highly magnetised plasmas, such as jets, inflow velocities of $\sim 0.1v_{\m{A}}$ are typical, \citealt{2014ApJ...783L..21S}. From figure \ref{spectra} we have already constrained the jet plasma to be highly magnetised, in which case the Alfv\`{e}n speed is approximately the speed of light $v_{\m{A}}\approx c$ (e.g. \citealt{1993PhRvE..47.4354G}).

We can parameterise the reconnection rate in terms of the power dissipated in a region in which magnetic reconnection occurs, multiplied by the volume filling factor of such regions. The power dissipated by a reconnecting sheet of surface area, $S$, with inflow velocities into the reconnecting surface, $v_{\m{rec}}$, and an efficiency, $\eta_{\m{rec}}$, which we define as the fraction of inflowing magnetic energy which is dissipated by the reconnection region (since only the component of magnetic field which cancels between the two inflows can be dissipated) is 
\be
p'_{\m{rec}}=Sv_{\m{rec}}\eta_{\m{rec}}U'_{\m{B}} =Sc\beta_{\m{rec}}\eta_{\m{rec}}U'_{\m{B}},
\ee
Since reconnection appears in nature to be a sudden, stochastic event, only a relatively small fraction of the total plasma volume will contain reconnecting regions at any given time and so we introduce the average surface area of reconnecting regions per unit volume of jet plasma measured in the plasma rest frame, $S'_{\m{rec}}$. This encapsulates our ignorance of both the initial size and rate of generation of perturbations to a uniform magnetic field (which can subsequently be dissipated) and the effective duty-cycle of the reconnection process. We expect, $\beta_{\m{rec}}\eta_{\m{rec}}S'_{\m{rec}}\lesssim1/r_{\m{s}}$ (see equation \ref{zrec}), because rapid dissipation of the available magnetic energy at the jet base would be incompatible with our constraints on the minimum magnetisation of the jet shown in figure \ref{spectra}. 

It is important to note that although the reconnecting regions must occupy a relatively small fraction of the total jet volume, jet emission models of large emission regions which assume homogeneous small-scale tangled, isotropic magnetic field and particle velocity distributions are able to accurately reproduce the jet emission. This tells us that the reconnection regions are well distributed throughout the jet volume, since otherwise if high energy (fast cooling) electrons were confined to only a small fraction of the total jet volume, close to concentrated regions of reconnection, their high energy optically thin emission would not be well reproduced by simple, large homogeneous blob emission models. This is because for the same observed synchrotron emission, a compact emission region would have a higher synchrotron photon energy density than a large region the size of the jet radius. The SSC emission from such a compact region of the jet would then be distinguishable from the SSC emission and Compton-dominance of the existing successful large-scale homogeneous emission models. For this reason we suggest that the reconnection regions and the associated high energy accelerated leptons are well distributed through the jet and our treatment of the jet as essentially 1D is appropriate.   

Using the average reconnecting area per unit volume in the rest frame, $S'_{\m{rec}}$, the expected power injected by reconnection per unit volume of jet plasma is given by
\be
p'_{\m{rec}}=c\beta_{\m{rec}}\eta_{\m{rec}}U'_{\m{B}}S'_{\m{rec}}.
\ee
The power dissipated per unit width of a cylindrical slab of plasma as measured in the jet rest frame is
\be
P'_{\m{rec}}=\pi R^{2}c\beta_{\m{rec}}\eta_{\m{rec}}U'_{\m{B}}S'_{\m{rec}}.
\ee
When calculating our numerical results we assume that each of the reconnection parameters can be adequately described by a simple power law of the form $g_{\m{rec}}(z)=g_{\m{rec}\,0}X^{h}$: a constant $g_{\m{rec}\,0}=g_{\m{rec}}(z=z_{0})$ defined at the base of the jet and a power-law dependence of the dimensionless distance $X=z/r_{\m{s}}$. Using this power-law approximation to the reconnection parameters and substituting in the value of $U'_{\m{B}}$ from equation \ref{UBprimed} we find
\be
P'_{\m{rec}}=\frac{3}{4}A_{\m{rec}}f_{\m{loss}}f_{\m{B}}X^{d-2c}, \qquad A_{\m{rec}}=\frac{S'_{\m{rec}\,0}\beta_{\m{rec}\,0}\eta_{\m{rec}\,0}P_{\m{j}}}{\gamma_{bulk\,0}^{2}}, \label{P'rec}
\ee
where the power-law exponent $d$, is the sum of the individual power-law exponents of the reconnection parameters $\beta_{\m{rec}}$, $\eta_{\m{rec}}$ and $S'_{\m{rec}}$. We might expect the value of $d$ to be dominated by the efficiency and effective surface area per unit volume, $\eta_{\m{rec}}S'_{\m{rec}}$, since the Alfv\`{e}n speed will not change significantly along the magnetically dominated sections of the jet. We have chosen to use this form for $A_{\m{rec}}$ so that our numerical results will be independent of black hole mass, which we discuss in more detail at the end of this subsection. In order to convert this expression into a convenient form, we wish to calculate the reconnection rate in terms of the reconnection term, $\pd(R^{2}\gamma^{2}_{\m{bulk}}U'_{\m{rec}})/\pd z$, in the equation for the evolution of the fractional magnetic energy (\ref{magevol}). To do this we follow the same method used in section \ref{section5} to calculate the radiated energy in a slab of constant lab frame width, $dw$, in terms of the radiated power per unit width in the rest frame, $P'_{\m{rad}}$, except replacing the radiative terms with the equivalent reconnection terms. We find the desired expression by using equation \ref{U'rad} and replacing $U'_{\m{rad}}$ with $U'_{\m{rec}}$ and $P'_{\m{rad}}$ with $P'_{\m{rec}}$. 
\be
\frac{\pd (R^{2}\gamma^{2}_{\m{bulk}}U'_{\m{rec}})}{\pd z}=\frac{\gamma_{\m{bulk}}P'_{\m{rec}}(z)}{\pi c}.
\ee
\begin{figure*}
	\centering
		\subfloat[Radiative energy losses for $f_{\m{Edd}}=10$.]{ \includegraphics[width=9cm, clip=true, trim=0cm 0cm 0cm 1cm]{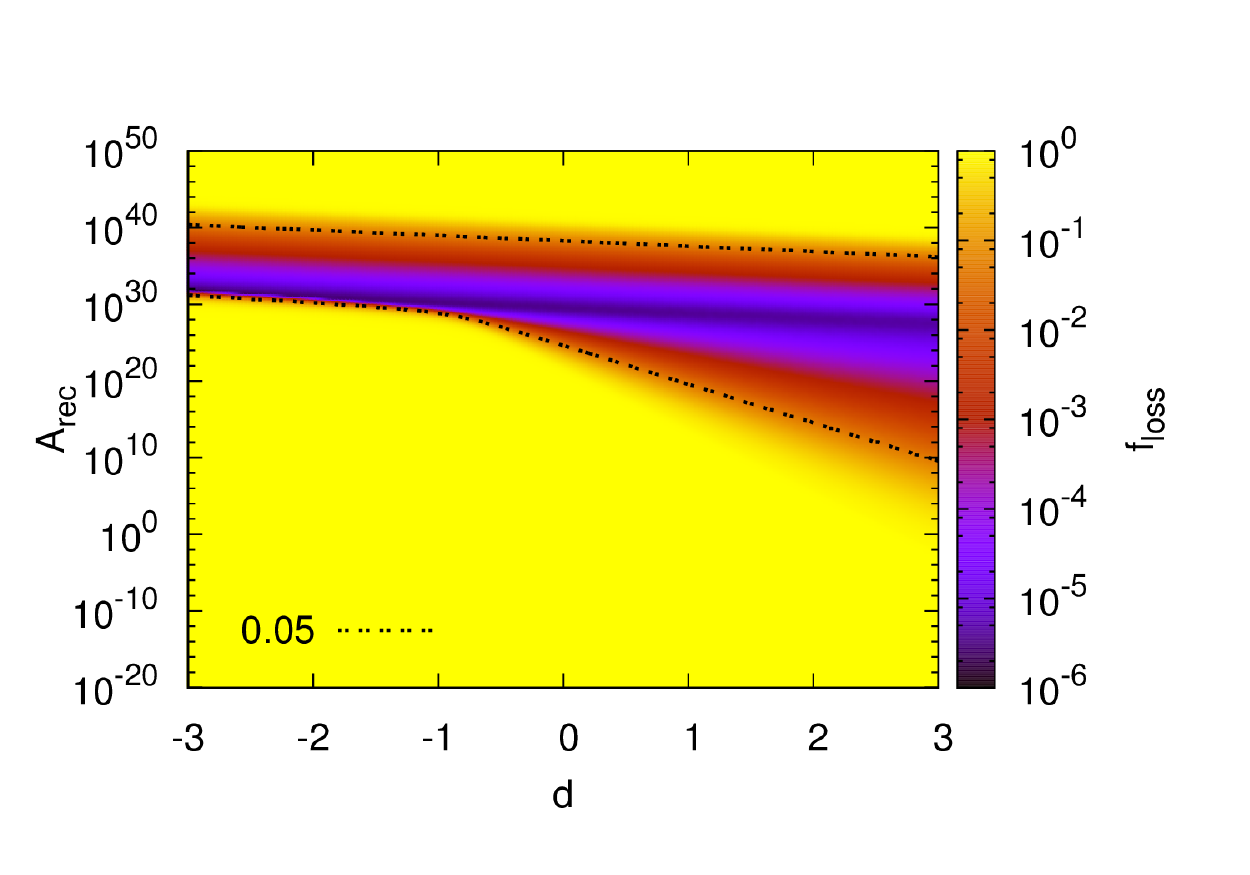} }
		\subfloat[Radiative energy losses for $f_{\m{Edd}}=1$.]{ \includegraphics[width=9cm, clip=true, trim=0cm 0cm 0cm 1cm]{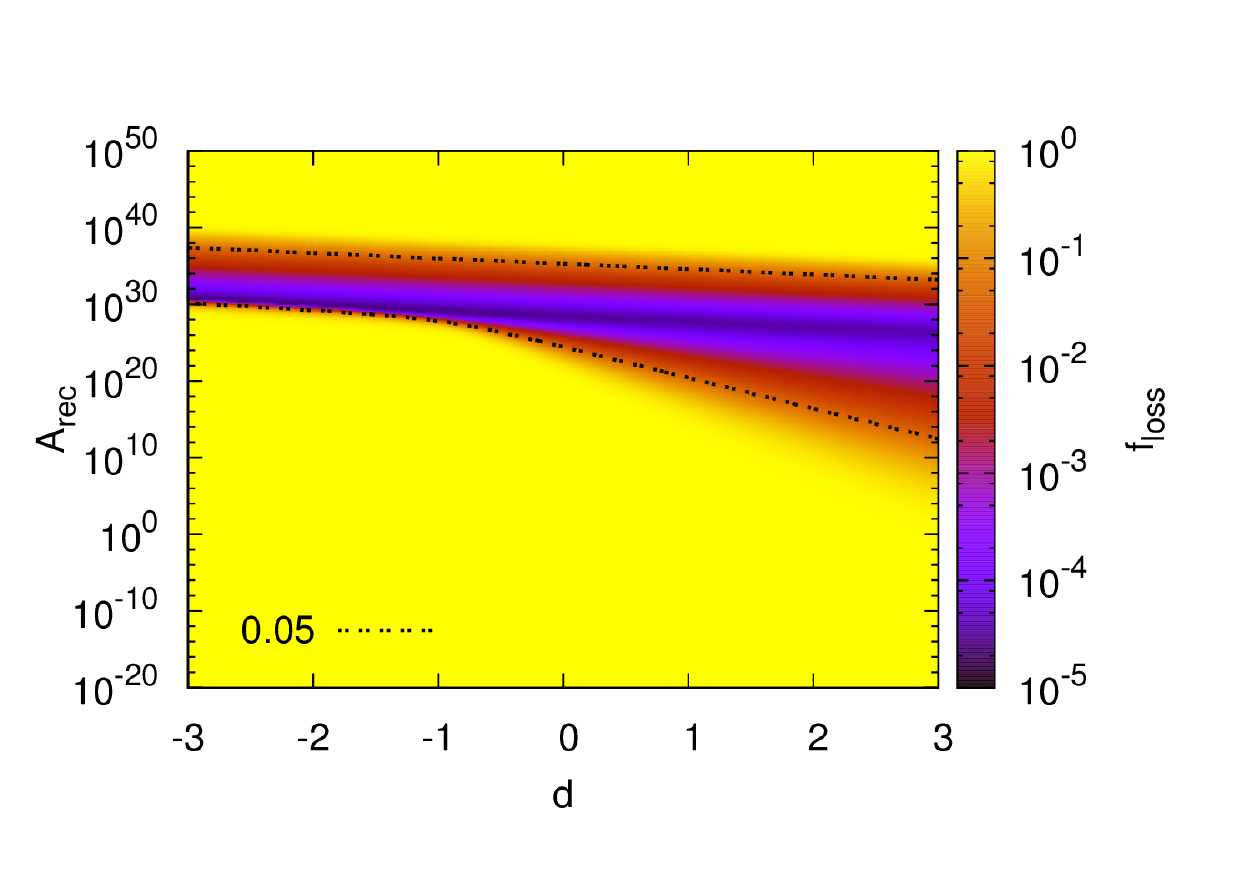} }
\\
		\subfloat[Radiative energy losses for $f_{\m{Edd}}=0.1$.]{ \includegraphics[width=9cm, clip=true, trim=0cm 0cm 0cm 1cm]{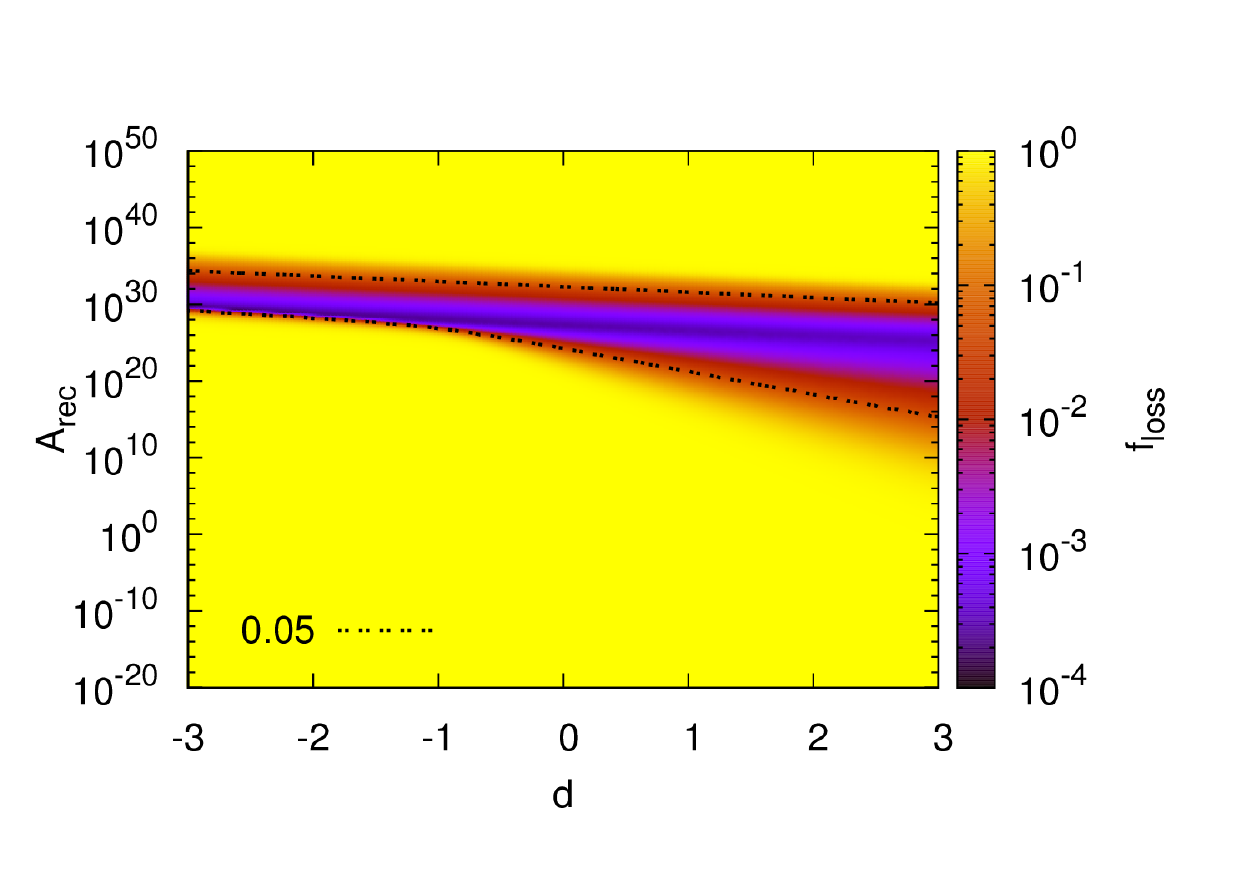} }
		\subfloat[Radiative energy losses for $f_{\m{Edd}}=0.01$.]{ \includegraphics[width=9cm, clip=true, trim=0cm 0cm 0cm 1cm]{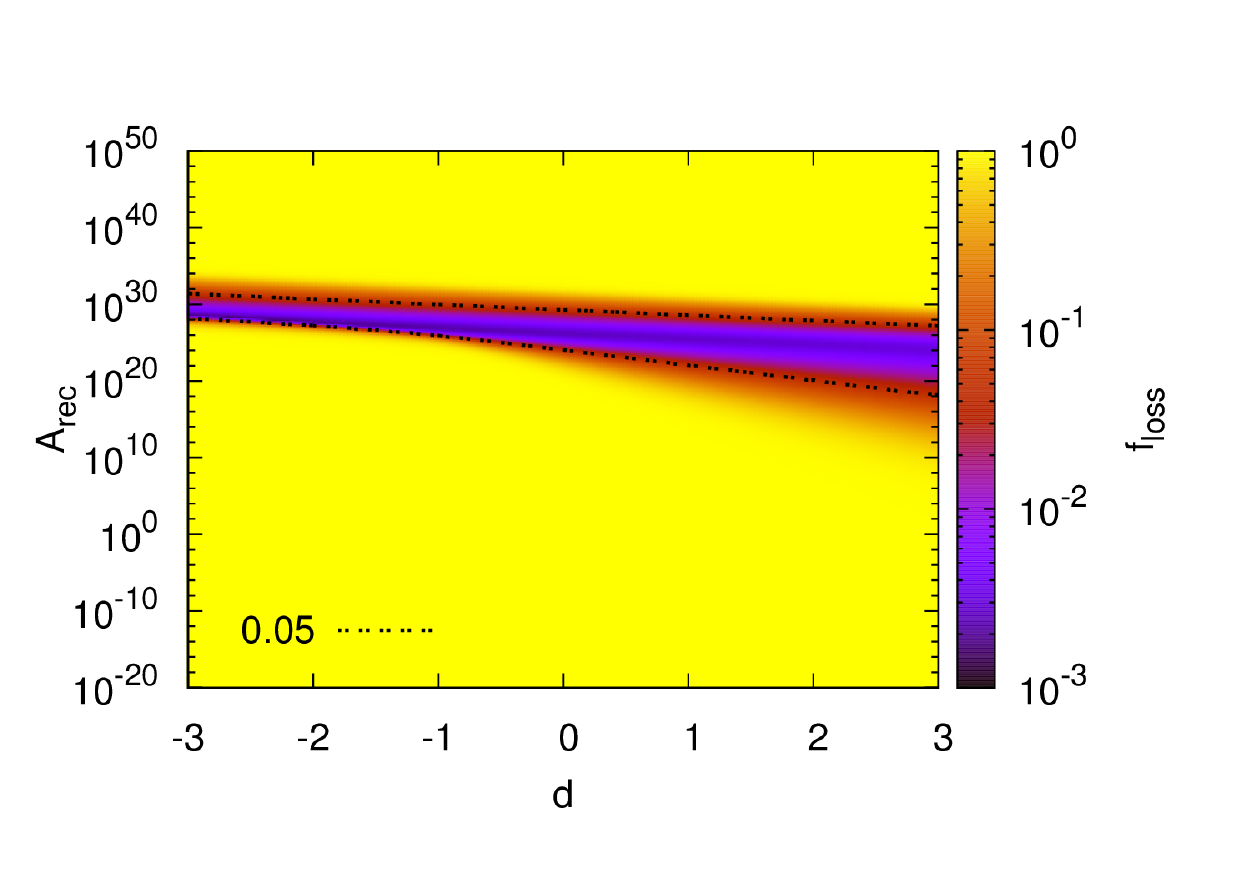} }
\\
		\subfloat[Radiative energy losses for $f_{\m{Edd}}=10^{-3}$.]{ \includegraphics[width=9cm, clip=true, trim=0cm 0cm 0cm 1cm]{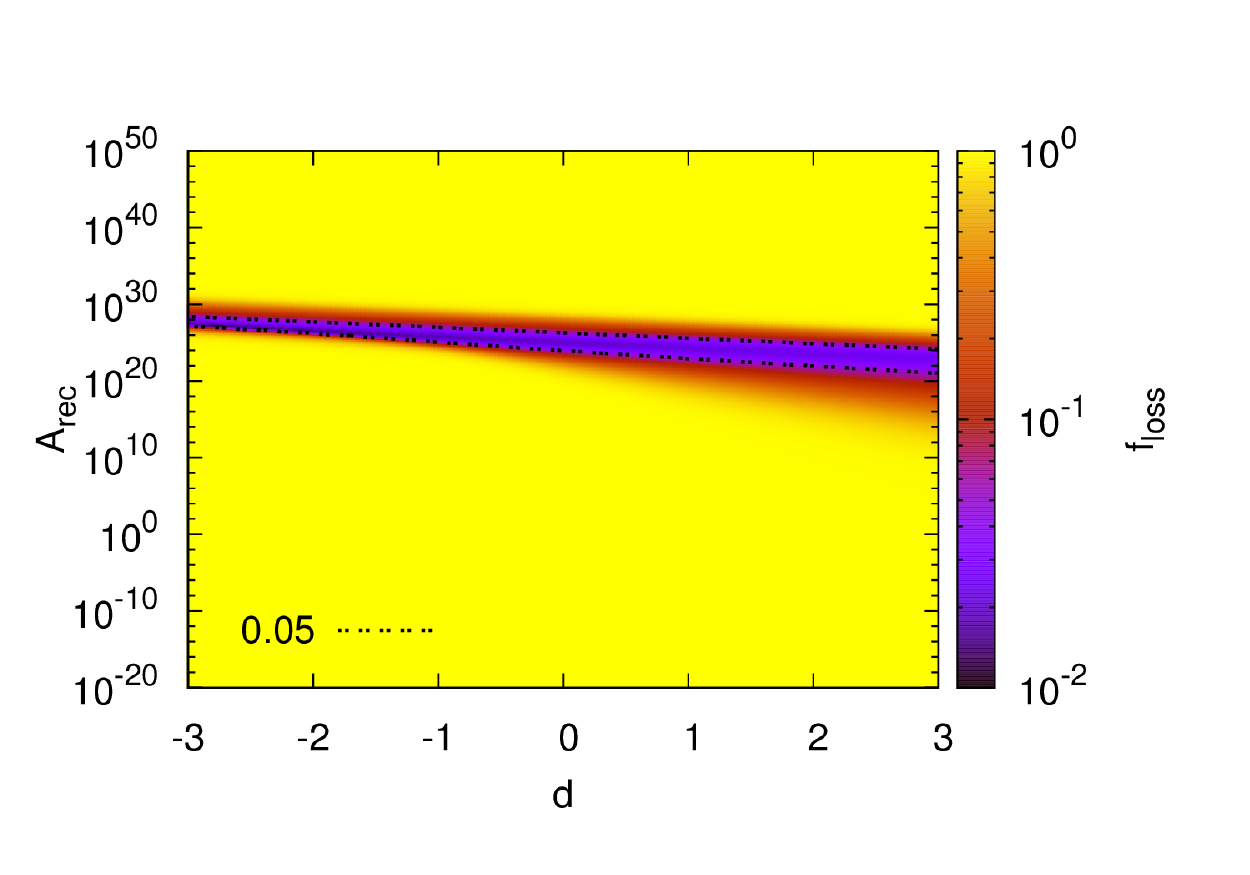} }	
		\subfloat[Radiative energy losses for $f_{\m{Edd}}=10^{-4}$.]{ \includegraphics[width=9cm, clip=true, trim=0cm 0cm 0cm 1cm]{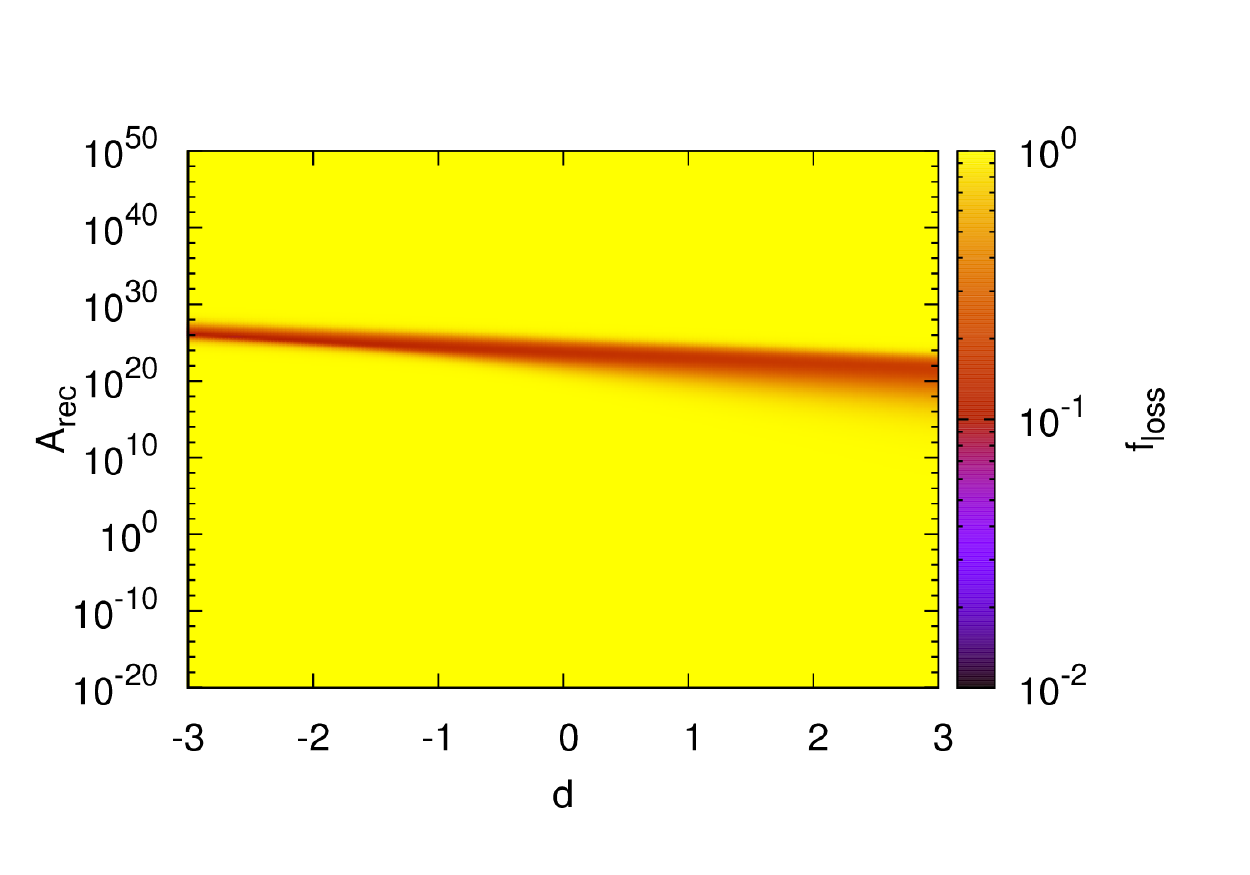} }
	\caption{Figures $a-f$ show the fractional radiative energy losses of the jet due to synchrotron and SSC emission at the distance where the jet first reaches equipartition, or $z=10^{8}r_{s}$ (at distances corresponding to those in figure \ref{rec_paramsz}), for different values of the reconnection parameters $A_{\m{rec}}$ and $d$ (see equation \ref{P'rec}) and fractional Eddington jet power $f_{\m{Edd}}$. The jet parameters used to obtain these results are shown under Model C in Table \ref{Table2}. As the Eddington fraction increases, the radiative energy losses become more severe (due to the larger initial magnetic field strengths) and this constrains the allowed values of parameters. The black dashed contour at, $f_{\m{loss}}=0.05$, corresponds to only $5\%$ of the initial jet power being retained up to the region where the jet first reaches equipartition (i.e. $95\%$ of the total jet power has been emitted as radiation). We use $f_{\m{loss}}>0.05$ as a conservative constraint on the total radiative losses. This is conservative compared with observations that suggest that a substantial amount of energy remains in the jet to large distances, $>85\%$, or $f_{\m{loss}}>0.85$, \citealt{2012Sci...338.1445N}. Losses are small for very large and very small reconnection rates $A_{\m{rec}}$. For small rates this is because the jet always remains highly magnetised out to large distances and so the radiative efficiency is low. For very high reconnection rates, reconnection proceeds so quickly that the timescale for reconnection is much faster than the radiative lifetime and so the jet reaches equipartition at a distance which is much shorter than the radiative lengthscale, over which substantial radiative energy losses would occur. These results are independent of black hole mass.}
\label{rec_params}
\end{figure*}
\begin{figure*}
	\centering
		\subfloat[$f_{\m{Edd}}=10$.]{ \includegraphics[width=6.0cm, clip=true, trim=0.3cm 0cm 0.5cm 1.0cm]{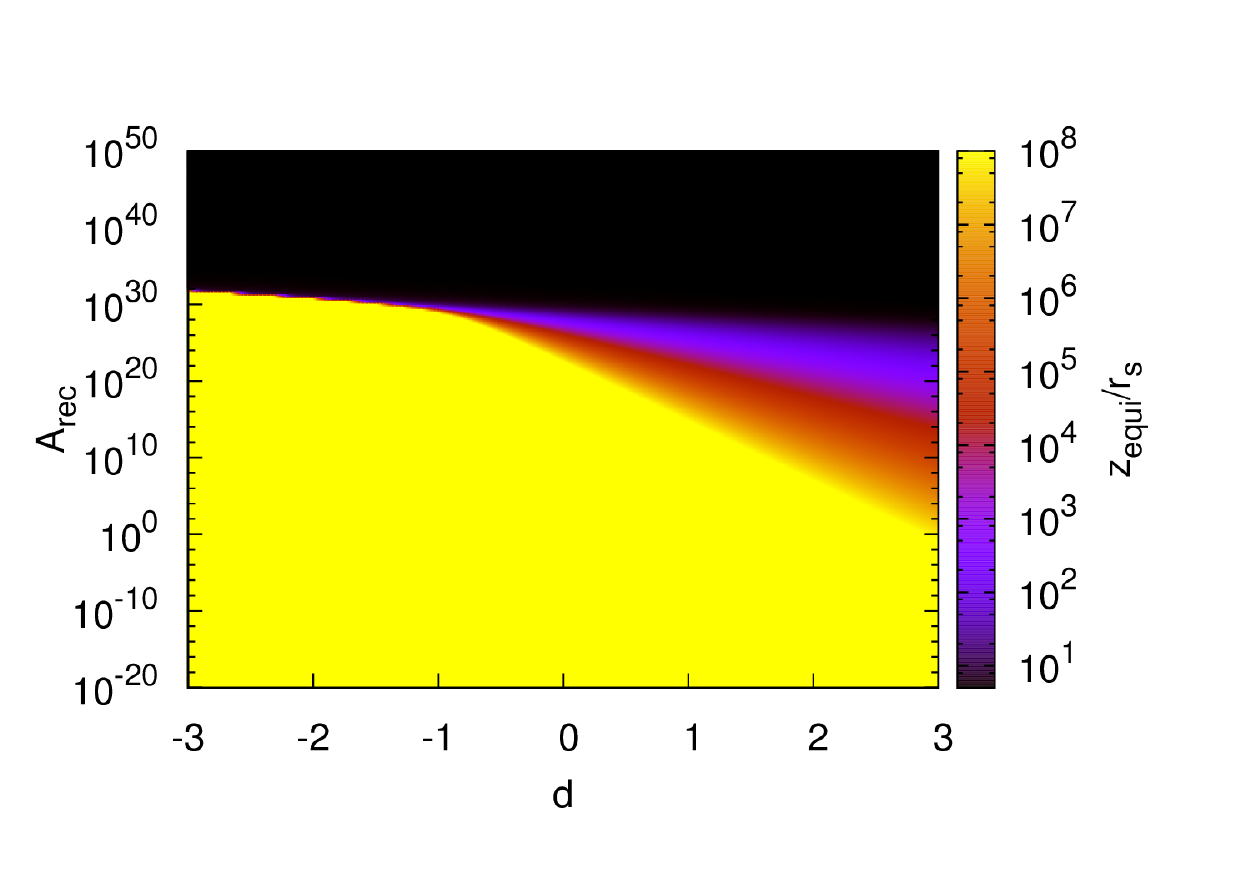} }
		\subfloat[$f_{\m{Edd}}=1$.]{ \includegraphics[width=6.0cm, clip=true, trim=0.3cm 0cm 0.5cm 1cm]{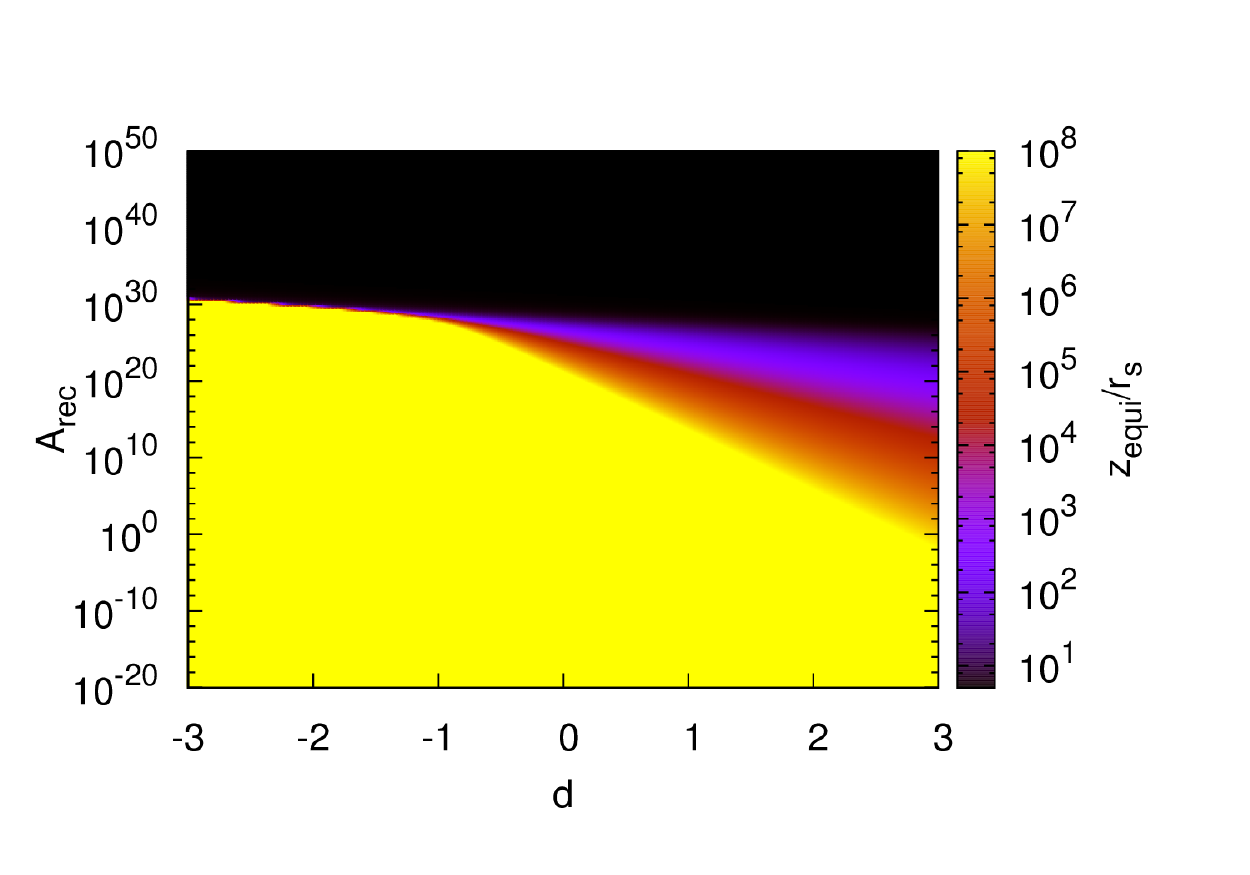} }
		\subfloat[$f_{\m{Edd}}=0.1$.]{ \includegraphics[width=6.0cm, clip=true, trim=0.3cm 0cm 0.5cm 1cm]{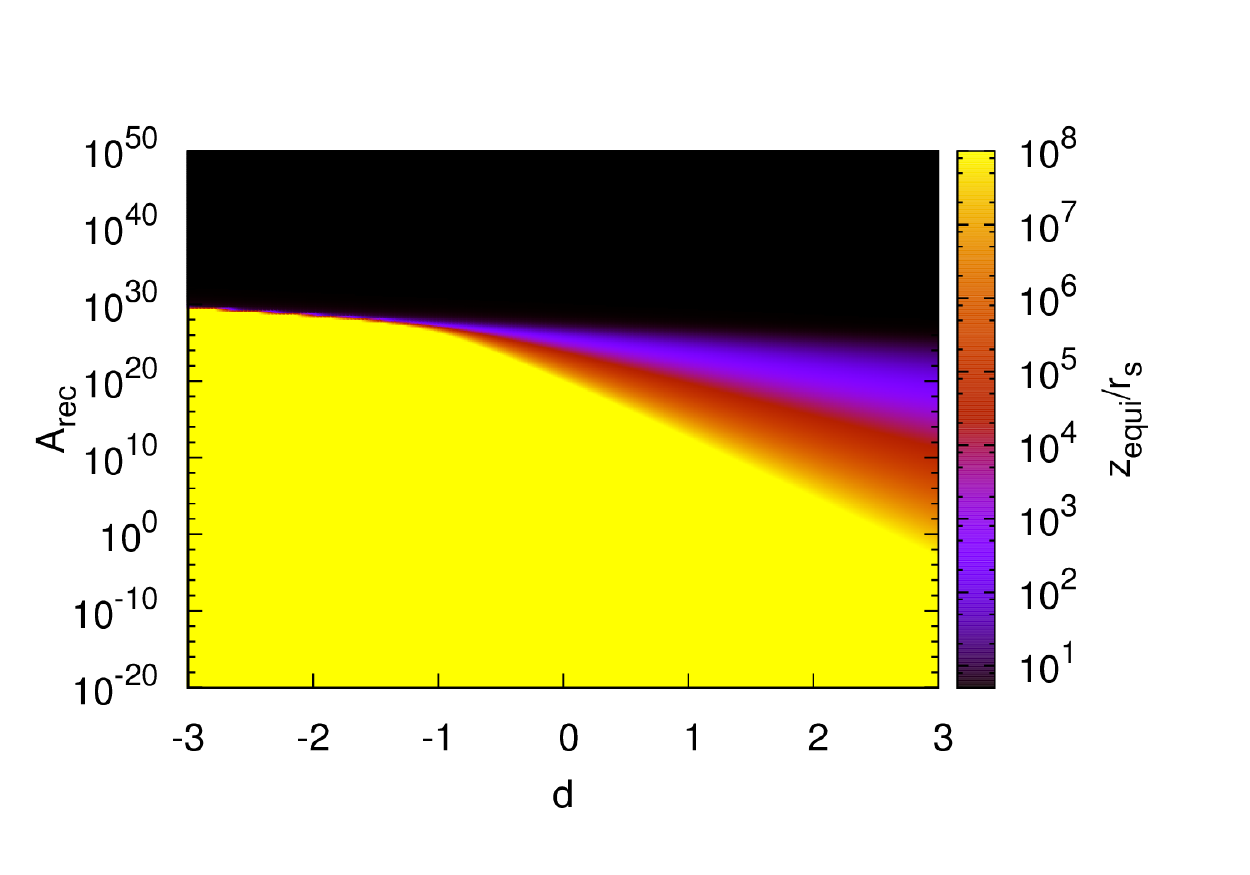} }
\\
		\subfloat[$f_{\m{Edd}}=0.01$.]{ \includegraphics[width=6cm, clip=true, trim=0.3cm 0cm 0.5cm 1cm]{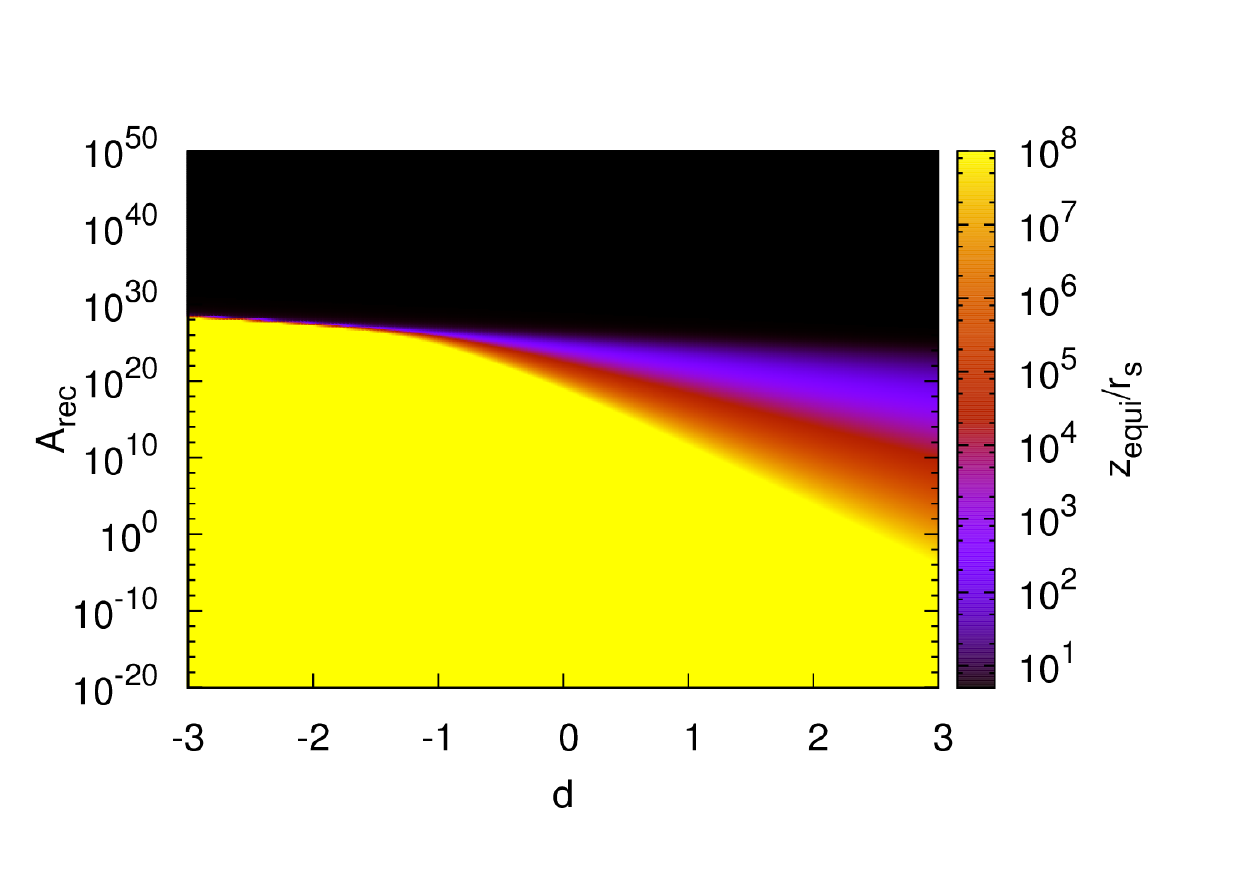} }
		\subfloat[$f_{\m{Edd}}=10^{-3}$.]{ \includegraphics[width=6cm, clip=true, trim=0.3cm 0cm 0.5cm 1cm]{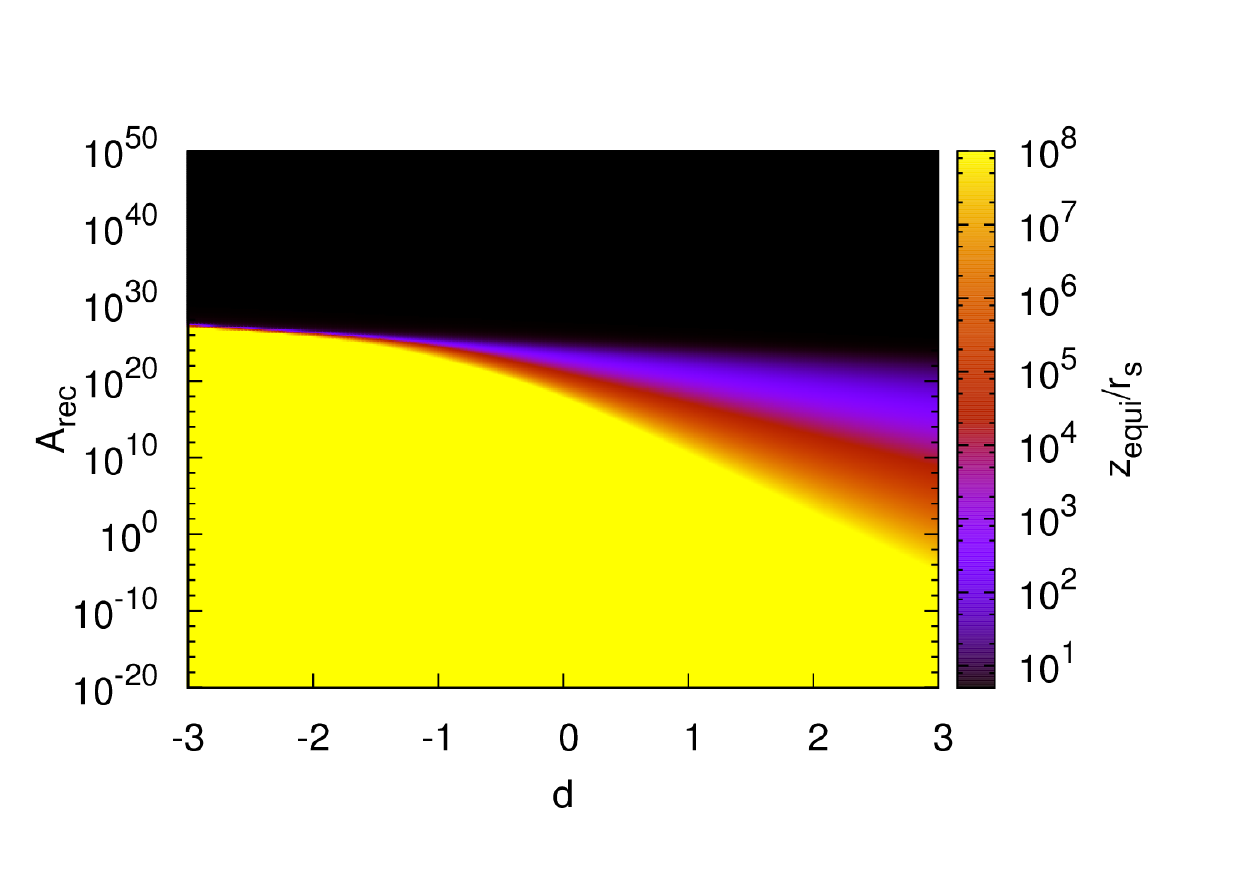} }	
		\subfloat[$f_{\m{Edd}}=10^{-4}$.]{ \includegraphics[width=6cm, clip=true, trim=0.3cm 0cm 0.5cm 1cm]{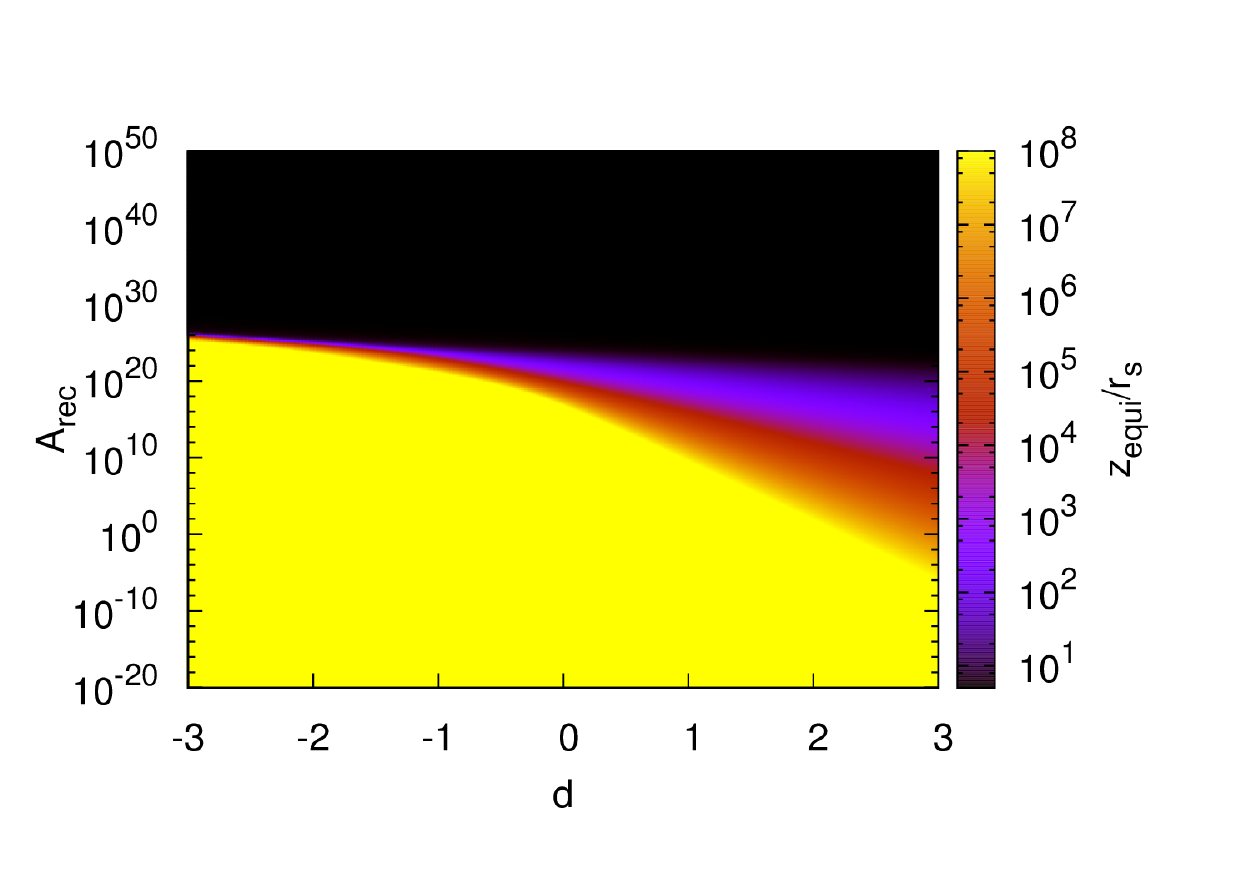} }
	\caption{Figures $a-f$ show the distance travelled by the jet before reaching equipartition for a range of values of the reconnection parameters $A_{\m{rec}}$ and $d$, from (\ref{P'rec}) and fractional Eddington jet power $f_{\m{Edd}}$. We stop integration of the jet evolution equations at a distance of $10^{8}r_{s}$ if the jet plasma has not yet reached equipartition, since radiative energy losses at larger distances than this are insignificant. The jet parameters used to obtain these results are shown under Model C in Table \ref{Table2} and the radiative losses corresponding to these models are shown in figure \ref{rec_params}. We find that the dimensionless distance to equipartition, $X_{\m{equi}}=z_{\m{equi}}/r_{\m{s}}$, depends weakly on the fractional Eddington accretion rate, $f_{\m{Edd}}$. The distance to equipartition decreases slowly with increasing $f_{\m{Edd}}$. These results are independent of black hole mass, $M$.   }
\label{rec_paramsz}
\end{figure*}
\begin{figure*}
	\centering
		\subfloat[$f_{\m{Edd}}=10$.]{ \includegraphics[width=9cm, clip=true, trim=0cm 0cm 0cm 1cm]{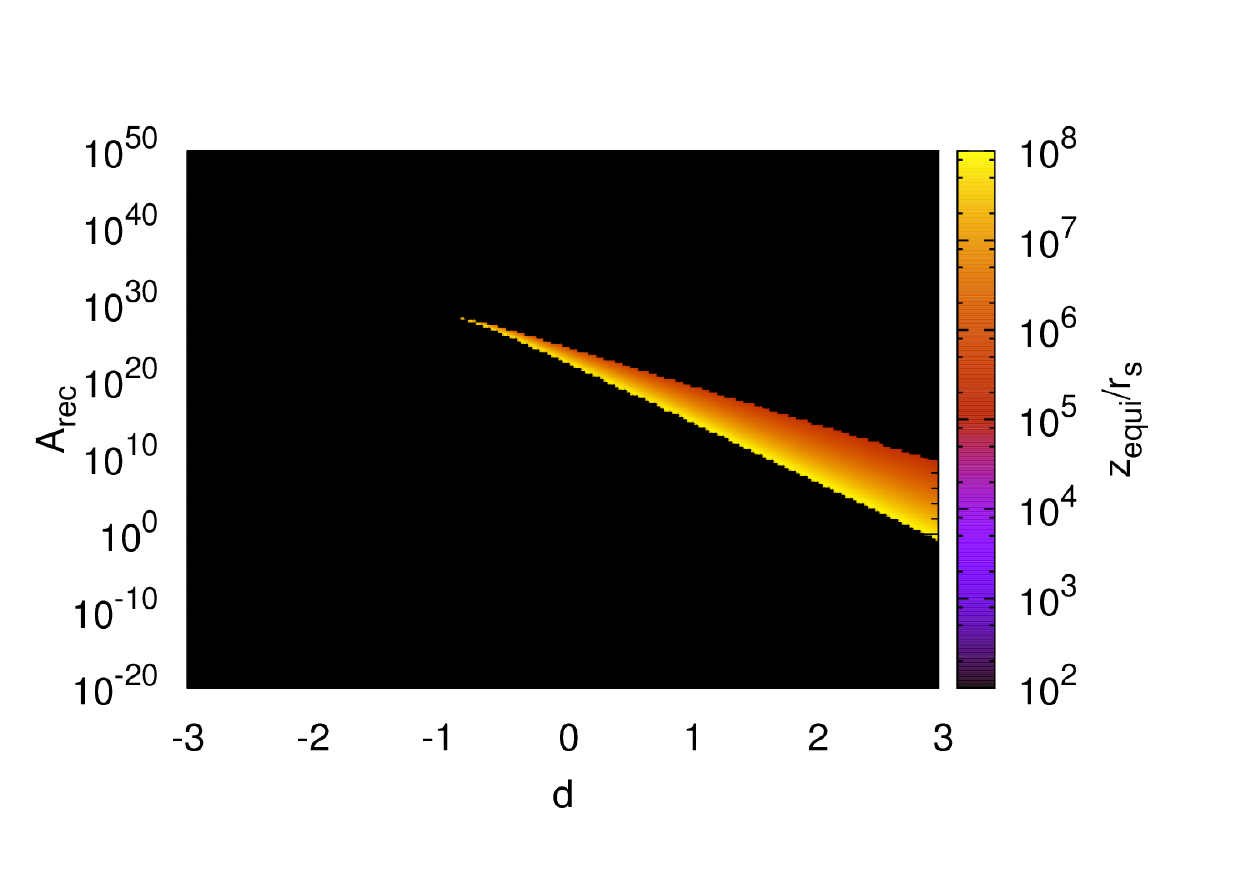} }
		\subfloat[$f_{\m{Edd}}=1$.]{ \includegraphics[width=9cm, clip=true, trim=0cm 0cm 0cm 1cm]{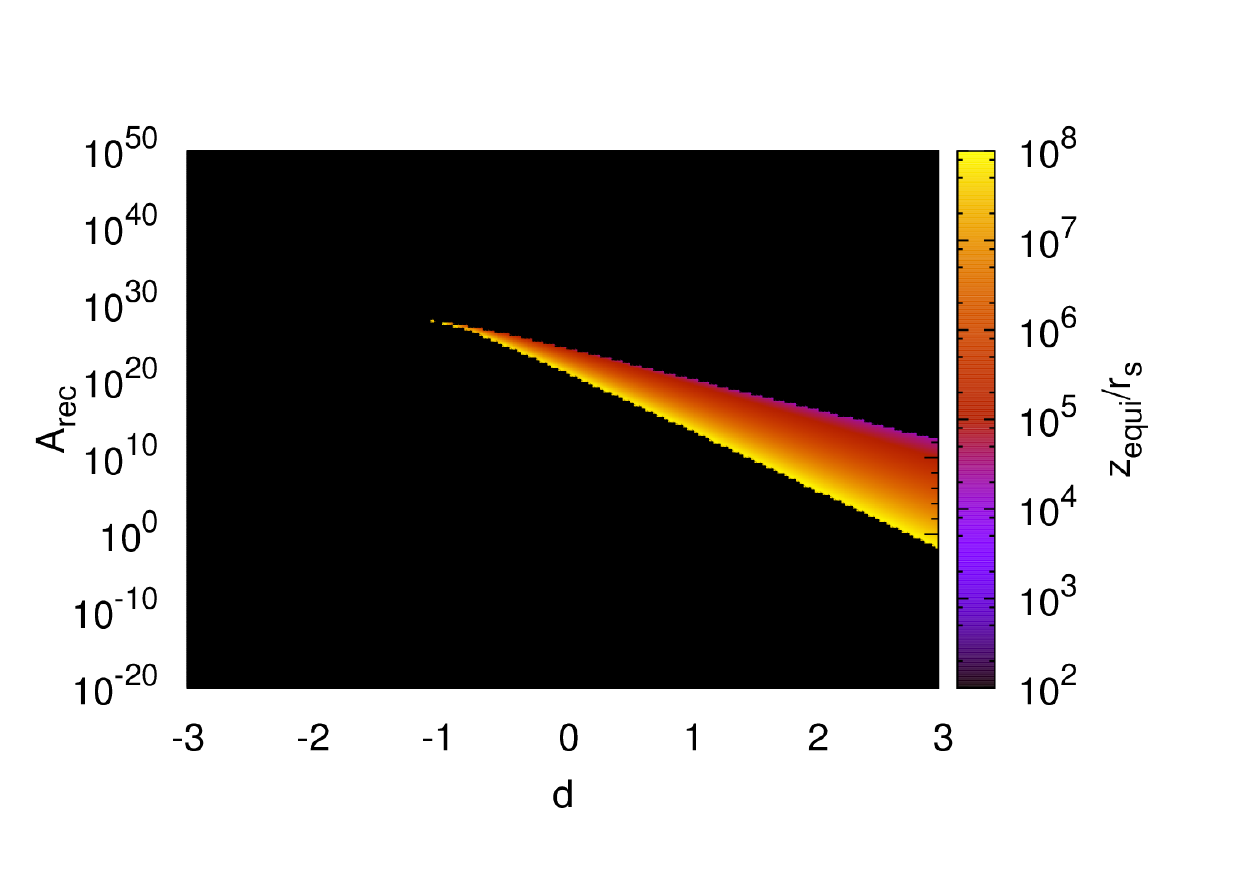} }
\\
		\subfloat[$f_{\m{Edd}}=0.1$.]{ \includegraphics[width=9cm, clip=true, trim=0cm 0cm 0cm 1cm]{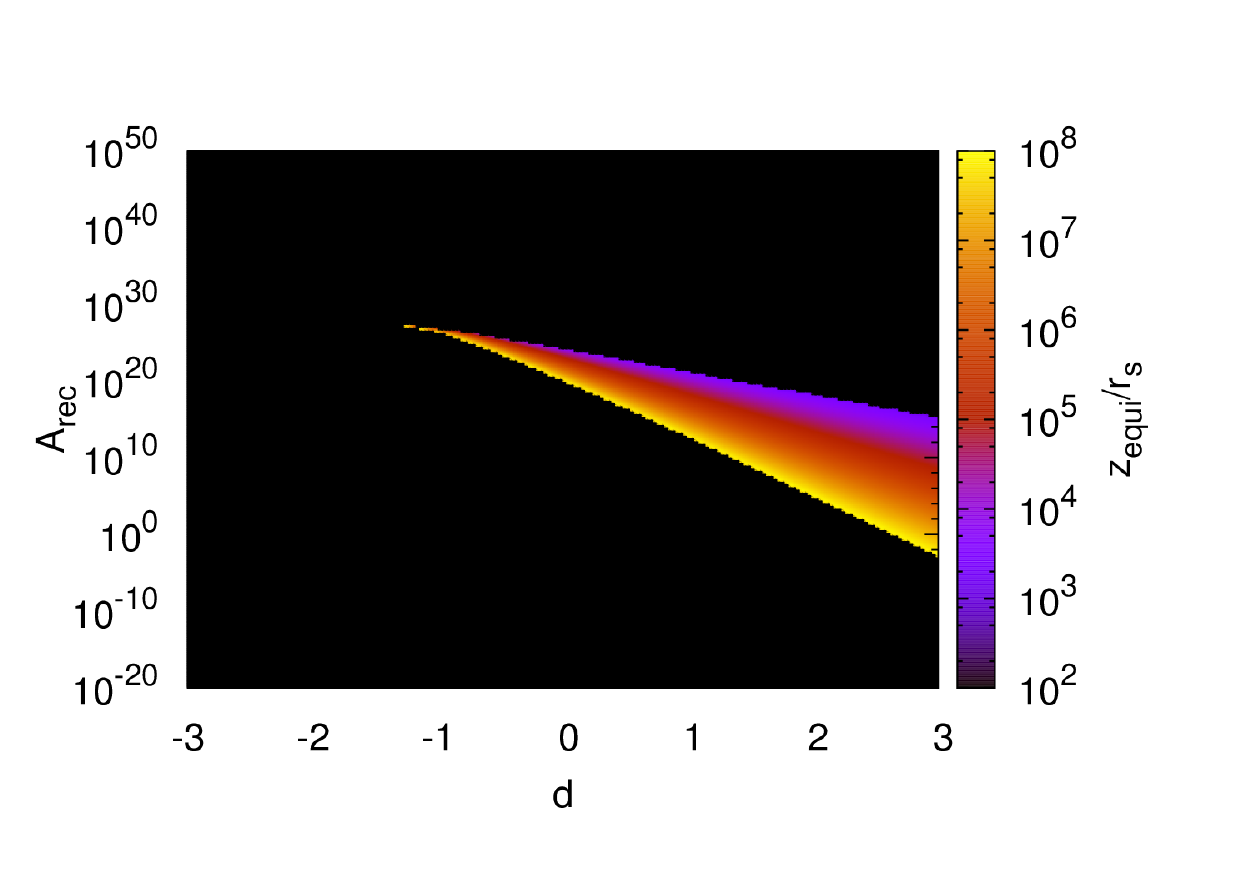} }
		\subfloat[$f_{\m{Edd}}=0.01$.]{ \includegraphics[width=9cm, clip=true, trim=0cm 0cm 0cm 1cm]{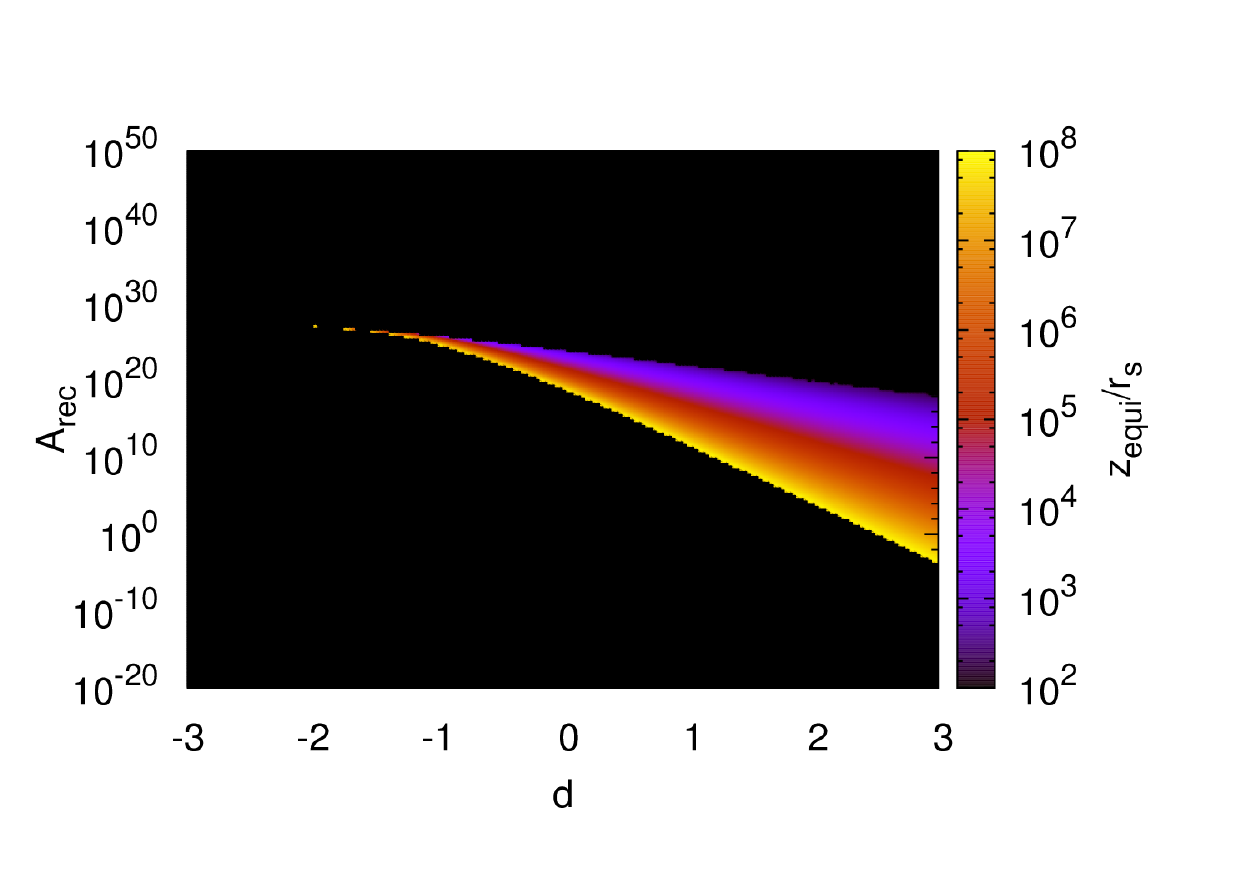} }
\\
		\subfloat[$f_{\m{Edd}}=10^{-3}$.]{ \includegraphics[width=9cm, clip=true, trim=0cm 0cm 0cm 1cm]{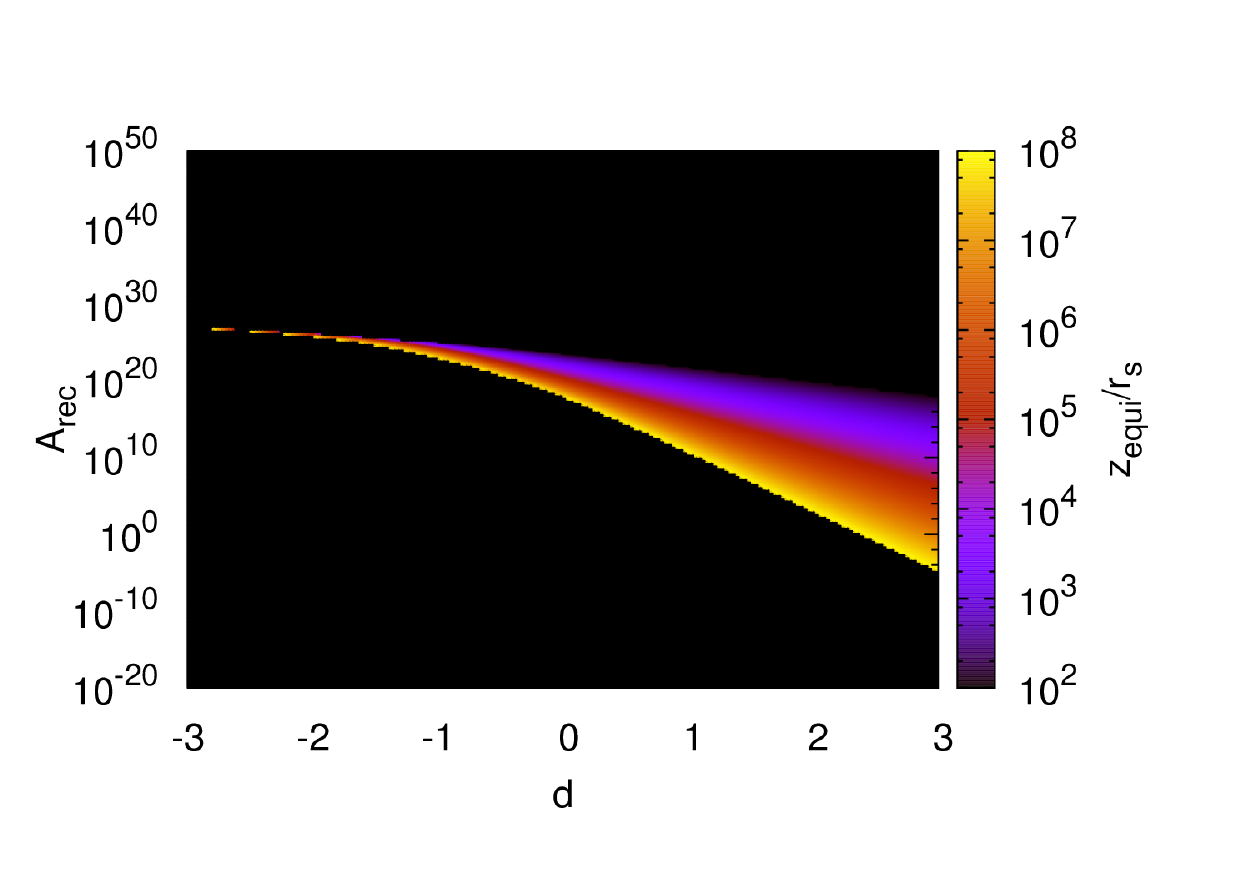} }	
		\subfloat[$f_{\m{Edd}}=10^{-4}$.]{ \includegraphics[width=9cm, clip=true, trim=0cm 0cm 0cm 1cm]{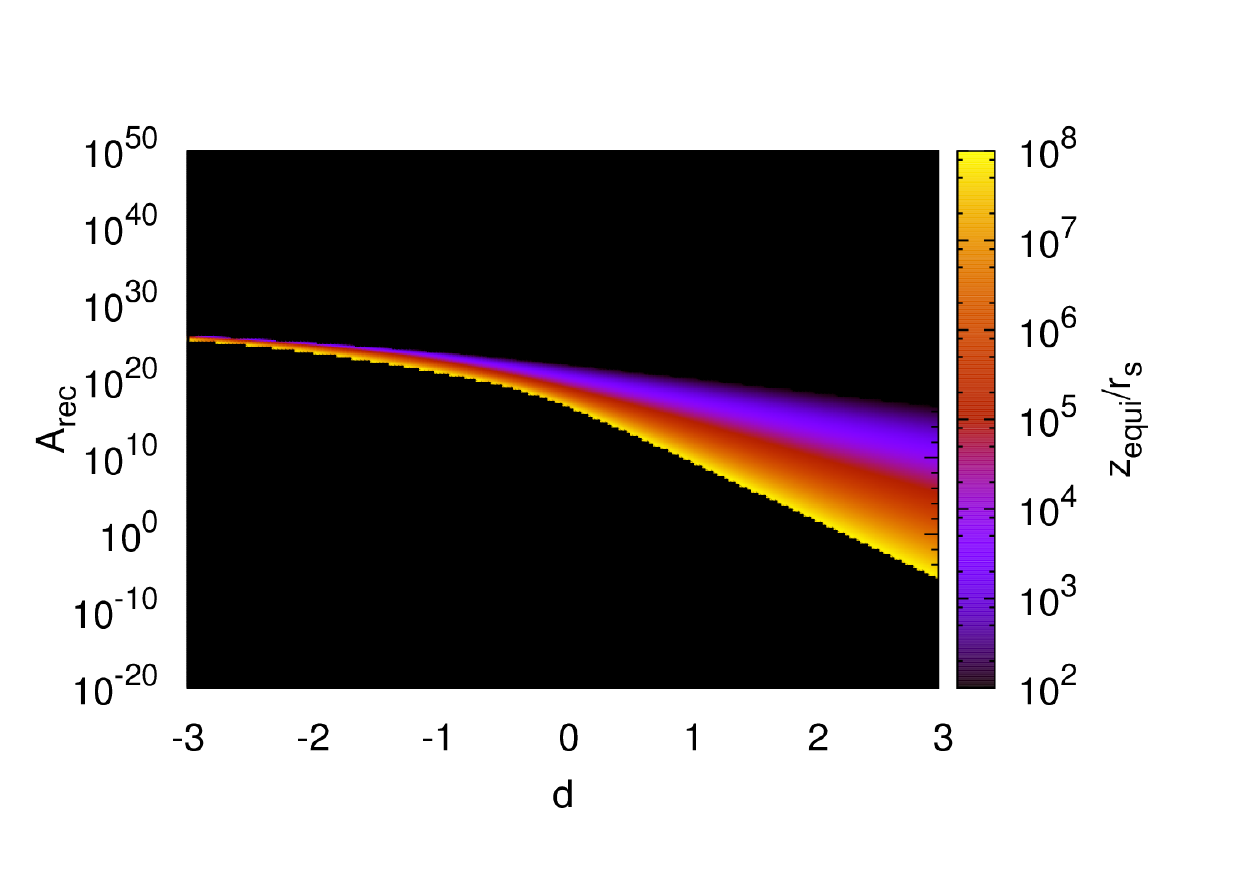} }	

	\caption{The desirable region of parameter space for jets with different fractional Eddington luminosities and reconnection parameters from Figures \ref{rec_params} and \ref{rec_paramsz}. Here {\lq}desirable{\rq} means that the jet reaches equipartition at a distance, $z_{\m{equi}}$, between $100r_{\m{s}}-10^{8}r_{\m{s}}$, whilst not radiating more than $95\%$ of its initial energy in reaching this point. The unfavoured regions of parameter space which do not fulfill these criteria are coloured in black. The jet parameters used to obtain these results are shown under Model C in Table \ref{Table2}. Due to more severe radiative losses at higher fractional Eddington jet powers ($f_{\m{Edd}}$), the desirable region becomes more restrictive for higher values of $f_{\m{Edd}}$. This means that lower power jets are able to reach equipartition at smaller distances whilst retaining at least $5\%$ of their initial energy. For jet powers close to the Eddington luminosity ($f_{\m{Edd}}>0.1$) the strong radiative losses essentially prohibit the jet from coming into equipartition at a distance smaller than $\sim10^{4}r_{\m{s}}$, since otherwise the jet radiates away more than $95\%$ of its initial power in the compact base region with high magnetic field strengths. This is consistent with our earlier results in Figure \ref{spectra}, in which we found that the base of the jet must be highly magnetised in order not to suffer from devastating radiative energy losses. We also see that large negative values of the reconnection power law $d$ are effectively prohibited since reconnection occurring predominantly at the base of the jet will result in heavy radiative losses in the strong magnetic fields. These results are independent of black hole mass.  }
\label{rec_params2}
\end{figure*}
\begin{figure*}
	\centering
		\subfloat[The fraction of initial total energy radiated.]{ \includegraphics[height=6cm, clip=true,  trim=0cm 0cm 0cm 0cm]{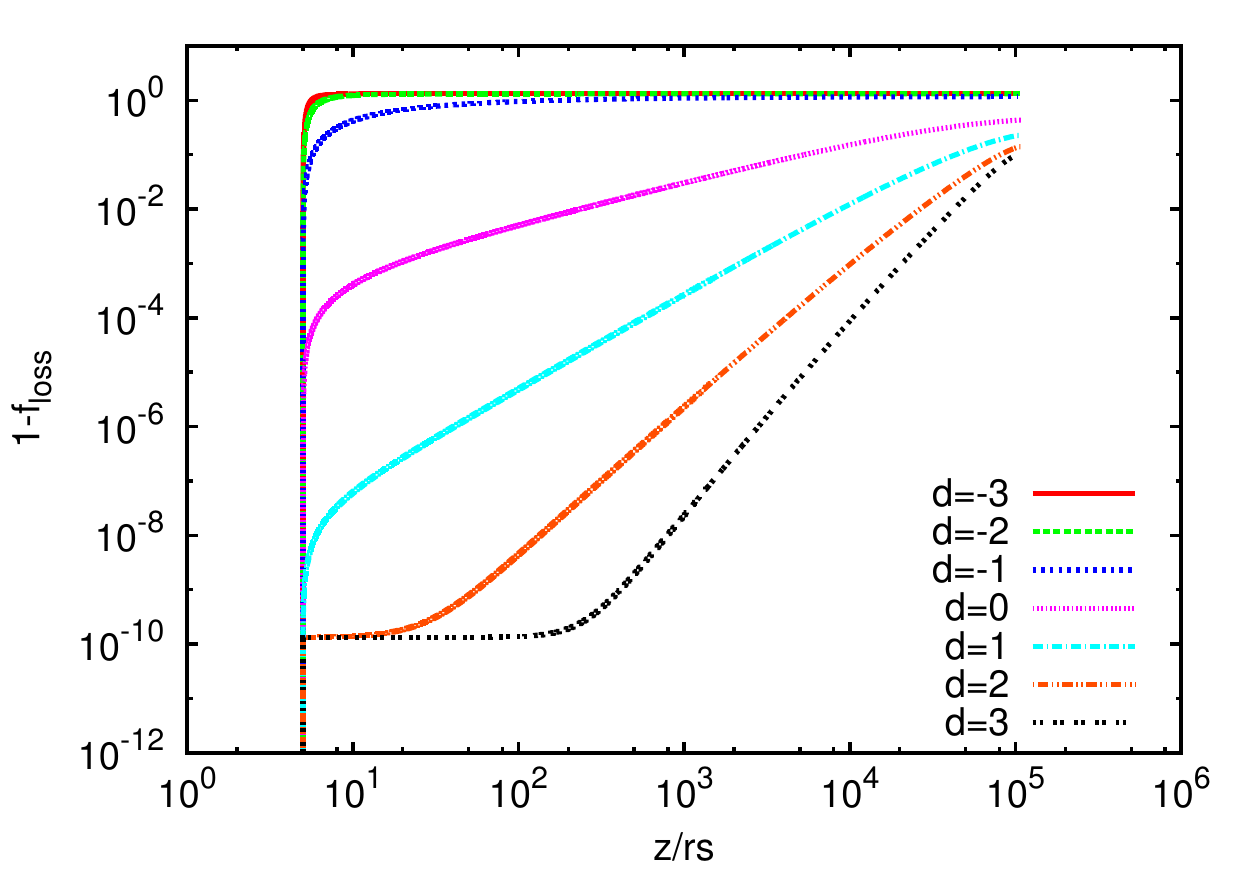} }\,\,\,\,\,\,
		\subfloat[The ratio of magnetic to non-thermal electron-positron energy densities.]{ \includegraphics[height=6cm, clip=true,  trim=0cm 0cm 0cm 0cm]{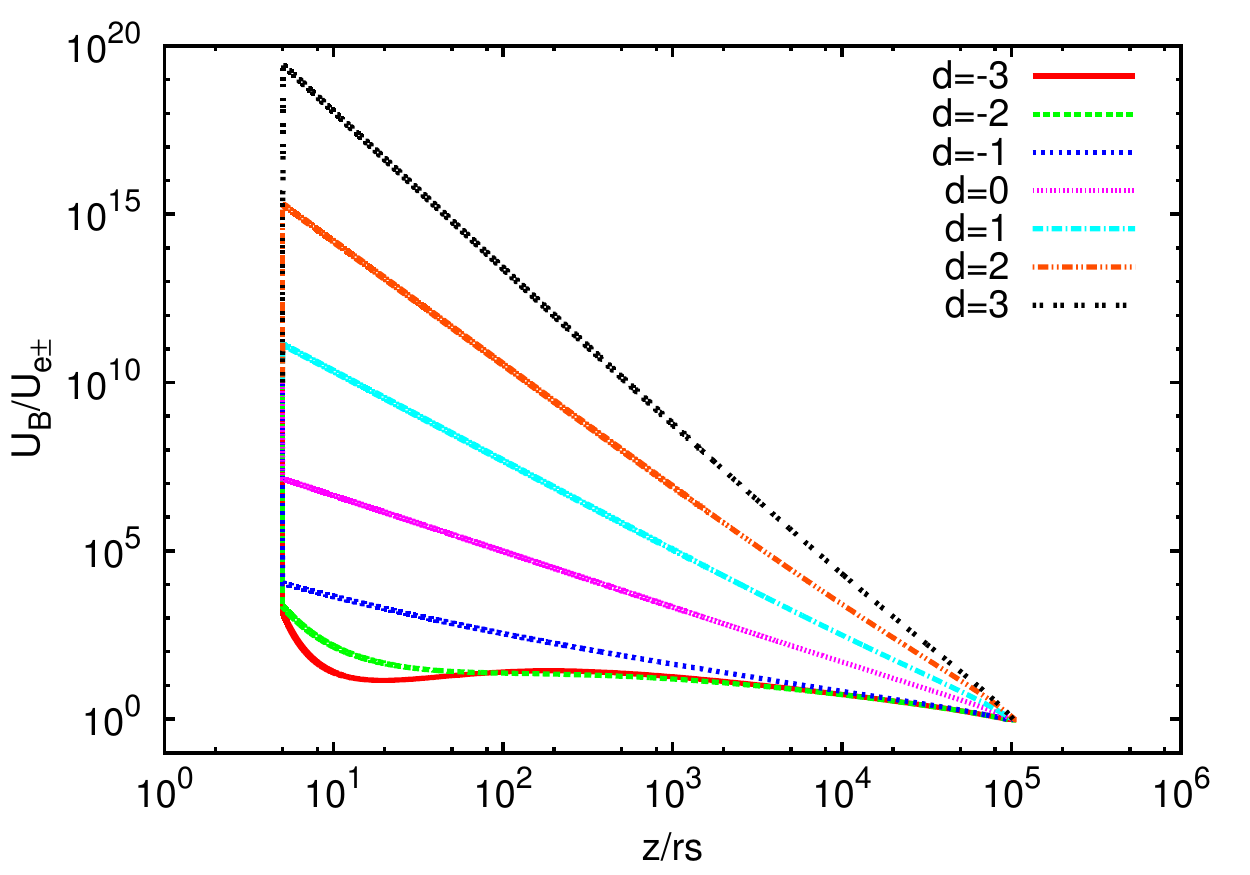} }
\\
		\subfloat[Cumulative reconnection energy density as a fraction of the remaining total energy density.]{ \includegraphics[height=6cm, clip=true,  trim=0cm 0cm 0cm 0cm]{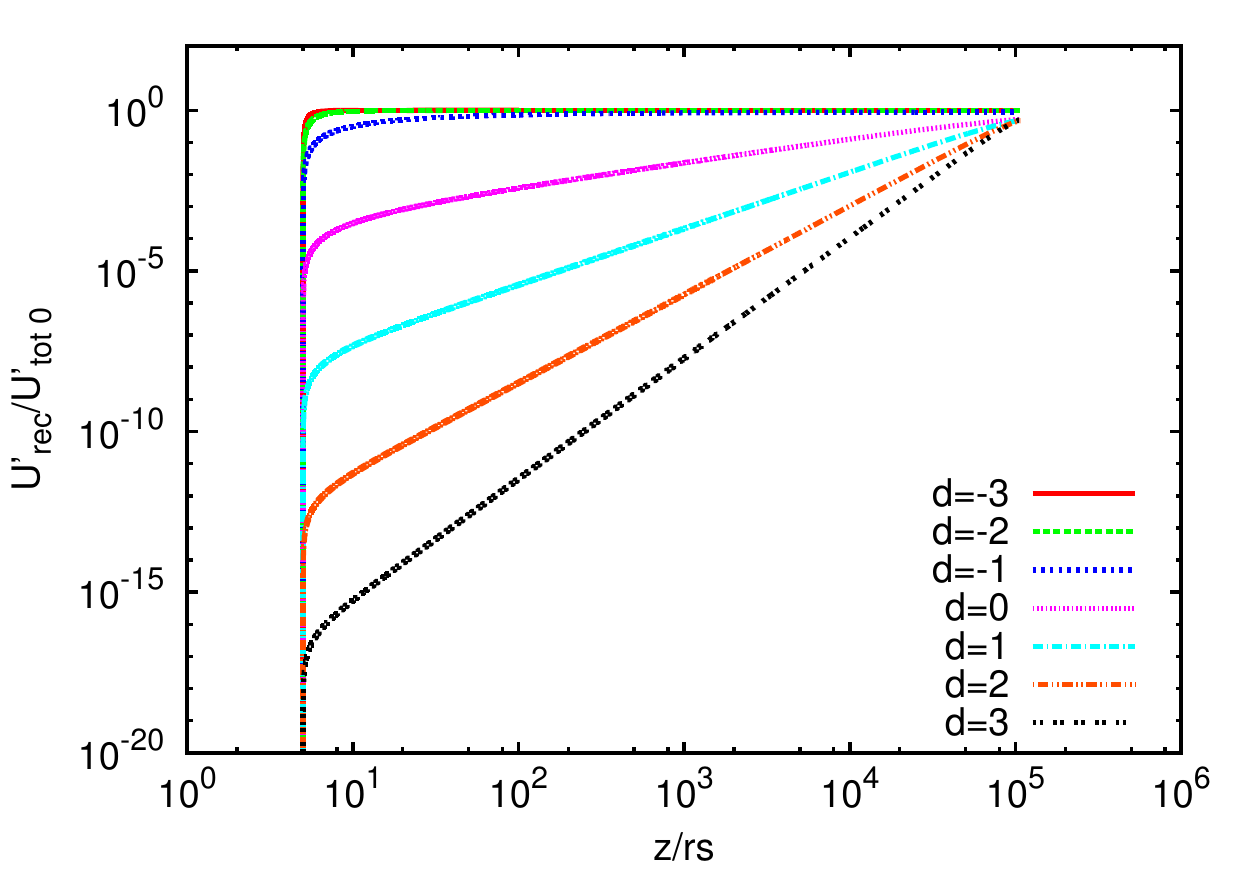} } \,\,\,\,\,\,
		\subfloat[Cumulative reconnection energy contained in a cylindrical shell of unit width, measured in the lab frame.]{ \includegraphics[height=6cm, clip=true, trim=0cm 0cm 0cm 0cm]{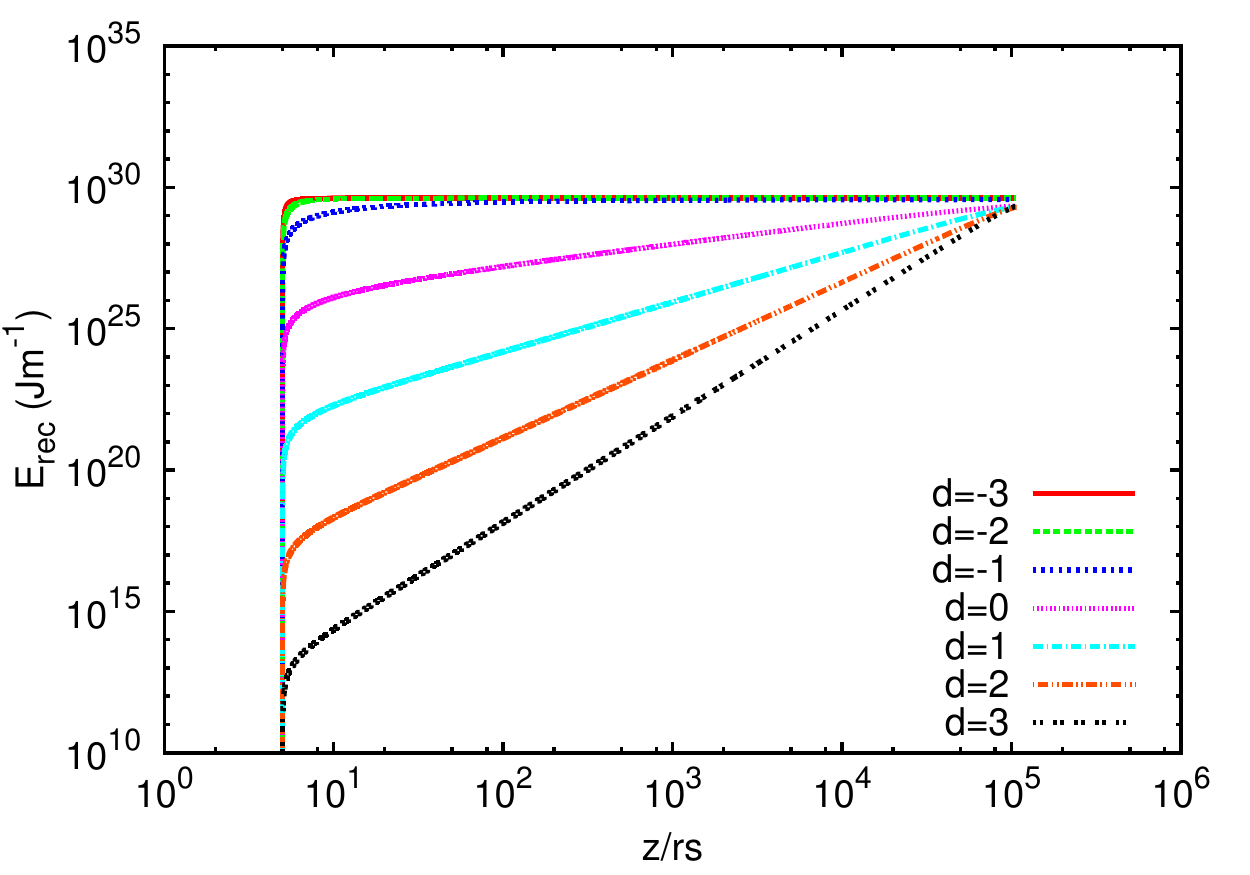} }
	\caption{Figures a-d show the spatial dependence of the radiative energy losses, magnetisation and reconnection rate along the jet for different values of the reconnection power-law distance dependence, $d$. The value of, $A_{\m{rec}}$, has been chosen such that all the models first come into equipartition at a similar distance $z\sim 10^{5}r_{\m{s}}$, to facilitate an easy comparison for the reader. The jet parameters used to obtain these results are shown under Model D in Table \ref{Table2}. From figures c and d we see that the models for which $d\ll 0$ have reconnection occurring mainly at the base of the jet and, apart from different values of $A_{\m{rec}}$, have very similar spatial dependencies because all reconnection occurs in approximately the same region. For $d>0$ the amount of energy from reconnection increases along the jet. With the distance to equipartition fixed, the fractional radiative energy losses of the different models decrease with increasing $d$ as shown in figure a. This is because larger values of $d$ correspond to less non-thermal particle acceleration occurring at the very base of the jet where radiative losses are most severe. In order to balance having enough particle acceleration at the jet base to emit the observed radio synchrotron, whilst not accelerating too many non-thermal electrons such that radiative losses are too severe, these results favour a model with $-1<d<1$ corresponding to an approximately even energy injection of non-thermal electrons per logarithmic distance interval. Our assumed model for the dependence of the bulk Lorentz factor on distance $\gamma_{\m{bulk}}=0.8(z/r_{s})^{1/4}$ based on the blazar spectral fitting of \citealt{2015MNRAS.453.4070P}, gives jet bulk Lorentz factors of $\approx14$ at $z\approx10^{5}r_{s}$, where these models reach equipartition. }
\label{rec_individual}
\end{figure*}
We substitute the above expression into (\ref{magevol}) to find the equation for the evolution of the fractional magnetic energy. To calculate the evolution of the fractional magnetic energy, $f_{\m{B}}$, (\ref{UBprimed}) and radiative losses we need to solve the following set of simultaneous differential equations (we have also included the equation for the evolution of $f_{\m{loss}}$ (\ref{floss}) for convenience), subject to our chosen boundary conditions: $f_{\m{loss}}(z_{0})=1$, $f_{\m{B}}(z_{0})=(1-10^{-10})$ and $E'_{\m{rec}}(z_{0})=0$, and using equations \ref{P'synch}, \ref{P'SSC}, \ref{P'EC} and \ref{P'rec} for $P'_{\m{synch}}$, $P'_{\m{SSC}}$, $P'_{\m{EC}}$ and $P'_{\m{rec}}$, respectively. 
\be
\frac{\partial f_{\m{B}}}{\partial z}=-f_{\m{B}}\frac{\partial \ln f_{\m{loss}}}{\partial z} -2(1-f_{\m{B}})\frac{\partial  \ln \gamma_{\m{bulk}}}{\partial z} - \frac{4\gamma_{\m{bulk}}P'_{\m{rec}}}{3P_{\m{j}}f_{\m{loss}}}, \label{magevol2}
\ee
\be
\frac{\partial f_{\m{loss}}}{\partial z}=-\frac{\gamma_{\m{bulk}}(P'_{\m{synch}}+P'_{\m{SSC}}+P'_{\m{EC}})}{P_{\m{j}}},
\ee
In the next section we shall solve these equations numerically in order to understand the physical effect of the dissipated reconnection power, $P'_{\m{rec}}$, and to constrain the allowed values of the reconnection parameters $A_{\m{rec}}$ and $d$ defined in (\ref{P'rec}). It is worth noting that our choice of the parameter $A_{\m{rec}}$ allows our numerical results in Figures \ref{rec_params}, \ref{rec_paramsz} and \ref{rec_params2} to be independent of black hole mass (i.e. for a given numerical value of $A_{\m{rec}}$ the results will be independent of black hole mass). This can be seen by writing equation \ref{magevol2} out explicitly in terms of $A_{\m{rec}}$
\be
\frac{\partial f_{\m{B}}}{\partial z}=-f_{\m{B}}\frac{\partial \ln f_{\m{loss}}}{\partial z} -2(1-f_{\m{B}})\frac{\partial  \ln \gamma}{\partial z} - \frac{\gamma_{\m{bulk}}A_{\m{rec}}f_{\m{B}}X^{d-2c}}{P_{\m{j}}}, \label{magevol2}
\ee
Each term has the same, $1/M$, dependence on black hole mass since, $z\propto XM$, and $P_{\m{j}}\propto f_{\m{Edd}}M$, so the dependences cancel and the numerical results in figures \ref{rec_params}, \ref{rec_paramsz} and \ref{rec_params2} are independent of black hole mass. However, it can be seen from equation \ref{P'rec} that if we wish to interpret, $A_{\m{rec}}$, in terms of the reconnection parameters, then, $A_{\m{rec}}\propto P_{\m{j}} \propto f_{\m{Edd}}M$ and so although our numerical results are independent of black hole mass, their interpretation in terms of the values of the reconnection coefficients $\beta_{\m{rec}}$, $\eta_{\m{rec}}$ and $S'_{\m{rec}}$ are not. 

In terms of the equipartition fraction or magnetisation, which we define as, $\sigma=U_{\m{B}}/U_{\m{e}\pm}$, the equation becomes
\bea
&&\hspace{-0.5cm}\frac{\partial \sigma}{\partial z}=-\sigma(1+\sigma)\frac{\partial \ln f_{\m{loss}}}{\partial z} -2(1+\sigma)\frac{\partial  \ln \gamma}{\partial z}...\nonumber \\&&\hspace{2.5cm} -\sigma(1+\sigma) \frac{\gamma_{\m{bulk}}A_{\m{rec}}X^{d-2c}}{P_{\m{j}}}.
\eea
It can be more convenient to use this expression when solving the evolution equations numerically, if, $U_{\m{B}}/U_{\m{e}\pm}$, becomes very large. 
\subsection{Constraining the rate of reconnection in the jet}

Finally, we try to understand how the dependence of the reconnection rate influences the magnetisation and radiative energy losses in jets. We solve the jet fluid equations \ref{floss} and \ref{magevol2} numerically using a Bulirsch-Stoer alogrithm with adaptive step sizes for a range of values of $A_{\m{rec}}$ and $d$, choosing jet parameters given by Model C in Table \ref{Table2}. We integrate the fluid equations from the base of the jet at, $z=z_{0}$, up to the shortest distance out of either: the distance where the jet first comes into equipartition (where we expect the geometry to become conical and acceleration to cease e.g. \citealt{2006MNRAS.368.1561M}, \citealt{2007MNRAS.380...51K} and \citealt{2009MNRAS.394.1182K}), or $10^{8}r_{\m{s}}$. Figure \ref{rec_params} shows the total fractional radiative energy losses ($f_{\m{loss}}=1$ or $0$ correspond to no radiative energy losses and complete radiative energy losses respectively) experienced by the jet plasma when it first reaches either equipartition, or $z=10^{8}r_{s}$, for a variety of values of the reconnection parameters $A_{\m{rec}}$ and $d$. The corresponding distance at which the jet plasma first reaches equipartition is shown in figure \ref{rec_paramsz}. In Figure \ref{rec_params2} we show a conservative estimate of the viable range of reconnection parameters, $A_{\m{rec}}$ and $d$, for which jets reach equipartition at a distance similar to that found in blazars $100r_{\m{s}}-10^{8}r_{\m{s}}$, with maximum bulk Lorentz factors between $2.5-80$ (this is slightly conservative compared to the range values found by \citealt{2015MNRAS.453.4070P} $10^{3}r_{\m{s}}-10^{7}r_{\m{s}}$ and $8<\gamma_{\m{bulk}}<60$), whilst simultaneously not radiating away more than $95\%$ of their total energy, $f_{\m{loss}}>0.05$. This is again a conservative constraint since observations suggest $f_{\m{loss}}>0.85$, \citealt{2012Sci...338.1445N}). We find that larger fractional Eddington accretion rates lead to substantially increased radiative losses due to the higher magnetic field strengths, resulting in a more restrictive region of {\lq}desirable{\rq} reconnection parameters and requiring the jet to first come into equipartition at larger distances to avoid the heavy radiative losses closer to the jet base. 

In Figure \ref{rec_individual} we show the magnetisation and fractional radiative energy losses as a function of distance for a few values of the power-law $z$-dependence of the reconnection rate, $d$. If the power law exponent $d\ll0$, the majority of reconnection occurs predominantly at small distances along the jet, close to the base. If $d\sim c-1$ the power dissipated by reconnection is spread approximately evenly per log bin in distance. This is because the energy dissipated by magnetic reconnection is given by (\ref{dErad})
\be
\frac{\pd E_{\m{rec}}}{\pd z}=\frac{4\gamma_{\m{bulk}}P'_{\m{rec}}}{3c}\propto z^{d-c},\nonumber
\ee
\be
\Delta E_{\m{rec}} \propto \int_{X_{1}}^{X_{2}}X^{d-c}dX.
\ee
In the case where $d-c=-1$, the amount of reconnected energy, $\Delta E_{\m{rec}}$, is approximately proportional to, $\ln(X_{2}/X_{1})$, and so the reconnected energy will be evenly distributed per logarithmic distance interval i.e. the same energy is dissipated between a distance of $10r_{\m{s}}-100r_{\m{s}}$ as $100r_{\m{s}}-1000r_{\m{s}}$. Finally, if $d\gg0$, reconnection occurs predominantly at large distances along the jet. Since synchrotron emission is observed throughout the base of jets (e.g. \citealt{2013ApJ...775...70H}), where radiative lifetimes are much shorter than the travel time along the jet, we know that there must be some small, but non-negligable, amount of particle acceleration at the base. Power-laws $d\gg 0$, produce a negligible amount of particle acceleration at small distances and so are unlikely to be compatible with these radio observations. For values of $d\ll0$, reconnection occurs mostly at small distances along the jet and so radiative energy losses are so large that this case is also not viable, as can be seen from figures \ref{rec_params}, \ref{rec_params2} and \ref{rec_individual}. This suggests that values $-1<d<1$ are most likely to be appropriate if reconnection is responsible for accelerating non-thermal particles in jets. This corresponds to magnetic reconnection which dissipates a similar amount of energy into accelerating non-thermal electrons per logarithmic bin in distance. 

It is useful to interpret the constraints on $A_{rec}$ in terms of the reconnection parameters in (\ref{P'rec}). We expect the effective reconnecting surface area per unit volume $S'_{\m{rec}\,_0}$ and the efficiency of dissipation of magnetic energy flowing into the reconnecting surface $\eta_{\m{rec}\,0}$ to be the most interesting and poorly understood quantities. We assume that the inflow speed of the plasma into the reconnecting current sheet has $\beta_{\m{rec}\,0}\approx0.1$ and the bulk Lorentz factor of the flow at the base of the jet is only mildly relativistic $\gamma_{\m{bulk}\,0}\approx1$. Given these assumptions we find
\be
A_{rec}=1.3\times10^{30} \gamma_{\m{bulk}}^{-2}S'_{\m{rec}}\eta_{\m{rec}}f_{\m{Edd}}\frac{M}{M_{\odot}}
\ee
We have argued that favourable parameters for producing a typical AGN jet would be $d\approx0$, for which the $f_{\m{Edd}}=1$ constraints suggest $A_{\m{rec}}\sim10^{22}$.
\be
S'_{\m{rec}}\eta_{\m{rec}}\sim 7.7\times10^{-9} \gamma_{\m{bulk}}^{2}\left(\frac{M}{M_{\odot}}\right)^{-1}
\label{effectiverecarea}
\ee
This estimate suggests that reconnection regions occupy a relatively small but significant volume filling factor in the jets. This is consistent with our expectation, since we know that reconnection does not quickly dissipate all of the magnetic energy at the jet base (which would be implied if $S'_{\m{rec}}\eta_{\m{rec}}\gtrsim1/(\beta_{\m{rec}}r_{\m{s}}$), see equation \ref{zrec}) and that in models where it is responsible for in-situ particle acceleration it should be contributing significantly throughout the jet. From (\ref{effectiverecarea}) it is also clear that in order for jets originating from different mass black holes to have the same properties (scaled linearly with mass), the number/size of reconnection surfaces per unit volume should scale inversely with black hole mass. This would be consistent with supermassive black hole jets having the same magnetic field structure as stellar mass black holes but scaled with the mass of the black hole (or scaled with Schwarzschild radii). It is also useful to look at the dimensionless dissipation lengthscale over which a substantial fraction of the magnetic energy is dissipated by reconnection, $z_{\m{rec}}/r_{s}$. We estimate this using equation \ref{dErad}, replacing $\partial E_{\m{rad}}/\partial z$ by $E_{B}/z_{\m{rec}}$ and $P'_{\m{rad}}$ with $P'_{\m{rec}}$ to estimate the reconnection lengthscale
\be
\frac{z_{\m{rec}}}{r_{s}}\sim\frac{\gamma_{\m{bulk}}}{\beta_{\m{rec}}\eta_{\m{rec}}S'_{\m{rec}}r_{s}}\sim\frac{4.3\times10^{27}f_{\m{Edd}}}{\gamma_{0}A_{\m{rec}}X^{d-c}}\label{zrec}
\ee
Here we see that the dimensionless dissipation lengthscale does not explicitly depend on black hole mass and for $f_{\m{Edd}}=1$, $A_{rec}=10^{22}$, we find $z_{\m{rec}}/r_{s}\approx10^{5}$.
 
Our results suggest that in order for the jet base not to suffer catastrophic non-thermal radiative energy losses the plasma must be initially highly magnetised. Naively we might expect this high magnetisation should lead to the jet accelerating to very large terminal bulk Lorentz factors ($\sim1/\sigma_{0}$) much larger than those observed (this would be the case if high initial magnetisations $\sigma>10^{2}$ were used in current relativistic MHD simulations which ignore explicit magnetic reconnection and non-thermal radiative losses). In this paper we instead propose a scenario in which the magnetic energy is gradually dissipated by magnetic reconnection occurring along the length of the jet, which is responsible for producing the continuous in-situ acceleration of non-thermal electrons whose non-thermal emission is observed. This results in the magnetic energy being converted both into bulk acceleration and particle acceleration, with particles constantly radiating energy away. It is then the balance between these processes which determines the total radiative losses, evolution of the magnetisation and terminal bulk Lorentz factor of the jet. We have then investigated what range of initial magnetisations and magnetic reconnection rates are compatible with observations which constrain: the total fraction of initial jet power which should be retained to large distances, the likely range of maximum bulk Lorentz factors and the distances over which the jet accelerates before reaching equipartition. 

These results show that it is possible to understand and constrain the microscopic reconnection physics, which is likely to occur in black hole jets, by considering the effect of reconnection on the macroscopic jet properties. In particular, we have demonstrated that by only considering the effect of radiative energy losses on the jet plasma and the distance at which the jet plasma reaches equipartition we can constrain the large-scale average reconnection parameters. This is important because it is likely that the most important quantities determining the macroscopic reconnection rate in jets are the large-scale average efficiency of magnetic dissipation and the volume filling factor of reconnecting regions in the jet. It is not possible to obtain these quantities from performing a small-scale simulation of a single reconnecting region with prescribed initial conditions and so this work adds important, new information to enable us to better understand the large-scale distribution of reconnecting regions in jets.   
\section{Conclusions}

In this paper we consider the effects of the severe radiative energy losses acting at the base of black hole jets using an inhomogeneous fluid jet model. We demonstrate for the first time that the radiative energy losses from synchrotron and synchrotron self-Compton emission close to the base are so severe, and the corresponding electron lifetimes so short, that they impose strong constraints on the magnetisation of the jet plasma. The jet plasma at the base must be highly magnetised with only a small fraction of the total energy contained in non-thermal electrons in order for the initial energy in the jet plasma not to be rapidly depleted by radiative energy losses. In the case of a jet with a constant initial magnetisation we calculate easy-to-use analytic expressions for the fraction of the initial jet power which is radiated as a function of distance along the jet using a 1D relativistic fluid jet model. Our fluid jet model conserves relativistic energy-momentum and particle number flux whilst allowing for a variable shape, bulk acceleration profile and electron energy distribution for the jet base.

We derive analytic and numerical constraints on the the allowed ratio of magnetic to non-thermal particle energy at the jet base (the magnetisation or equipartition fraction). For typical black hole jet parameters we find the jet base must be very highly magnetised to avoid sustaining excessive radiative energy losses, with $U_{\m{B}}/U_{\m{e}\pm}>5\times 10^{4}f_{\m{Edd}}$, where $f_{\m{Edd}}$ is the fractional Eddington power of the jet. This conservative constraint comes from assuming that not more than $95\%$ of the initial total energy in the jet plasma should be radiated away as the plasma traverses the base of the jet. This requirement of a high magnetisation provides direct evidence in favour of an electromagnetic launching mechanism for jets. These results are independent of the black hole mass and the constraints we find from considering radiative energy losses require the jet plasma to be more highly magnetised at the base than is usually assumed in relativistic MHD simulations of jets. Our results demonstrate the importance of including a self-consistent calculation of radiative energy losses in jet simulations in determining the magnetisation and physical conditions at the base of the jet. This means that jet simulations in which non-thermal emission is calculated via post-processing, and the severe radiative energy losses are not taken into account, are likely to vastly overestimate the emitted synchrotron and inverse-Compton power. 

We then consider the in-situ acceleration process which must be acting along the jet to replenish the non-thermal electron population against the severe radiative energy losses. Since the base of the jet is expected to be magnetically dominated we consider the possibility of in-situ magnetic dissipation, whereby magnetic energy is converted into accelerating non-thermal particles, by magnetic reconnection. We derive a set of fluid equations which conserve energy and particle number flux taking into account radiative energy losses to the electron population, in-situ acceleration and allowing for a variable jet shape and acceleration. We solve these equations numerically allowing for a power-law form of the large-scale average rate of reconnection along the jet, motivated by simulations of collisionless reconnection, in order to place constraints on the reconnection process. 

We constrain the allowed parameter-space of the reconnection rate along the jet by imposing that the jet should not radiate away more than $95\%$ of its initial total energy and should come into equipartition at reasonable distances, $100r_{\m{s}}-10^{8}r_{\m{s}}$, compatible with the optically thick to thin break in synchrotron emission (found by modelling observations of blazar jets \citealt{2015MNRAS.453.4070P}). Furthermore, we find more generally that the mechanism leading to in-situ acceleration should deposit energy close to evenly in logarithmic distance along the jet (i.e. a similar amount of energy is injected into accelerating non-thermal electrons going from $10r_{\m{s}}-100r_{\m{s}}$ as from $100r_{\m{s}}-1000r_{\m{s}}$). This is because if the acceleration process occurs preferentially at the jet base it leads to severe radiative losses (since the magnetic field is stronger at the base radiative lifetimes are shorter) and if the process occurs predominantly at large distances there would be too few high-energy electrons to emit the radio synchrotron observed via VLBI at the base of jets (such as M87, \citealt{2012Sci...338..355D} and \citealt{2013ApJ...775...70H}). We find that for jets with total jet powers greater than $10\%$ of the Eddington luminosity the jet cannot first come into equipartition below a distance of $\sim 10^{4}r_{\m{s}}$ because of the severe radiative energy losses close to the jet base. These constraints help us to better understand the large-scale properties of magnetic reconnection in jets, such as the volume-filling factor of reconnecting regions.

These new and important constraints are the first realistic attempt to use radiative energy losses to constrain the physics at the jet base. In light of our findings it is clear that a self-consistent treatment of the severe radiative energy losses is necessary in order to perform accurate MHD simulations of black hole jets. We hope that the calculations and results in this paper will encourage such work in the near future.

\section*{Acknowledgements}

WJP acknowledges funding in the form of a Junior Research Fellowship from University College, University of Oxford. WJP would like to thank Steven Balbus, Charles Gammie and Alexander Schekochihin for helpful discussions and comments.

\bibliographystyle{mn2e}
\bibliography{Jetpaper2refs}
\bibdata{Jetpaper2refs}
%\bibstyle{mn2e}

\label{lastpage}

\end{document}